\documentclass[twocolumn]{aastex631}
\usepackage{amsmath}
\usepackage{graphicx}
\usepackage{longtable}
\usepackage{footnotehyper}
\makesavenoteenv{longtable}
\usepackage{breqn}
\usepackage{soul}

\definecolor{myblue}{RGB}{0, 160, 240} 
\definecolor{mygreen}{RGB}{0, 180, 0}

\newcommand{\corr}{\textcolor{red}}

\definecolor{newpurple}{rgb}{0.7,0,1}
\defcitealias{Breuval2021}{B21}
\defcitealias{Johnson2022}{J22}
\defcitealias{Pellerin2011}{PM11}
\defcitealias{Sarajedini2007}{S07}

\shorttitle{A 1.3\% distance to M33 from HST Cepheid photometry}
\shortauthors{L.~Breuval, et al.}

\begin{document}

\title{A 1.3\% distance to M33 from HST Cepheid photometry} 

\author[0000-0003-3889-7709]{Louise Breuval}
\affiliation{Department of Physics and Astronomy, Johns Hopkins University, Baltimore, MD 21218, USA}
\email{lbreuva1@jhu.edu}

\author[0000-0002-6124-1196]{Adam G.~Riess}
\affiliation{Department of Physics and Astronomy, Johns Hopkins University, Baltimore, MD 21218, USA}
\affiliation{Space Telescope Science Institute, 3700 San Martin Drive, Baltimore, MD 21218, USA}

\author[0000-0002-1775-4859]{Lucas M.~Macri}
\affiliation{Department of Physics \& Astronomy, Texas A\&M University, College Station, TX 77843, USA}

\author[0000-0002-8623-1082]{Siyang Li}
\affiliation{Department of Physics and Astronomy, Johns Hopkins University, Baltimore, MD 21218, USA}

\author[0000-0001-9420-6525]{Wenlong Yuan} 
\affiliation{Department of Physics and Astronomy, Johns Hopkins University, Baltimore, MD 21218, USA}

\author{Stefano Casertano}
\affiliation{Space Telescope Science Institute, 3700 San Martin Drive, Baltimore, MD 21218, USA}

\author[0000-0003-0452-9182]{Tarini Konchady}
\affiliation{Department of Physics \& Astronomy, Texas A\&M University, College Station, TX 77843, USA}

\author[0000-0001-5875-5340]{Boris Trahin}
\affiliation{Institut d’Astrophysique Spatiale, Universit\'e Paris-Saclay, CNRS, Batiment 121, F-91405 Orsay Cedex, France}

\author[0000-0001-7531-9815]{Meredith J.~Durbin}
\affiliation{Department of Astronomy, University of Washington, Box 351580, U.W., Seattle, WA 98195-1580, USA}

\author[0000-0002-7502-0597]{Benjamin F.~Williams}
\affiliation{Department of Astronomy, University of Washington, Box 351580, U.W., Seattle, WA 98195-1580, USA}

\begin{abstract}

We present a low-dispersion period-luminosity relation (PL) based on 154 Cepheids in Messier 33 (M33) with \textit{Hubble} Space Telescope (HST) photometry from the PHATTER survey. Using high-quality ground-based light curves, we recover Cepheid phases and amplitudes for multi-epoch HST data and we perform template fitting to derive intensity-averaged mean magnitudes. HST observations in the SH0ES near-infrared Wesenheit system significantly reduce the effect of crowding relative to ground-based data, as seen in the final PL scatter of $\sigma=0.11 \, \rm mag$.

We adopt the absolute calibration of the PL based on HST observations in the Large Magellanic Cloud (LMC) and a distance derived using late-type detached eclipsing binaries to obtain a distance modulus for M33 of $\mu = 24.622 \pm 0.030 \, \rm mag$ ($d = 840 \pm 11 \, \rm kpc$), a best-to-date precision of 1.3\%. We find very good agreement with past Cepheid-based measurements. Several TRGB estimates bracket our result while dissagreeing with each other.

Finally, we show that the flux contribution from star clusters hosting Cepheids in M33 does not impact the distance measurement and we find only $\sim$ 3.7\% of the sample is located in (or nearby) young clusters. M33 offers one of the best sites for the cross-calibration of many primary distance indicators. Thus, a precise independent geometric determination of its distance would provide a valuable new anchor to measure the Hubble constant.\\ 

\end{abstract}

%\tableofcontents
%\newpage
%%%%%%%%%%%%%%%%%%%%%%%%%%%
%%%%%%%%%%%%%%%%%%%%%%%%%%%

\section{Introduction}

Cepheid variables are the best-calibrated primary distance indicators and are commonly used to form the first rung of the empirical distance ladder \citep[e.g.,][]{Riess2022}. Their Period-Luminosity (PL) relation, also known as the ``Leavitt Law'' \citep{Leavitt1912}, is calibrated geometrically in the Milky Way (MW) from \textit{Gaia} DR3 parallaxes \citep{Riess2021}, in the Large Magellanic Cloud (LMC) from detached eclipsing binaries \citep{Riess2019, Pietrzynski2019}, and in NGC$\,$4258 with water masers \citep{Reid2019}. Cepheid distances are used to calibrate the second rung of the distance ladder, type Ia supernovae (SNe Ia), which allows us to measure the distance to further galaxies in the Hubble flow and to derive the value of the Hubble constant, $H_0$.

\begin{figure*}[t]
\centering
\includegraphics[width=15.0cm]{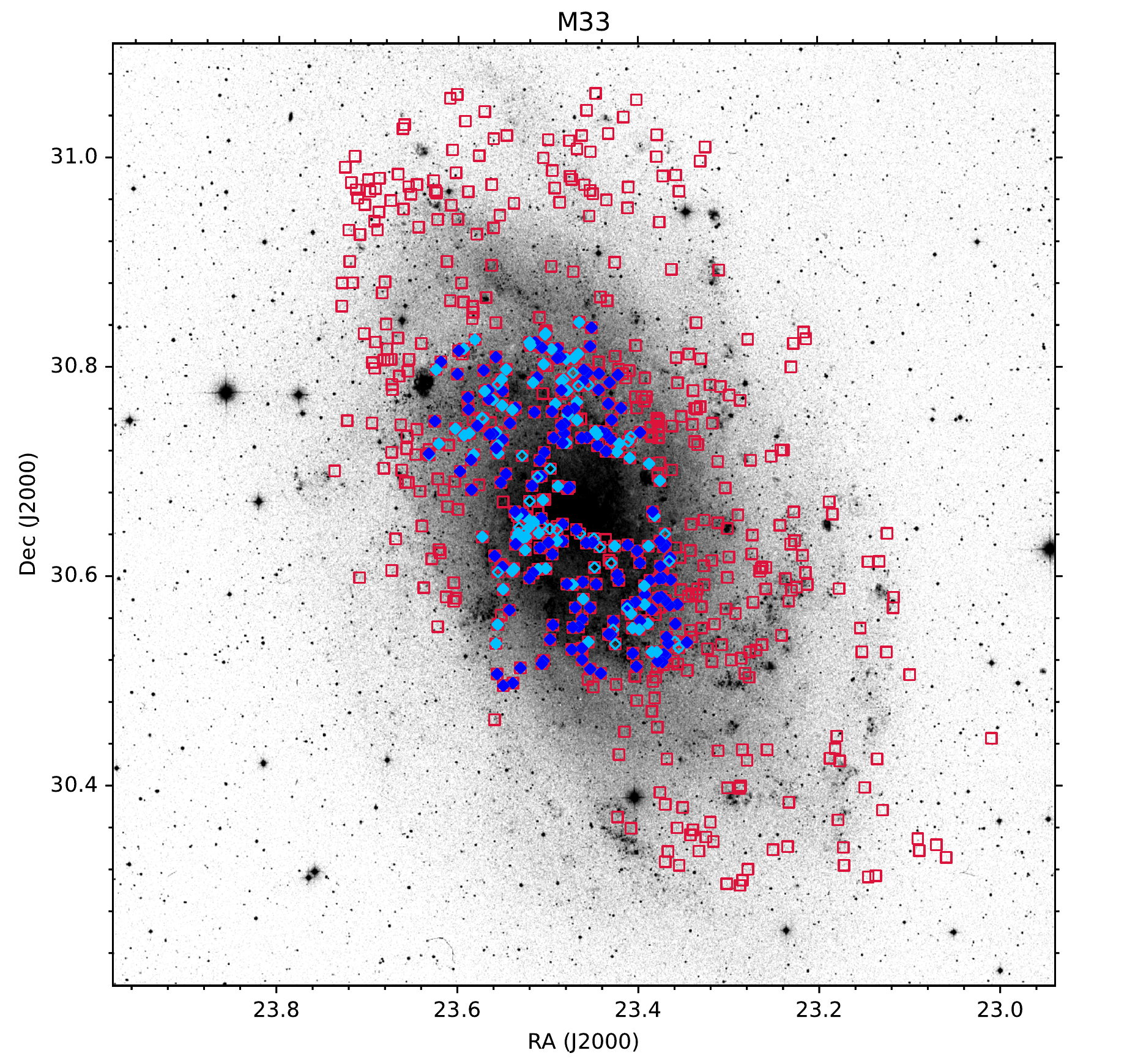} 
\caption{Map of M33: the template (i.e. ground-based) sample is shown in red while the HST (i.e. PHATTER) sample is shown in blue. Dark blue and light blue markers are Cepheids from the gold and silver sample respectively. Empty blue markers are excluded Cepheids (see \S\ref{sec:template_fitting_procedure}). Note the HST sample is a subset of the template sample.\\}
\label{fig:map_M33}
\end{figure*}

Messier 33 (hereafter M33) is a nearby type Sc II-III spiral galaxy and the third largest member of the Local Group. As early as 1926, Edwin Hubble used this galaxy as one of the \textit{spiral nebulae} to learn about the structure of the Universe and observed 35 Cepheid variables to measure its distance \citep{Hubble1926}. Since then, it has been extensively studied, and is still a crucial object for the distance scale \citep{Freedman1991, Lee2022}: M33 has intermediate inclination \citep[$i=57\pm4^{\circ}$,][]{Kourkchi2020b}\footnote{\href{http://edd.ifa.hawaii.edu}{http://edd.ifa.hawaii.edu}, Table ``CF4 Initial Candidates''}, which limits the effects of reddening and of geometry that can produce additional scatter in the PL relation. Additionally, M33 is known for its steep metallicity gradient, which was measured using red giant branch (RGB) stars \citep{Tiede2004}, planetary nebulae \citep{Magrini2009}, and \ion{H}{2} regions \citep{Bresolin2011, Toribio2016, Rogers2022}.

Cepheids are numerous in M33, and large samples have been obtained by various programs \citep{Macri2001, Hartman2006, Pellerin2011}. Recently, the PHATTER collaboration (PI: J.~Dalcanton) published a detailed catalog\footnote{\href{http://archive.stsci.edu/hlsp/phatter}{https://archive.stsci.edu/hlsp/phatter}} of UV to NIR photometry for 22 million stars in the central disk of M33 \citep{Williams2021} using the \textit{Hubble} Space Telescope (HST). Although they are not time-series observations, serendipitous overlaps between the PHATTER fields of view in a given filter provide multiple data points randomly spread across the phase of M33 Cepheids.  Out of 250 Cepheids in the HST sample (defined in \S2.2), 225 variables have more than one epoch in $F475W$ and $F814W$, and 66 objects have more than one epoch in $F160W$ (due to smaller overlaps of the WFC3/IR fields). Knowledge of the date and time of observation for each HST exposure, combined with periods previously measured from other surveys, enable the correction of these random-phase observations to mean magnitude. Finally, past studies \citep[e.g.][]{Macri2001, Riess2012, WagnerKaiser2015, Kodric2018} have revealed the advantages of space-based observations such as HST in limiting crowding effects and their impact for the PL dispersion, as well as providing homogeneous photometry including in the near-infrared. In this paper we aim to take advantage of the recently published high-quality PHATTER catalog in order to provide a new PL calibration for M33 Cepheids in HST filters and to improve the M33 distance measurement.

The outline of this paper is the following. In \S\ref{sec:photometric_data} we present the samples of M33 Cepheids used in this study. In \S\ref{sec:template_fitting} we describe the construction of template light curves from ground-based data and the procedure to recover mean magnitudes from random-epoch photometry. In \S\ref{sec:PL_distance} we calibrate the Cepheid PL relation and determine the M33 distance modulus. Lastly, in \S\ref{sec:clusters} we investigate the effects of Cepheids located in star clusters, we estimate their occurrence rate in M33 and compare it with that of other Local Group galaxies.\\

%%%%%%%%%%%%%%%%%%%%%%%%%%%
%%%%%%%%%%%%%%%%%%%%%%%%%%%
\section{Photometric data}
\label{sec:photometric_data}

In order to recover intensity-averaged mean magnitudes from random-phase HST data (hereafter the HST sample), we use templates obtained by compiling a large number of well-sampled ground-based light curves of M33 Cepheids (hereafter the \textit{template} sample). Both samples are described below. \\

\subsection{The template sample}

\begin{figure}[t!]
\centering
\includegraphics[width=8.2cm]{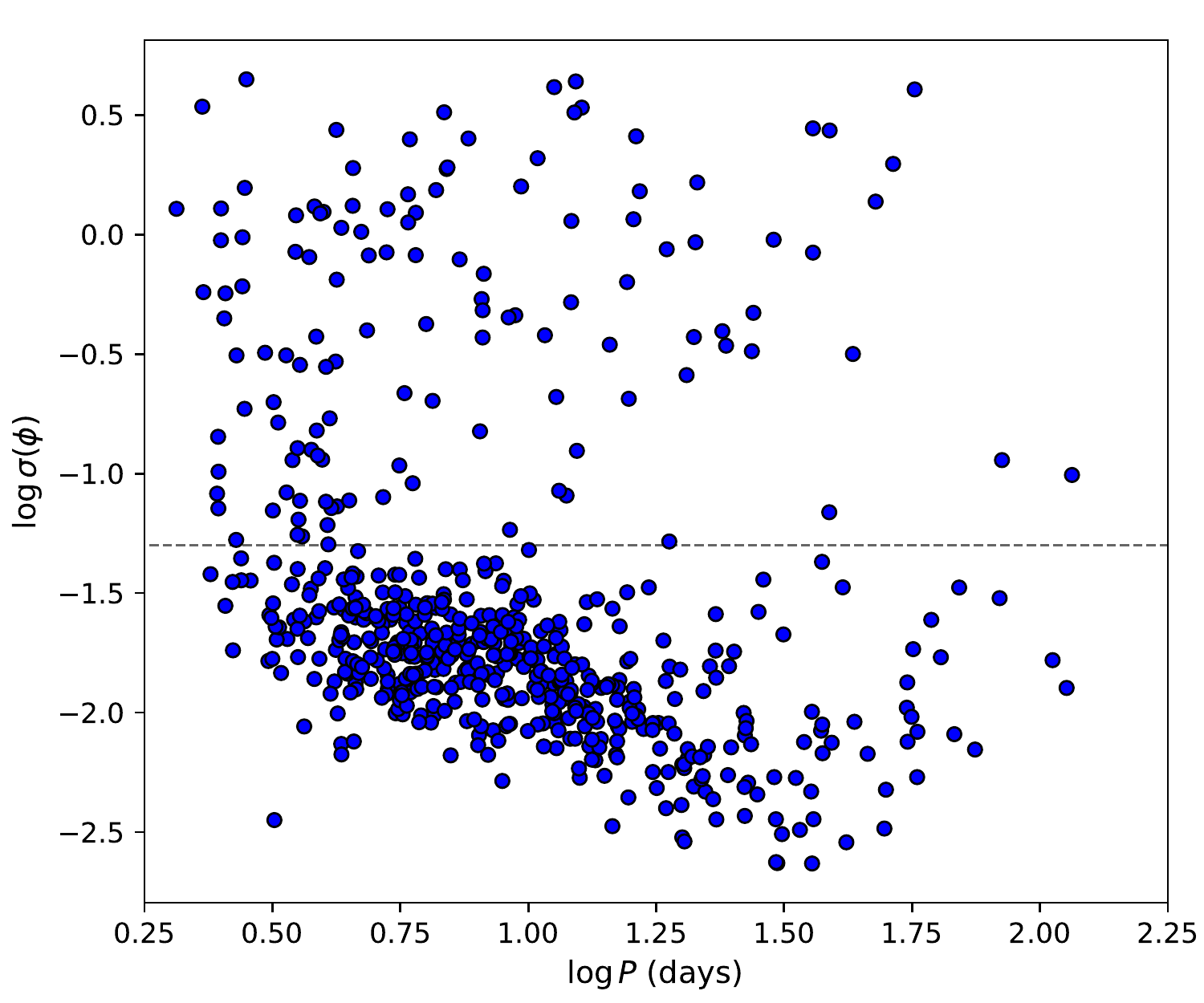} 
\caption{Distribution of phase uncertainties for Cepheids of the \textit{template} sample at the epoch of HST observations, estimated from the period uncertainty and the interval between the midpoint of the HST and of the ground observations. The dashed horizontal line represents our threshold for the gold sample (see below): we only keep Cepheids with $\sigma(\phi) < 0.05$. }
\label{fig:period_uncertainties}
\end{figure}

We used a sample of 609 previously-known Cepheids \citep{Macri2001, Pellerin2011} with homogeneous {\it gri} light curves obtained by Konchady et al. (in prep.) using archival CFHT/MegaCam observations (proposal ID 04BF26, PI Beaulieu; proposal ID 04BH98, PI Hodapp). They are represented in red in Fig.~\ref{fig:map_M33} and constitute the \textit{template} sample. 

The majority of the original CFHT observations (associated with proposal ID 04BF26) are extensively described in \citet{Hartman2006}; they span roughly one-and-a-half years (2003 August to 2005 January) and were obtained on 27 separate nights. We supplemented these with an additional four nights of $i$ observations obtained in 2004 August and September (associated with proposal ID 04HB98). Konchady et al. (in prep.) performed an independent analysis of these images, carrying out time-series PSF photometry that was calibrated against Pan-STARRS DR1 \citep{Chambers2016}. The periods and phases of the Cepheids were redetermined by simultaneously fitting the CFHT {\it gri} photometry and the WIYN {\it BVI} photometry of \citet{Pellerin2011} using the \citet{Yoachim2009} templates. We solved for a common period and phase across the six bands, and independent mean magnitudes and light curve amplitudes in each band.

Cepheid light curves of the \textit{template} sample have two purposes: they are used to build templates thanks to their complete phase coverage (see \S\ref{sec:build_templates}) and to recover the amplitudes and phases of the HST light curves (see \S\ref{sec:template_fitting_procedure}). For this reason their periods must be known precisely. From the period uncertainty, we estimate the uncertainty in the phase-shift between the mid-date of the ground observations (MJD = 52170) and of PHATTER observations (MJD = 57989), and we flag Cepheids for which this uncertainty $\sigma (\phi)$ is larger than 0.05 (or $\log \sigma (\phi) > -1.3$, dashed horizontal line in Fig.~\ref{fig:period_uncertainties}). They constitute the "silver" sample (see Sect. \ref{sec:template_fitting_procedure}). Additionally we only keep Cepheids which have optimal ground-based light curves \citep[Table 3 of][]{Pellerin2011}. This leaves a total of 420 Cepheids, for which we perform a visual inspection of each light curve's quality. Cepheids have an average of 45, 31 and 44 data points per light curve in $g$, $r$, and $i$, respectively.

We note that our ground-based sample is minimally affected by blending given the relatively high image quality of the CFHT and WIYN observations and the rejection of outliers by \citet{Pellerin2011}. \\

\subsection{The HST sample}
\label{sec:HST_sample}

The PHATTER survey \citep{Williams2021} contains photometric measurements for 22 million stars in M33 with 6 UV to NIR filters (Advanced Camera for Surveys and Wide Field Camera 3) on the \textit{Hubble} Space Telescope (HST). The survey focuses on the inner disk of the galaxy and covers $\sim 300 \sq\arcmin$ (equivalent to a de-projected area of $\sim 38 \, \rm kpc^2$), extending up to $\sim 14\arcmin$ from the center (equivalent to a distance of $\sim3.5\, \rm kpc $). The observations were taken between 2017 February 21 and 2018 February 25. The catalog reaches 26 to 28 mag in $V$ depending on crowding. It is the largest and most complete catalog to date for stellar populations in M33. We identified 250 Cepheids from the template sample in the PHATTER catalog: they are represented in blue in Fig.~\ref{fig:map_M33} and are hereafter referred to as the HST sample.

\begin{figure}[t!]
\centering
\includegraphics[width=8.2cm]{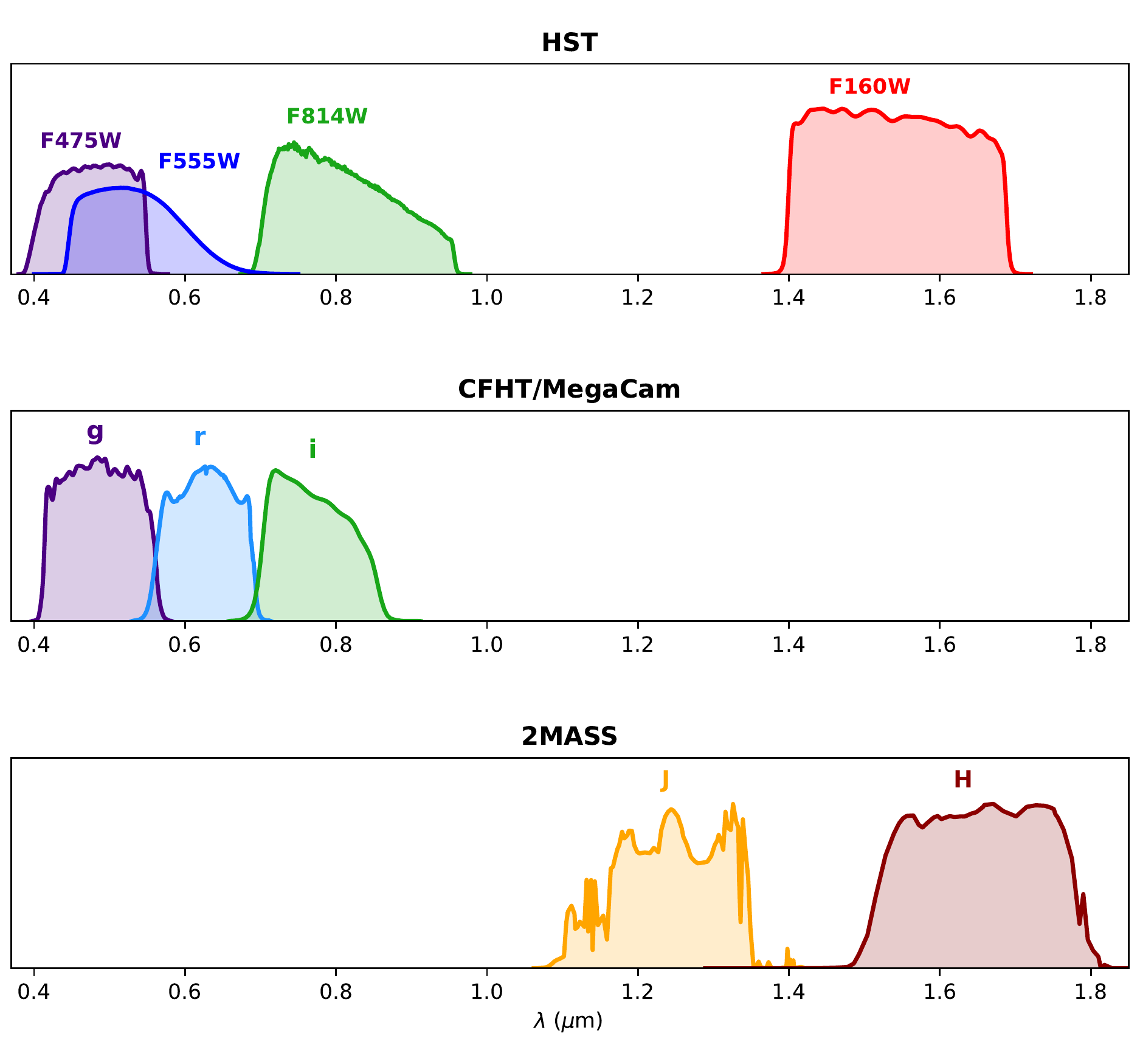} 
\caption{Wavelength coverage of HST filters used in this analysis (top panel), CFHT filters adopted to build optical template light curves (middle panel) and 2MASS filters used for NIR templates \citep[][bottom panel]{Inno2015}. We adopt $g$, $i$ and $H$ templates to fit light curves in $F475W$, $F814W$ and $F160W$, respectively. }
\label{fig:All_filters_coverage}
\end{figure}

We matched the Cepheid coordinates to the full-frame PHATTER catalogs using an initial search radius of 0.1 arcsec, and found that all had matches within $< 0.5$ mas with expected magnitudes ($19 < F475W < 23$). We then used the pixel coordinates from these catalogs to retrieve the exposure-level photometry from the original DOLPHOT \texttt{.phot} outputs. These files contain columns with photometry from each individual input frame in addition to the combined measurements.

The PHATTER survey does not provide time-series observations, which might make it \textit{in principle} poorly suited to study variable stars such as Cepheids. However, in the optical $F475W$ and $F814W$ filters, successive PHATTER pointings show a significant overlap, and therefore up to 4 epochs can be available for a given Cepheid. In the NIR, the pointings have smaller overlaps, which gives one to two epochs per Cepheid. Each epoch can be decomposed into 4 or 5 separate dithers/exposures and the phase-coverage of each epoch is random. We note that the first exposure of each $F475W$ and $F814W$ visit sequence is significantly shallower than the rest, as these are short exposures targeting the brightest stars (M.~Durbin, 2023, private communication). They were therefore excluded as they are not useful for Cepheids. The date and time of a given HST observation provide the relative phase of the corresponding measurement. Then, mean magnitudes can be recovered from sparse data by applying a template-fitting procedure (\S\ref{sec:template_fitting}).  \\

%%%%%%%%%%%%%%%%%%%%%%%%%%%
%%%%%%%%%%%%%%%%%%%%%%%%%%%
\section{Template fitting}
\label{sec:template_fitting}

In this section we describe the construction of template light curves from ground-based data (\S\ref{sec:build_templates}) and the procedure to recover mean magnitudes from PHATTER photometry in HST filters (\S\ref{sec:template_fitting_procedure}). The HST and \textit{template} samples were observed in different filters. The HST $F475W$ filter is very similar to the $g$ one from CFHT/MegaCam, and $F814W$ corresponds to the $i$ filter (see Fig.~\ref{fig:All_filters_coverage}). Finally, the template sample does not cover the NIR up to the $F160W$ filter, therefore we use the 2MASS $H$-band templates by \citet{Inno2015}, based on a large sample of LMC Cepheid light curves. We adopt $g$-band and $i$-band templates to derive HST mean magnitudes in $F475W$ and $F814W$ respectively, and $H$-band templates to derive $F160W$ mean magnitudes. \\

\subsection{Building template light curves}
\label{sec:build_templates}

\begin{figure}[t!]
\centering
\includegraphics[width=8.3cm]{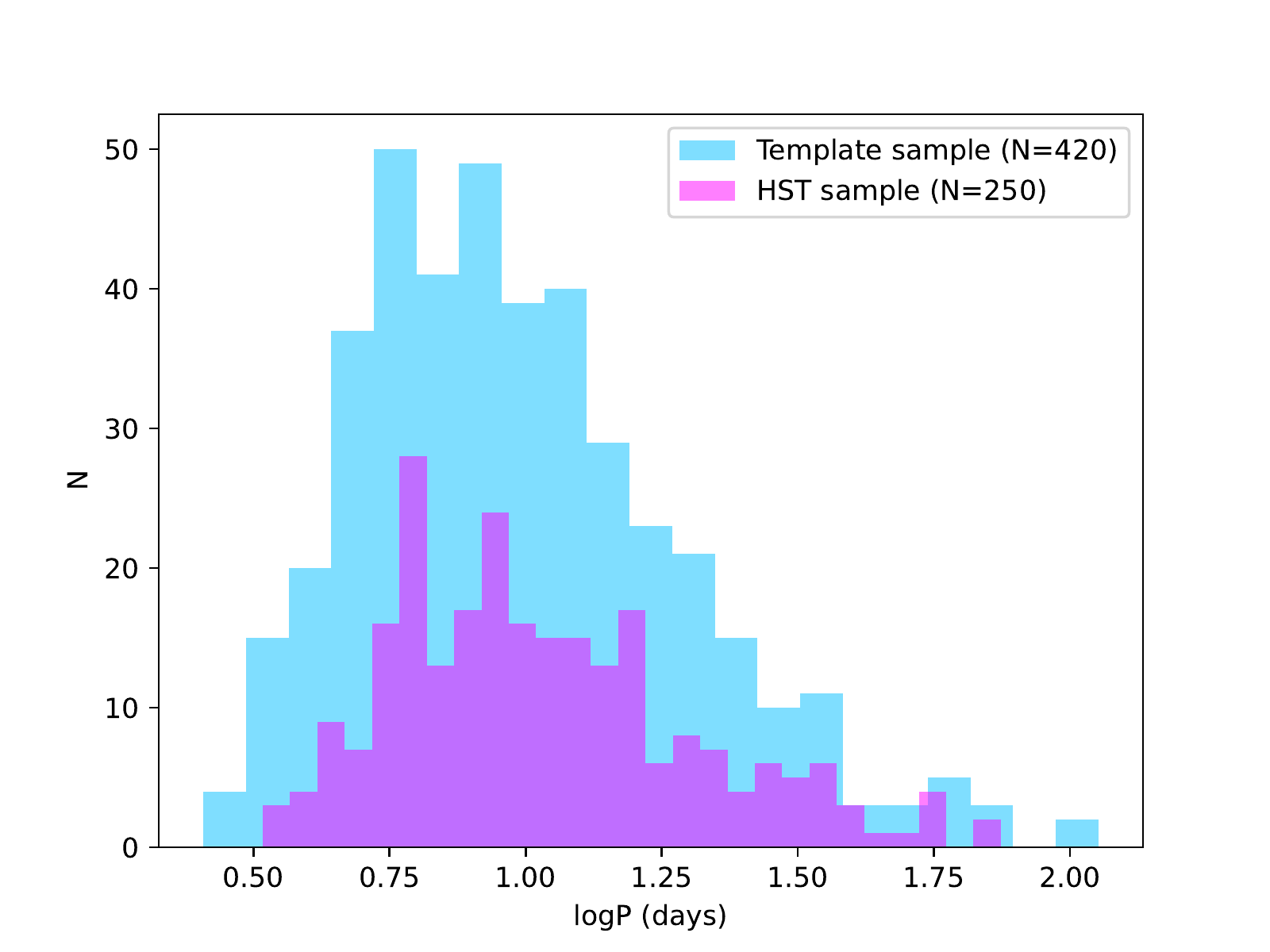} 
\caption{Period distribution of Cepheids from the template sample and from the HST sample.}
\label{fig:histogram_periods}
\end{figure}

We use the well-sampled light curves from the \textit{template} sample to build template light curves in the $g$, $r$ and $i$ filters of CFHT/MegaCam. These ground-based light curves are ideal to build templates and to recover the mean magnitudes from HST random phase observations: they are representative of Cepheids from the HST sample as they belong to the same host galaxy and have a very similar period distribution (Fig.~\ref{fig:histogram_periods}). Other templates from the literature \citep[e.g.][from LMC Cepheids]{Yoachim2009} could have been used instead of creating new ones. However, adopting templates built from a population similar to the HST sample avoids possible differences in light curve shapes for Cepheids from different galaxies \citep[possibly due to metallicity effects,][]{Antonello2000}. 

In order to account for changes in light curve shape as a function of period \citep{Hertzsprung1926}, we split the sample into four different period bins. They are described in Table~\ref{table:calibration_sample}. The number of bins was determined by the size and by the distribution of our calibrating sample: having a larger sample would have allowed us to use more bins.

\begin{table}[!t]
\caption{Number of Cepheids in each period bin for the template sample.}
\begin{tabular}{c c | c c c }
~\\
\hline
\hline
Bin & $\log P$ & $N_g$ & $N_r$ & $N_i$ \\
\hline
1 & $0.3-0.9$ & 148 & 136 & 143   \\
2 & $0.9-1.2$ & 91  & 84  & 81   \\
3 & $1.2-1.5$ & 46  & 47  & 44  \\
4 & $1.5-2.0$ & 20  & 19  & 16 \\
% \hline
% \multicolumn{2}{c}{All periods}  & 305 & 286 & 284  \\
\hline
~ \\
\end{tabular}
\label{table:calibration_sample}
\end{table}

\begin{figure*}[t]
\centering
\includegraphics[width=4.4cm]{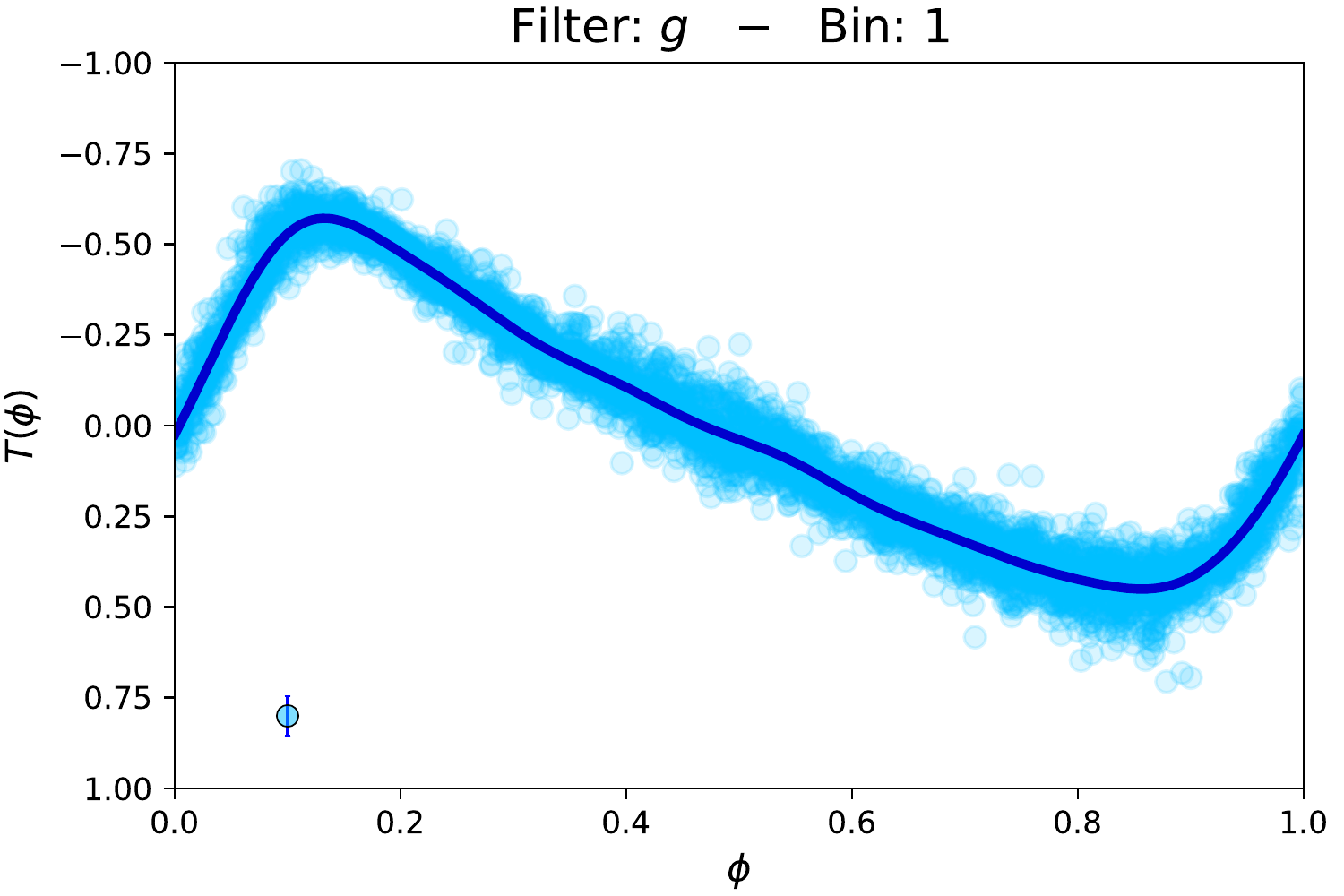} 
\includegraphics[width=4.4cm]{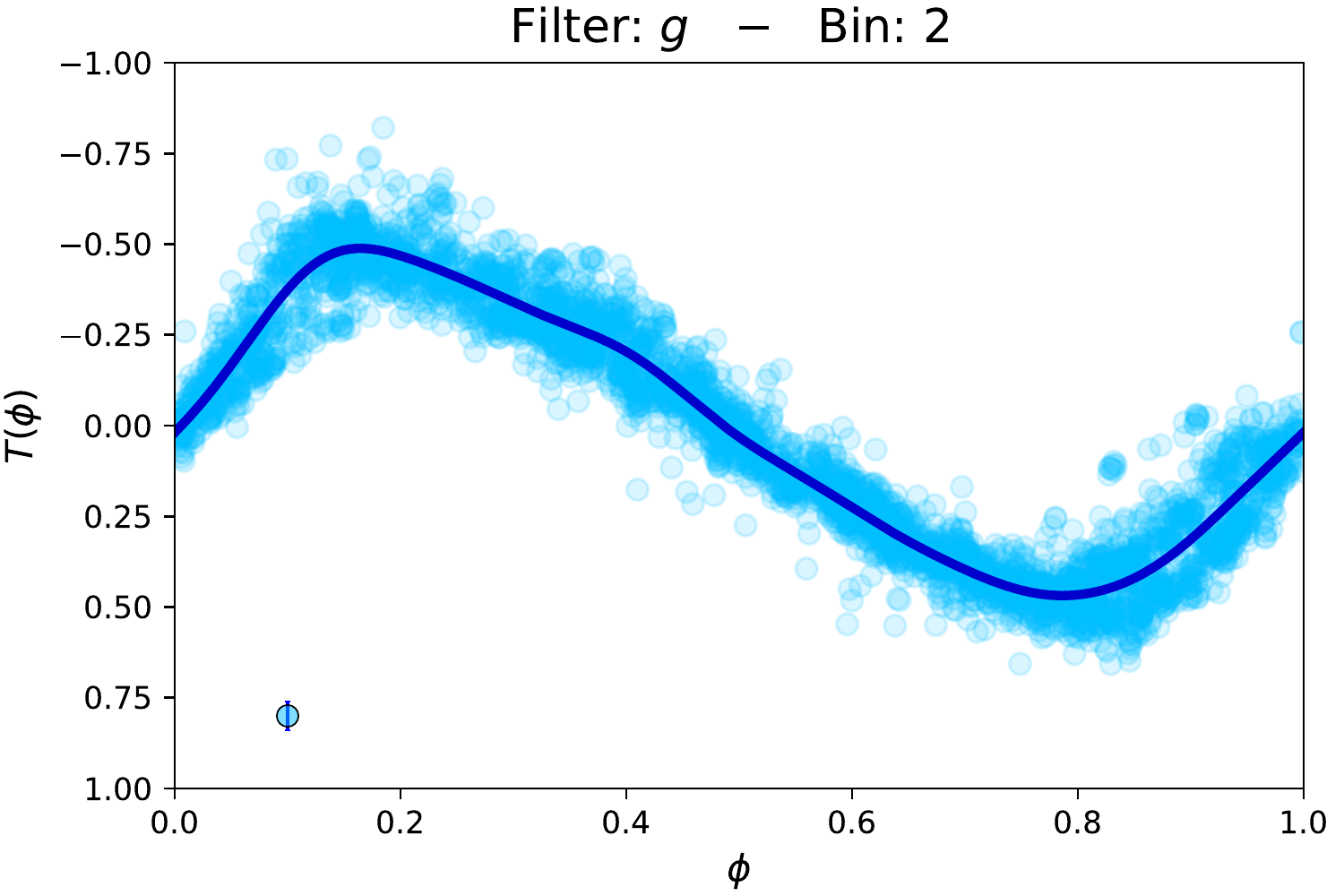} 
\includegraphics[width=4.4cm]{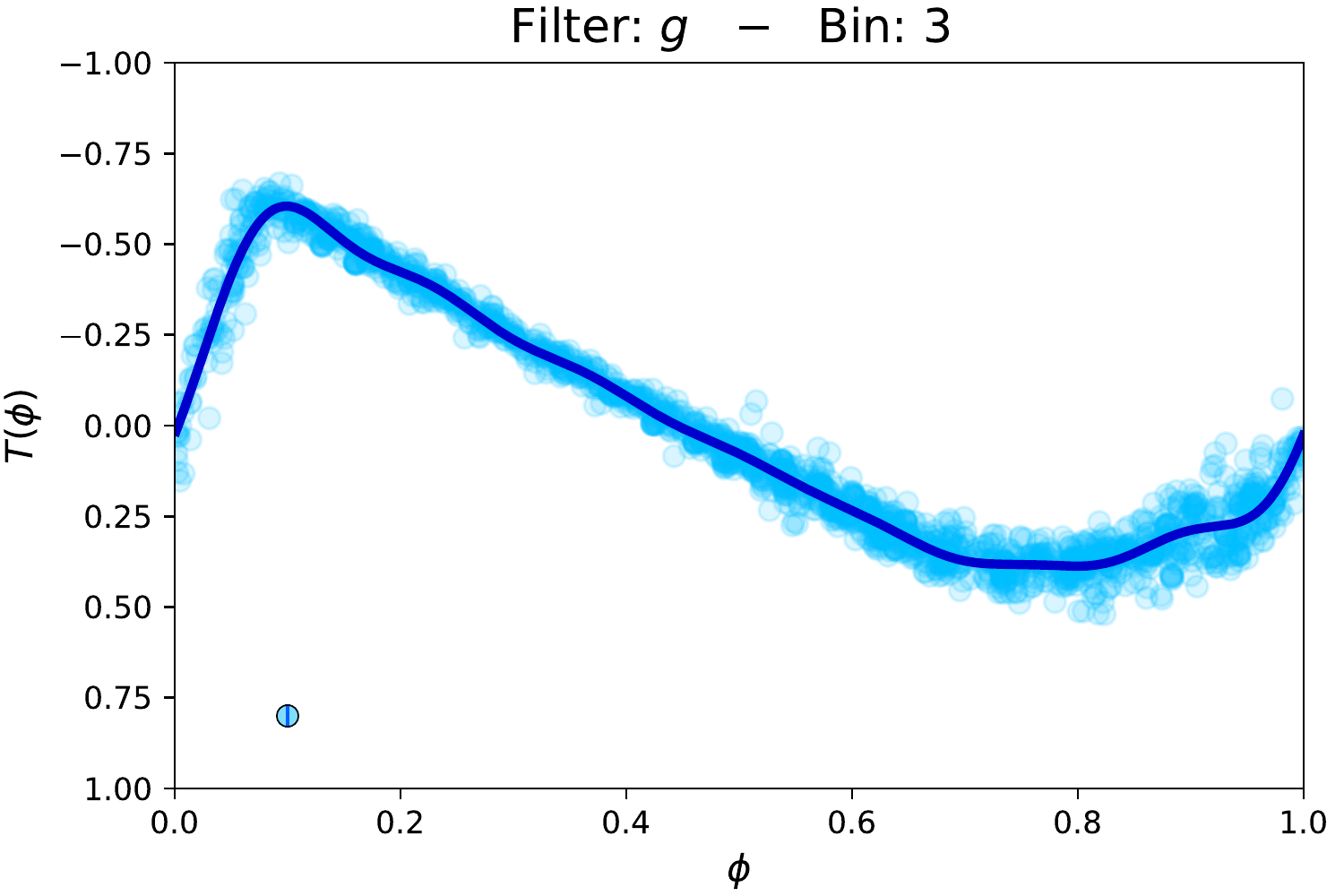} 
\includegraphics[width=4.4cm]{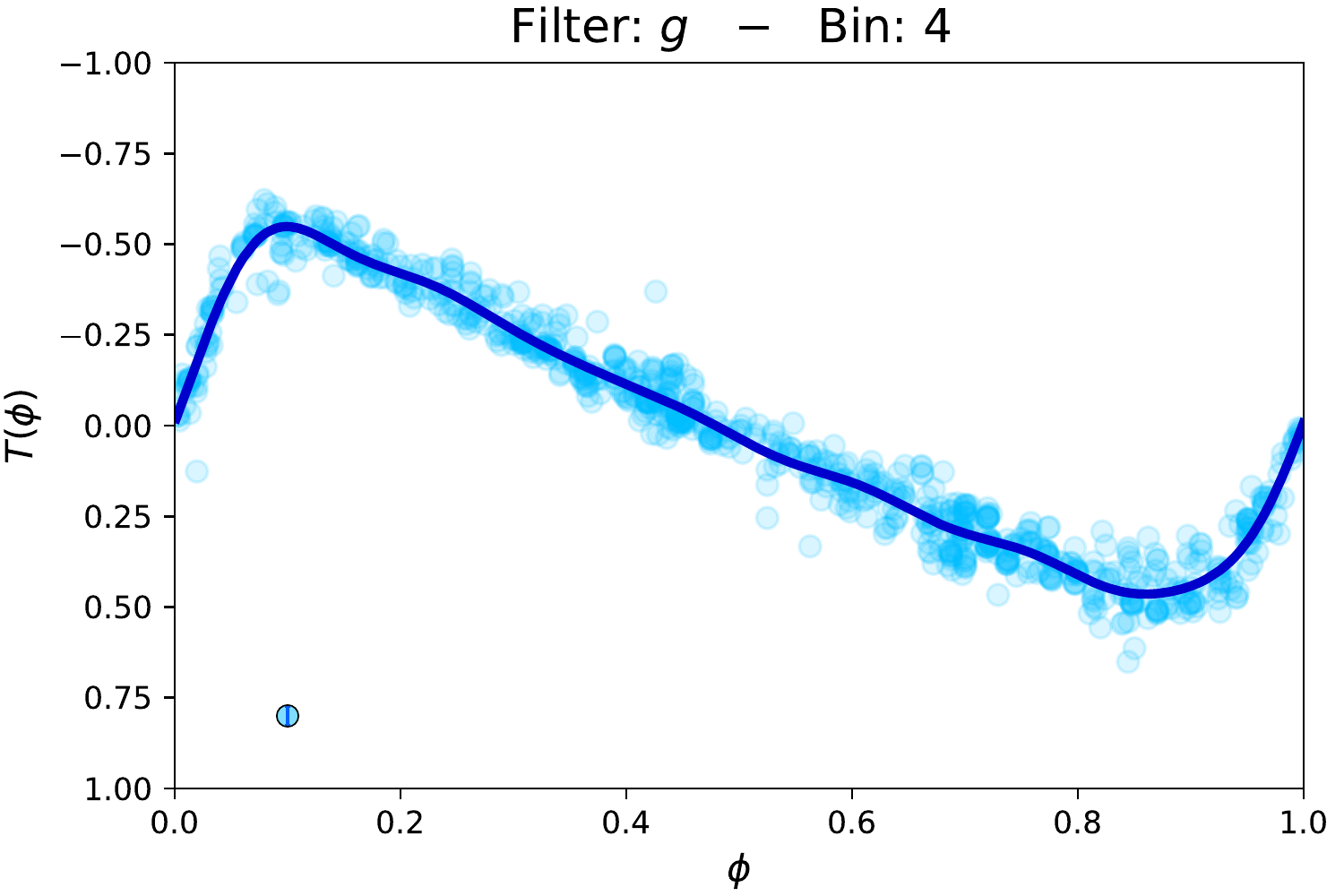} 
\includegraphics[width=4.4cm]{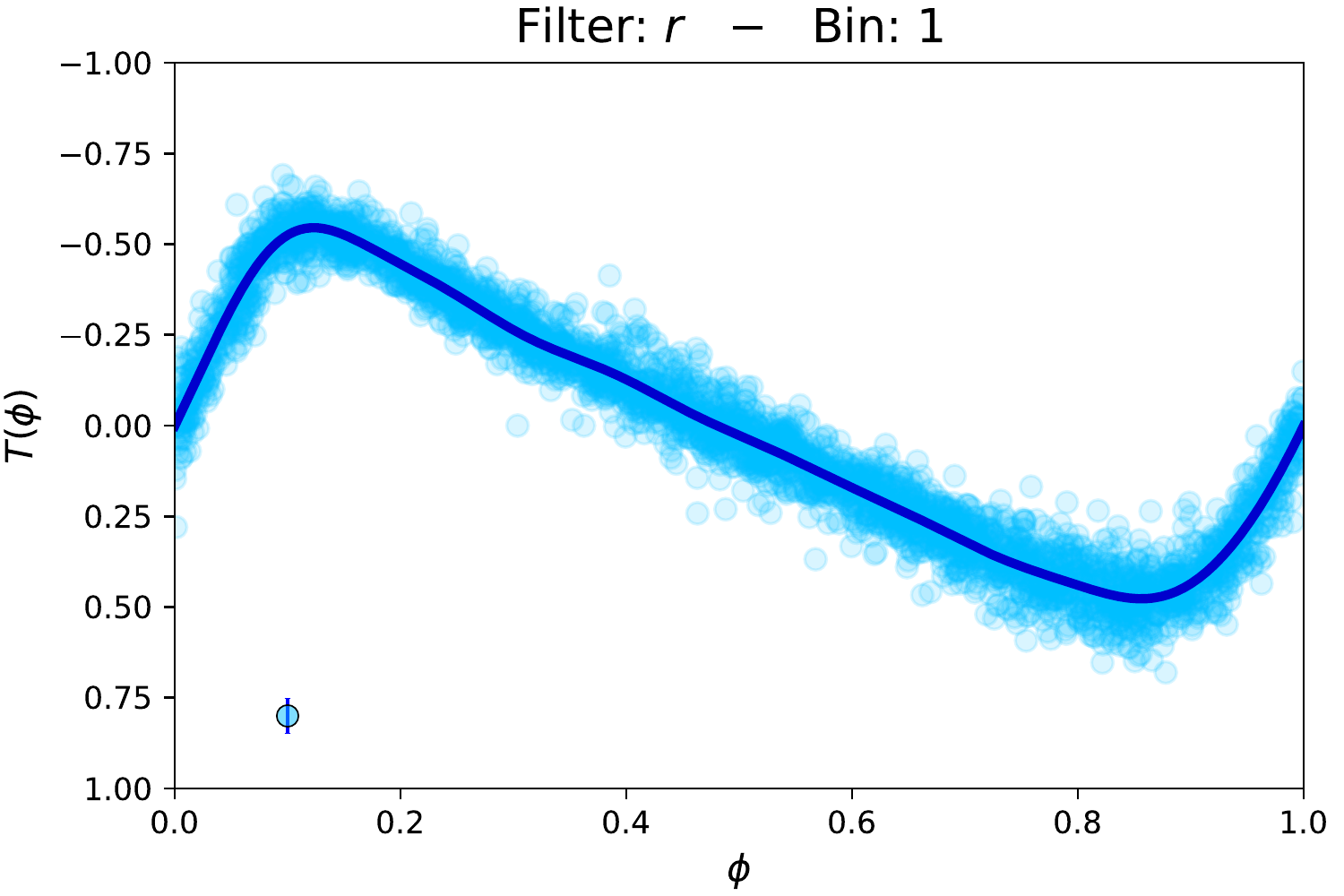} 
\includegraphics[width=4.4cm]{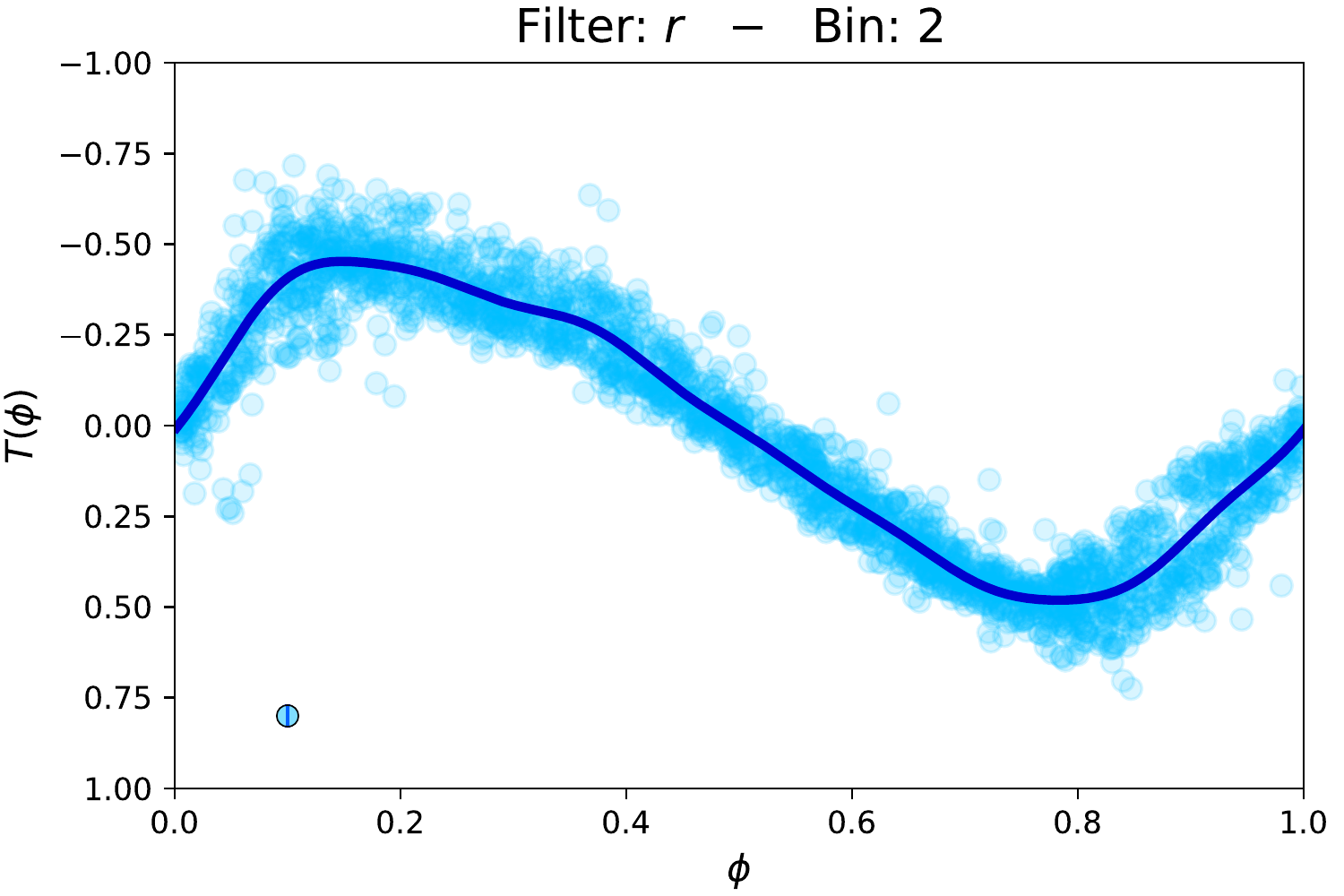} 
\includegraphics[width=4.4cm]{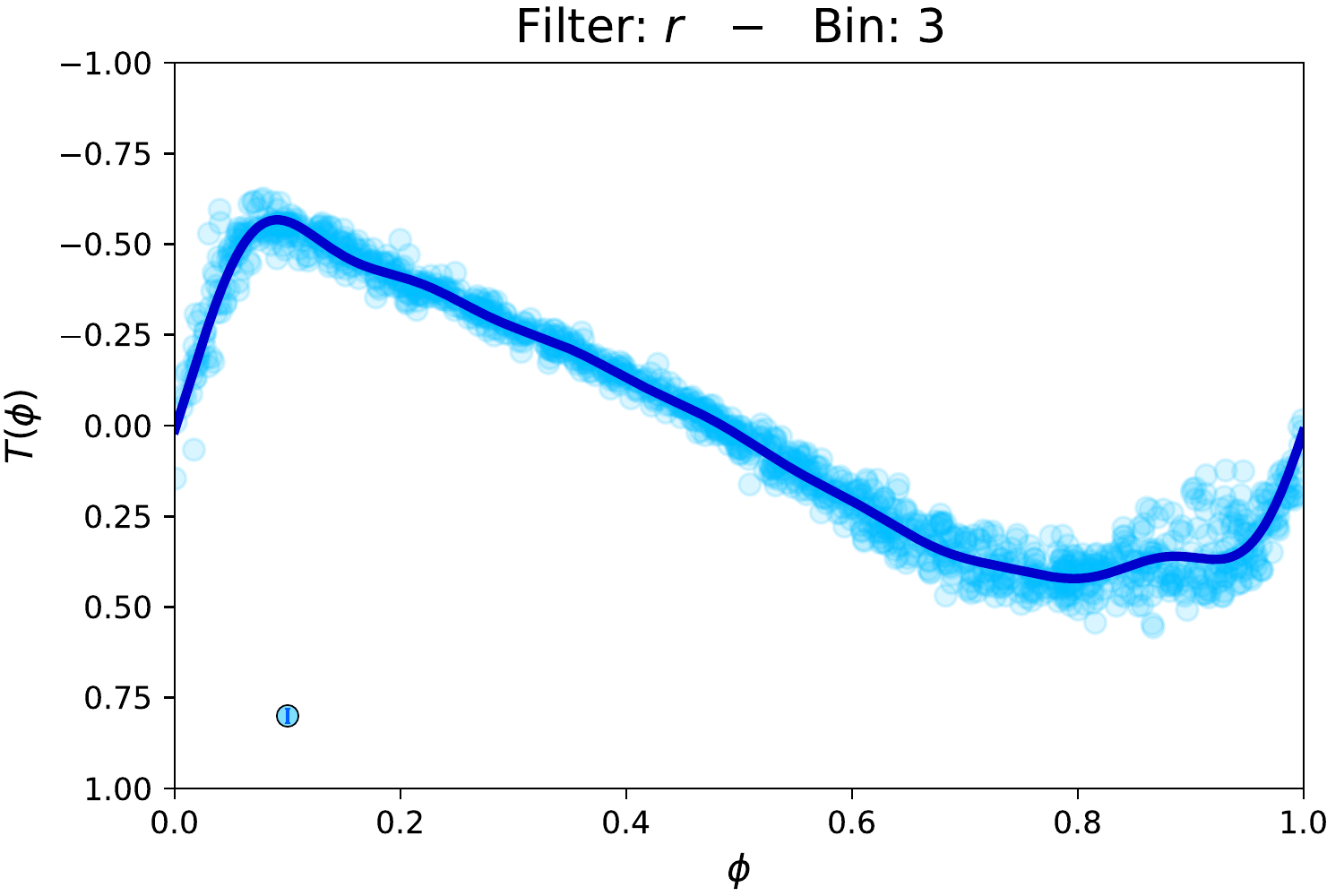} 
\includegraphics[width=4.4cm]{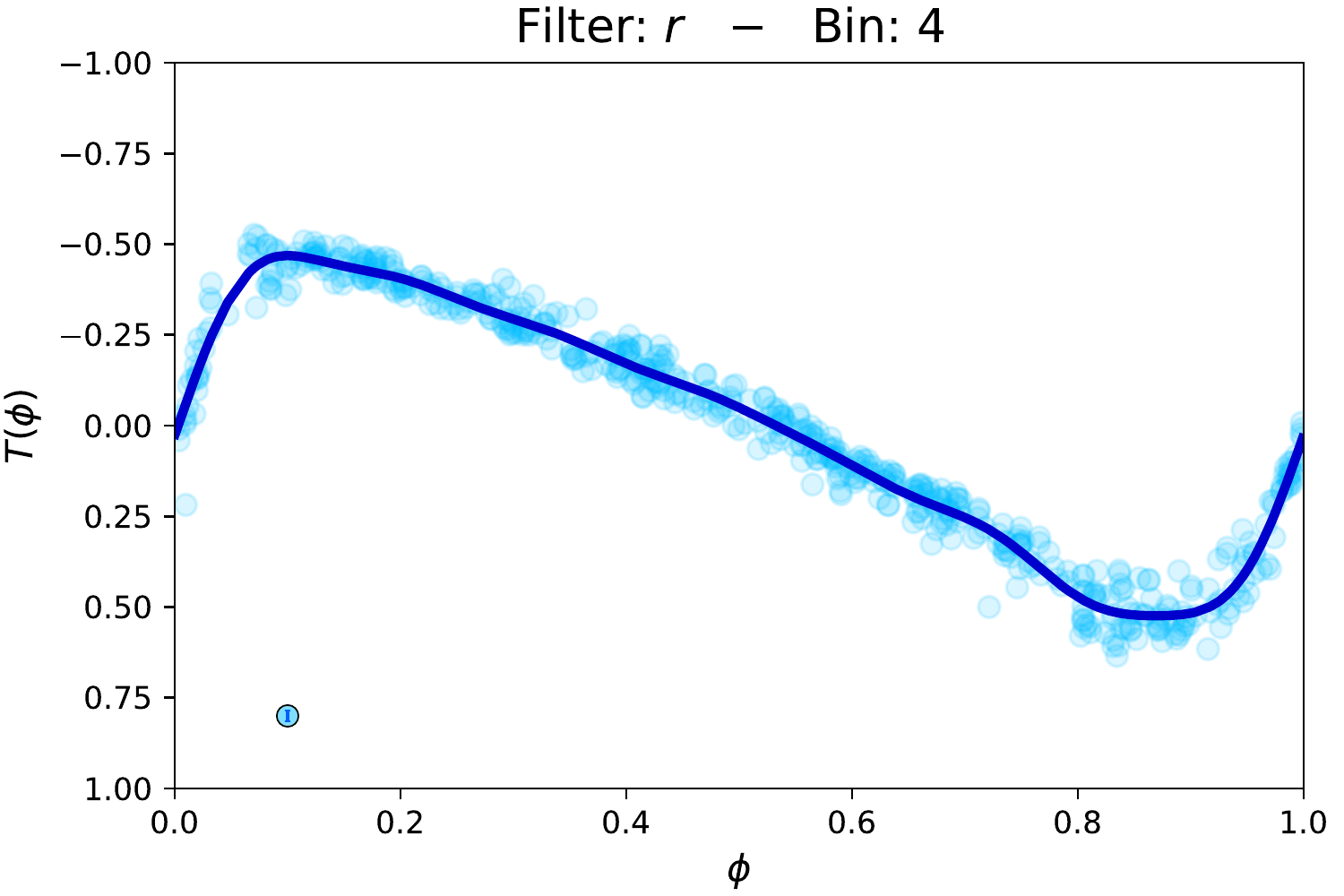} 
\includegraphics[width=4.4cm]{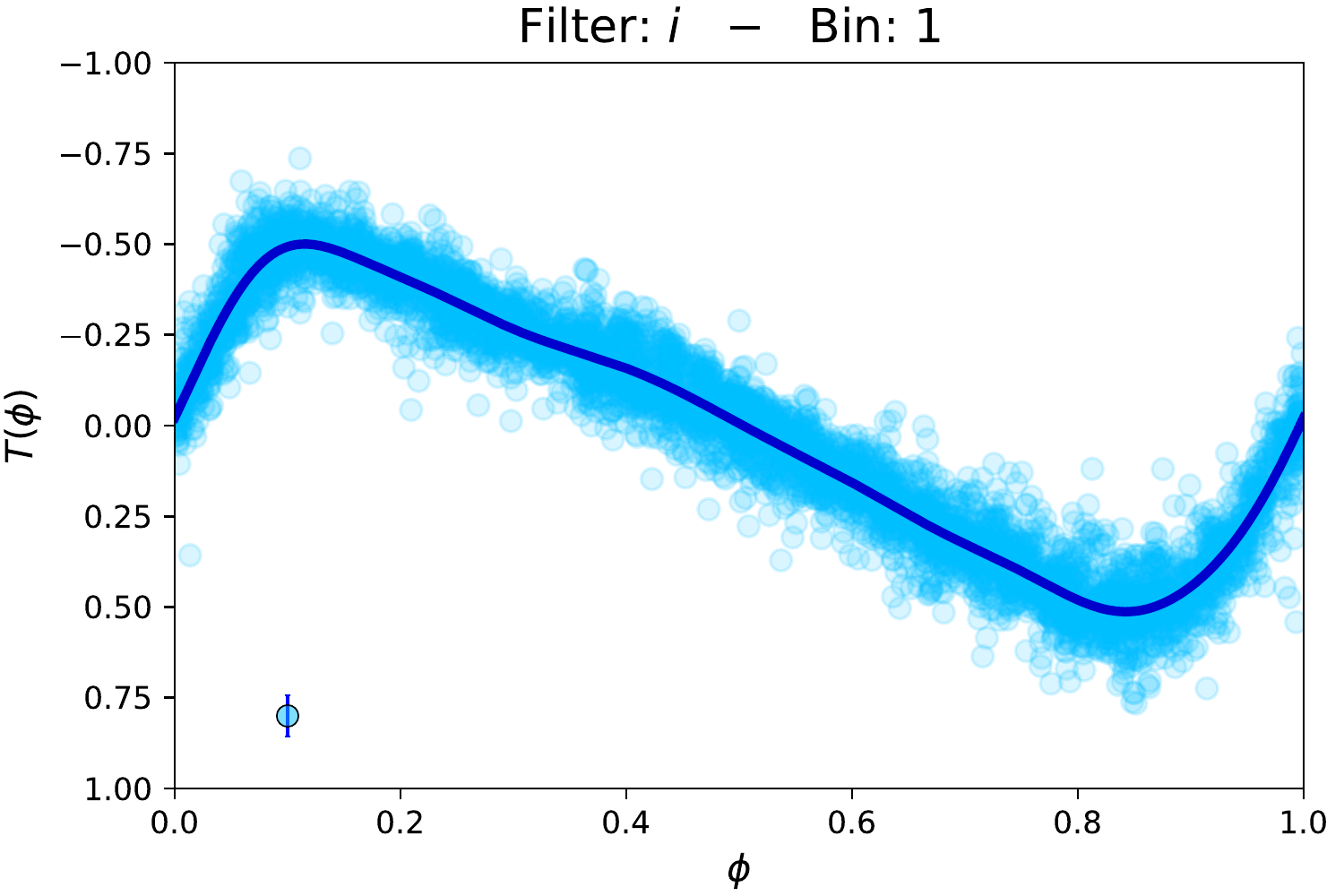} 
\includegraphics[width=4.4cm]{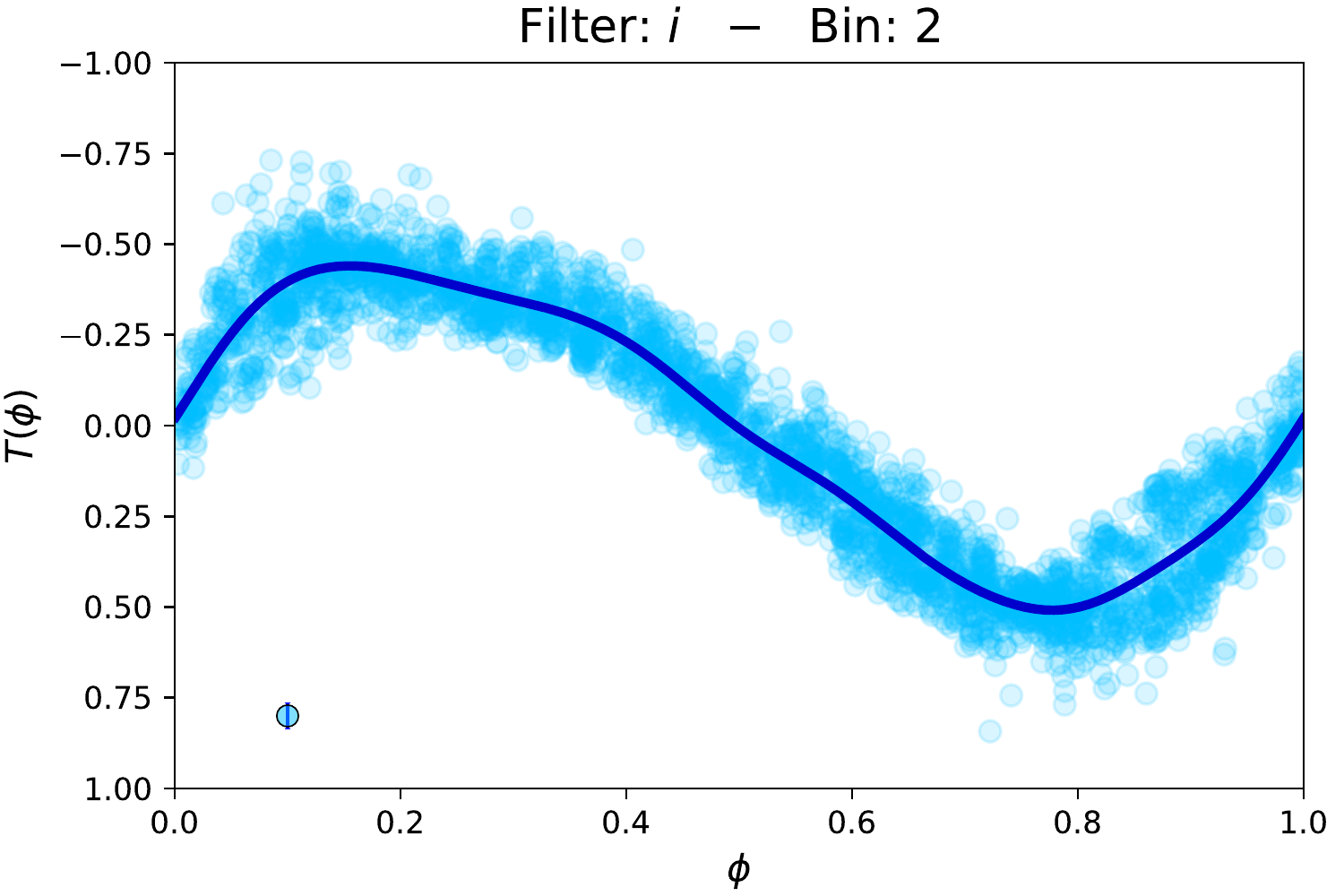} 
\includegraphics[width=4.4cm]{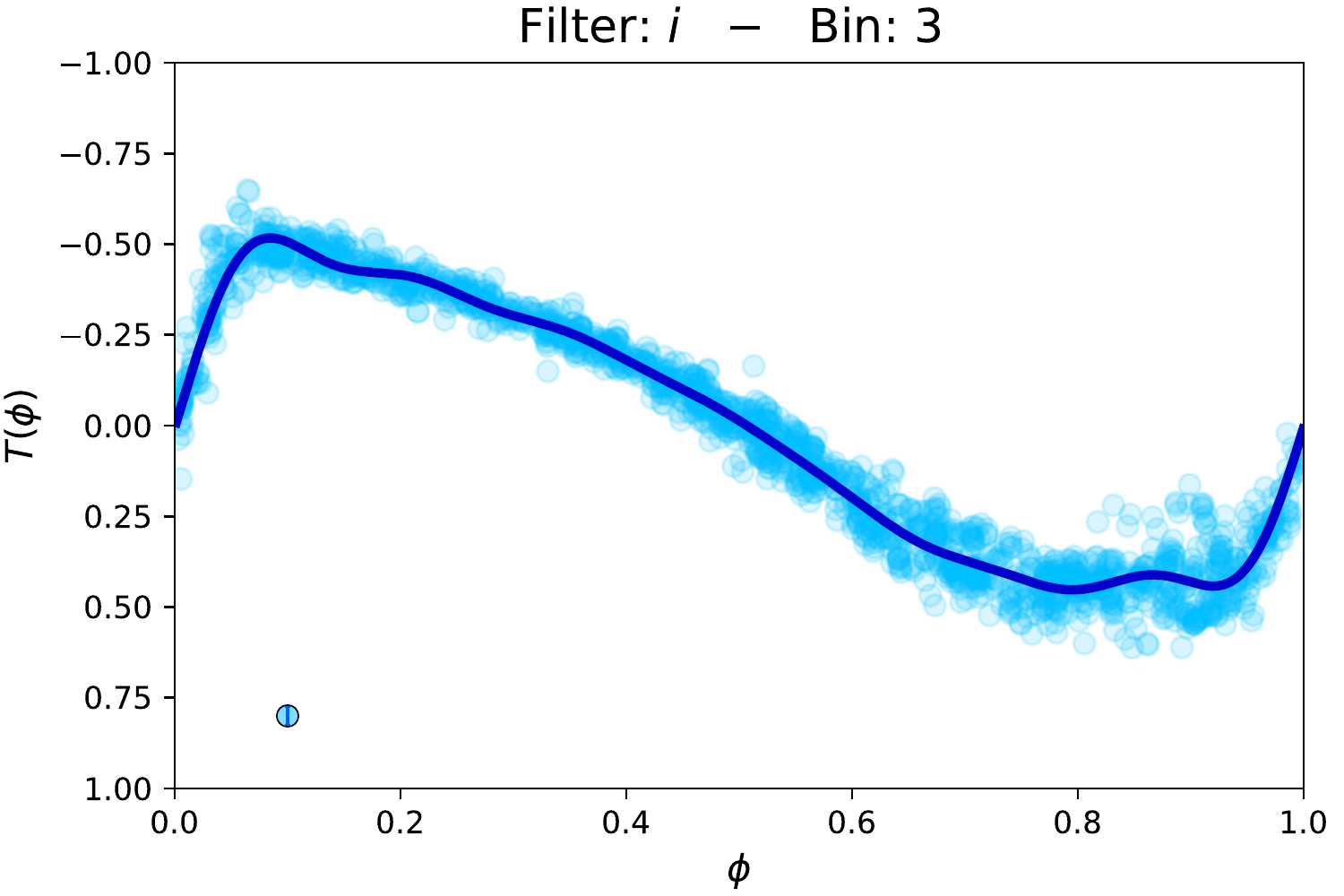} 
\includegraphics[width=4.4cm]{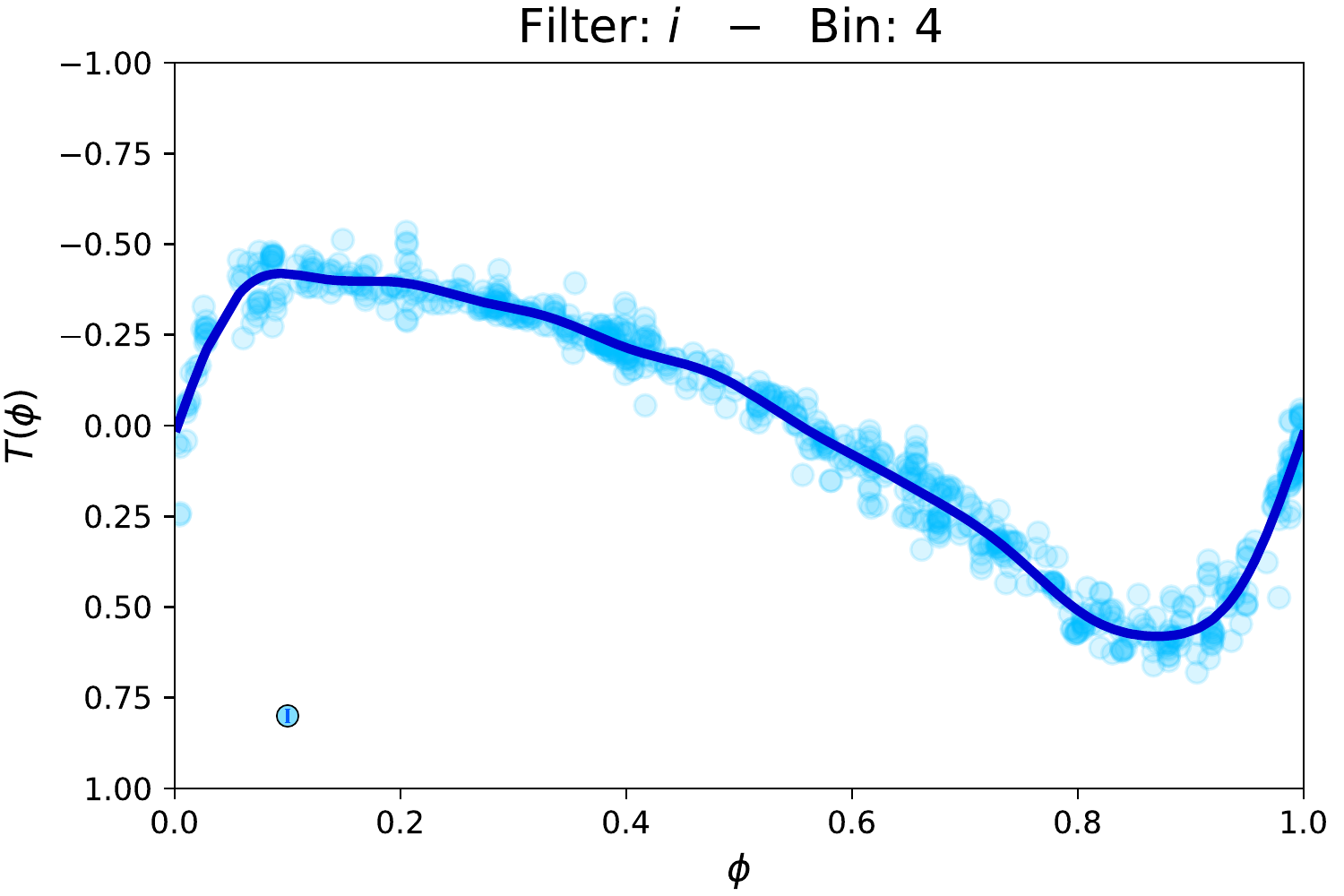} 
\caption{Merged light curves of M33 Cepheids (amplitude-scaled) used for the templates in the $g$, $r$ and $i$ band (1$\rm^{st}$, 2$\rm^{nd}$ and 3$\rm^{rd}$ lines respectively). The four columns correspond to the four period bins listed in Table~\ref{table:calibration_sample}. The point in the bottom left corner shows the typical error bar, multiplied by a factor of 3 for better visibility.}
\label{fig:templates}
\end{figure*}

While the reference phase of a Cepheid is often defined by the epoch of maximum brightness, this quantity can be biased by the presence of a bump in the light curve that varies in shape and phase as a function of period along the \citet{Hertzsprung1926} progression. This bump coincides with maximum light for Cepheids with periods around 10 days. To overcome this issue, \citet{Inno2015} adopted another feature to determine the phase of a Cepheid light curve: the mean magnitude along the rising branch (MRB). As mean magnitudes are known with great precision for our template sample, this approach is more reliable than using the maximum to set the phase \citep[see more details in ][]{Inno2015} and we adopt it in our analysis:
\begin{equation}
\phi_{obs} = \rm mod \left( \frac{JD_{obs} - JD_{MRB}}{P}  \right)
\end{equation}
For a filter $\lambda$, we normalize the magnitude values $m_i$ by deriving the quantity:
\begin{equation}
T_{\lambda} =  \frac{m_i - \langle m_i \rangle}{A_{\lambda}}
\end{equation}
where $\langle m_i \rangle$ is the mean magnitude and $A_{\lambda}$ is the amplitude. Finally, we merge all phased and normalized light curves into a single template for each period bin. The final templates and compiled light curves are shown in Fig.~\ref{fig:templates} in the $g$, $r$ and $i$ filters and for each of the four period bins. We follow \citet{Inno2015} and fit the merged light curves with a seventh order Fourier series of the form:
\begin{equation}
F_7 (\phi) = A_0 + \sum_{i=1}^7 A_i \,  cos(2\pi i \phi + \Phi_i )
\end{equation}

The resulting coefficients are listed in Table~\ref{table:fourier_parameters}. \\

\begin{table*}[t]
\caption{Fourier parameters for template light curves obtained with the calibrating sample of Cepheids.}
\footnotesize
\centering
\begin{tabular}{c c c c c c c c c c c c c c c c}
\hline
\hline
Bin & $A_0$ & $A_1$ & $A_2$ & $A_3$ & $A_4$ & $A_5$ & $A_6$ & $A_7$ & $\Phi_1$ & $\Phi_2$ & $\Phi_3$ & $\Phi_4$ & $\Phi_5$ & $\Phi_6$ & $\Phi_7$ \\
\hline
\multicolumn{16}{c}{$g$} \\
\hline
1 & 0.004 & 0.431 & 0.161 & 0.062 & 0.022 & 0.007 & 0.004 & -0.005 & 1.665 & 1.502 & 1.132 & 0.918 & 0.043 & 0.036 & 1.649 \\ 
2 & 0.007 & 0.450 & -0.073 & 0.019 & 0.020 & 0.008 & 0.005 & -0.004 & 1.615 & -1.554 & 0.747 & -0.567 & -1.423 & -13.477 & 1.027 \\ 
3 & 0.004 & 0.424 & 0.134 & 0.074 & 0.052 & 0.032 & 0.025 & 0.014 & 1.802 & 1.615 & 1.113 & 0.924 & 0.714 & 0.511 & 0.289 \\ 
4 & -0.001 & 0.415 & 0.169 & 0.082 & -0.040 & -0.023 & 0.014 & 0.008 & 1.656 & 1.587 & 1.555 & -1.682 & -1.854 & 0.464 & 0.749 \\ 
\hline
\multicolumn{16}{c}{$r$} \\
\hline
1 & 0.003 & 0.429 & -0.162 & 0.064 & 0.023 & -0.007 & 0.005 & -0.005 & 1.642 & -7.820 & 1.409 & 1.078 & 3.979 & -0.036 & -4.029 \\ 
2 & 0.005 & 0.451 & -0.076 & 0.019 & 0.022 & 0.009 & 0.007 & 0.008 & 1.616 & -1.273 & 1.251 & 0.046 & -0.583 & -0.073 & -0.397 \\ 
3 & 0.002 & 0.431 & -0.135 & 0.081 & 0.057 & 0.039 & 0.024 & 0.013 & 1.694 & -1.447 & 1.322 & 1.181 & 1.090 & 1.003 & 0.952 \\ 
4 & -0.001 & 0.422 & -0.173 & -0.082 & -0.037 & 0.022 & -0.017 & -0.009 & 1.477 & 4.767 & -7.740 & -7.727 & 1.217 & -1.784 & -1.636 \\ 
\hline
\multicolumn{16}{c}{$i$} \\
\hline
1 & 0.005 & 0.429 & -0.156 & -0.065 & 0.019 & 0.008 & 0.008 & -0.003 & 1.595 & -1.378 & -1.505 & 1.399 & 1.025 & 0.346 & -2.680 \\ 
2 & 0.007 & 0.462 & -0.080 & 0.024 & 0.022 & 0.005 & -0.005 & 0.002 & 1.599 & -1.124 & 1.579 & 0.527 & -4.821 & 16.604 & 14.585 \\ 
3 & 0.004 & 0.448 & -0.131 & 0.082 & 0.058 & 0.039 & 0.029 & 0.012 & 1.601 & -1.400 & 1.422 & 1.424 & 1.365 & 1.382 & -11.207 \\ 
4 & -0.002 & 0.440 & -0.171 & -0.082 & -0.037 & 0.023 & -0.015 & 0.010 & 1.382 & -1.433 & -1.285 & -1.166 & 1.450 & -1.479 & 1.365 \\ 
\hline
~ \\
~ \\
\end{tabular}
\label{table:fourier_parameters}
\end{table*}

\subsection{Template fitting procedure}
\label{sec:template_fitting_procedure}

Before performing the fit, we set the first-guess $V$-band ($F475W$) amplitude $A_V$ to that of the ground-based $g$-band light curve. Then, we fix the amplitude ratios to $A_I = 0.58 \, A_V$%\label{eq:amp_ratio_VI}
%\end{equation}
~from \citet{Yoachim2009} and to:
\begin{equation}
A_H = \left\{
    \begin{array}{ll}
        0.34 \, A_V & \mbox{if $P \leq 20$ d}  \\
        0.40 \, A_V & \mbox{if $P > 20$ d}
    \end{array}
\right.
\end{equation}
from \citet{Inno2015}. The first-guess phase in $F475W$ is set to the phase in $g$. By comparing CFHT $g$ and $i$ light curves, we derive a small phase lag between $V$ and $I$ of:
%\begin{equation}
$\phi_I = \phi_V +0.027$,
%\label{eq:phase_lag_VI}
%\end{equation}
%which we assume in the following
and we adopt the $H$-band phase lag from \citet{Inno2015} for $F160W$:
%\begin{equation}
$\phi_H = \phi_V + 0.080 - 0.002 \, \log P.$
%\end{equation}
We note that \citet{Soszynski2005} derived a different phase lag of about 0.3 between $H$ and $V$, but that is largely due to the choice of a different reference phase (maximum brightness). In \S\ref{sect:dist} we discuss the sensitivity of our results to the phase lag.

\begin{figure}[t]
\centering
\includegraphics[width=8.6cm]{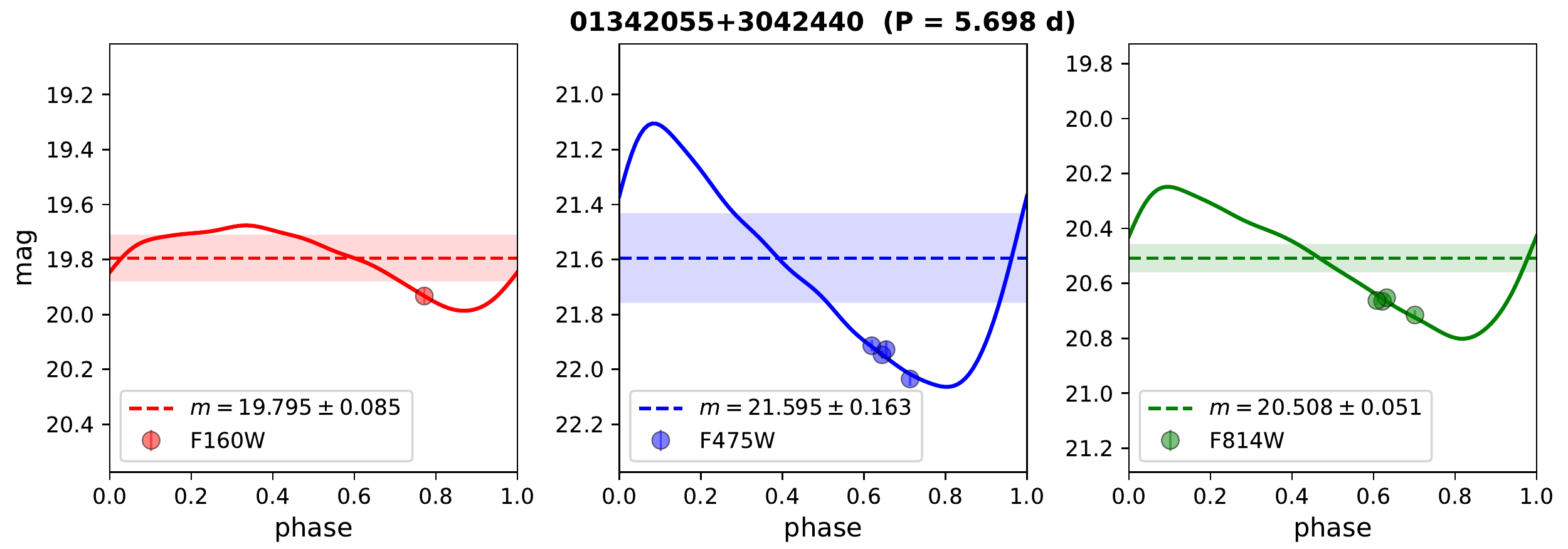} 
\includegraphics[width=8.6cm]{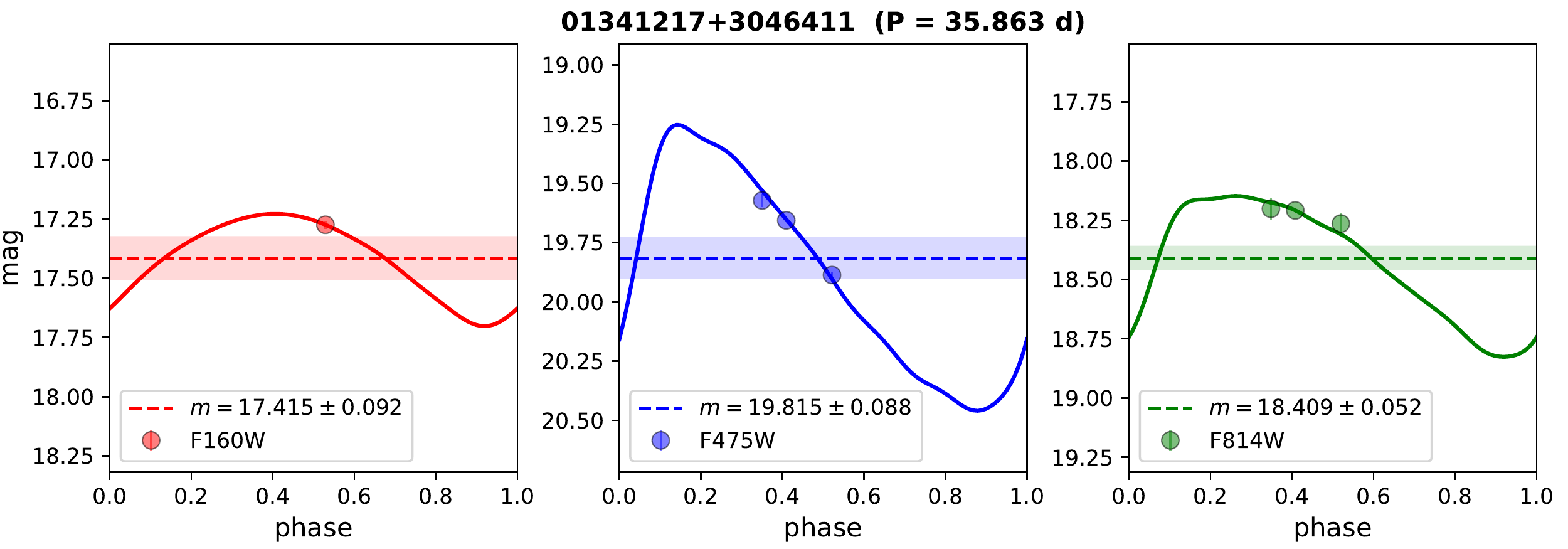} 
\includegraphics[width=8.6cm]{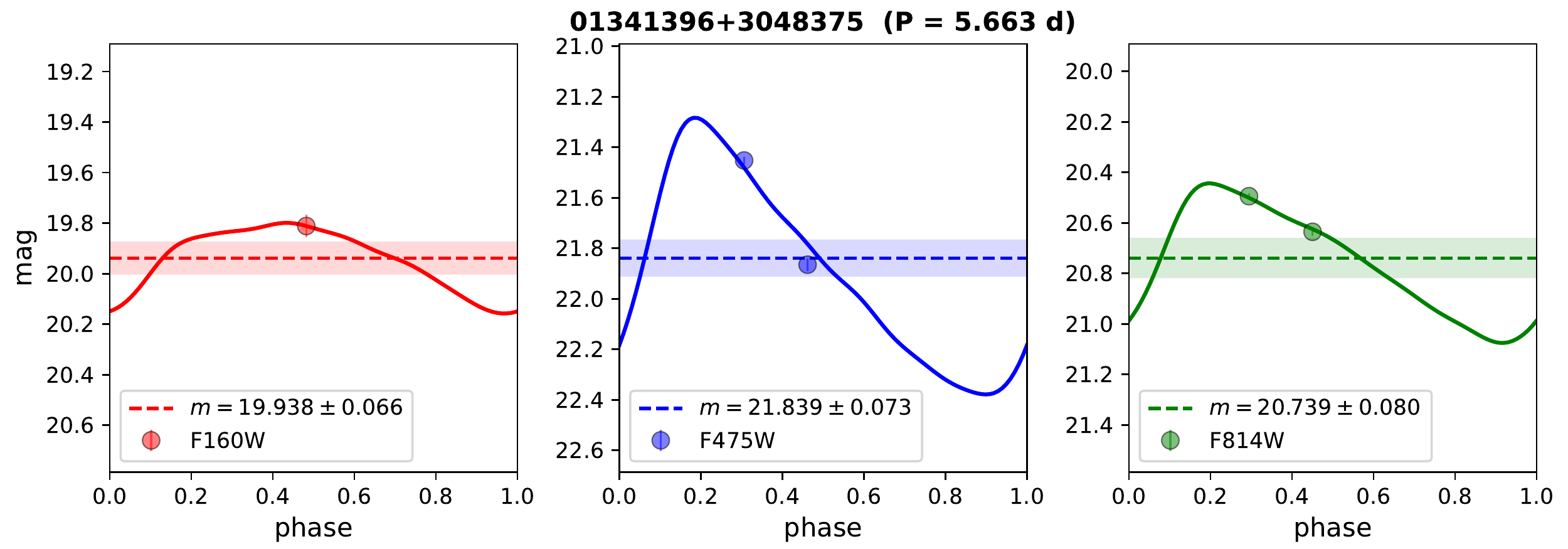} 
\includegraphics[width=8.6cm]{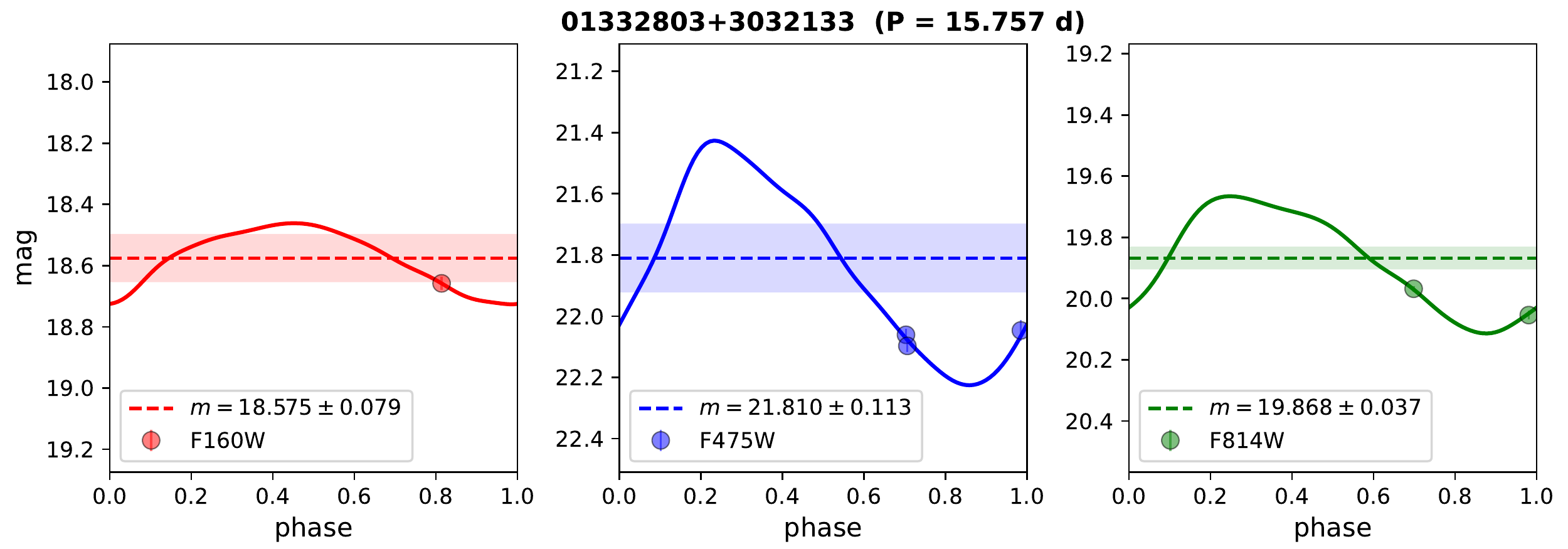} 
\includegraphics[width=8.6cm]{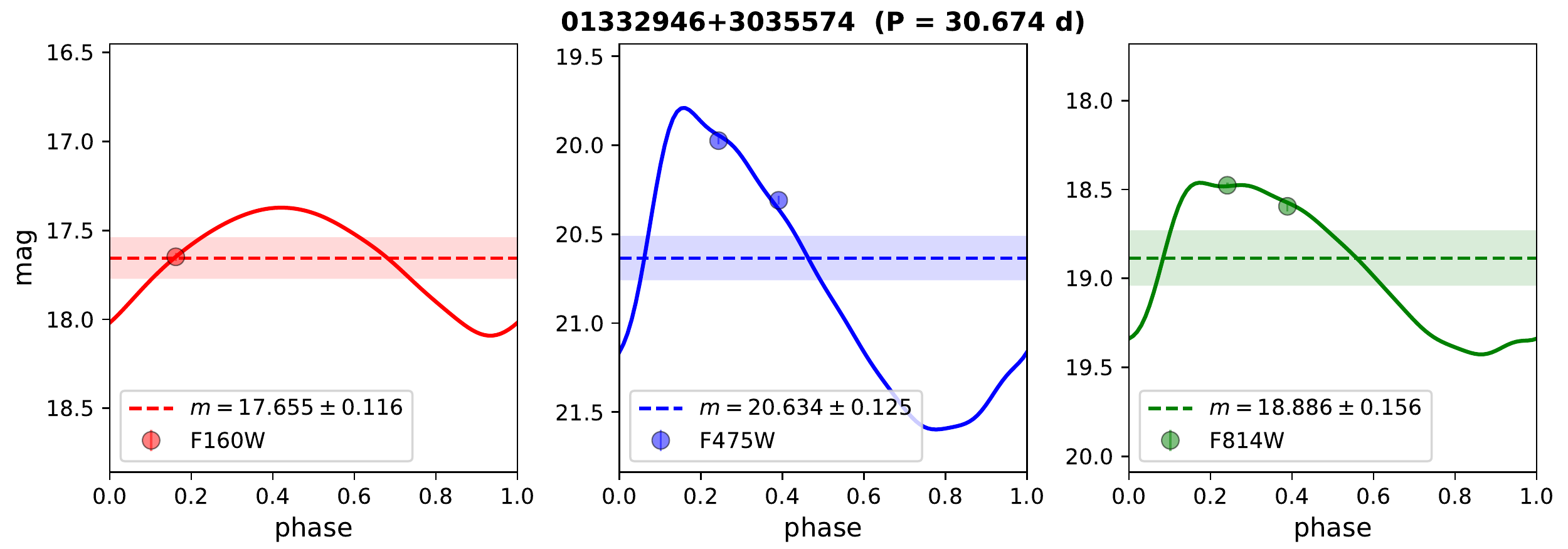} 
\caption{Example of light curves obtained from the template fitting procedure. Their quality is representative of that of the entire sample. The shaded area represents the mean magnitude error in each filter. For a given star, the three filters are shown with the same scale in magnitude.}
\label{fig:example_LCs}
\end{figure}

We fit the templates to the HST measurements in the three filters simultaneously 
% with five free parameters: $A_V$, $\phi_V$, and mean magnitudes in $F160W$, $F475W$ and $F814W$. 
by performing a grid-search on $A_V$ and $\phi_V$, where $\phi_V$ has a narrow, informative prior from the template sample. The amplitude ratios are fixed throughout the procedure and we retain as final parameters the solution that minimizes the quantity:
\begin{equation}
Z = \chi^2_{\rm tot} + Q(A_V)
\end{equation}
where $\, \chi^2_{\rm tot} = \chi^2_H + \chi^2_V + \chi^2_I \,$ and each $\chi^2$ is defined as:
\begin{equation}
\chi^2 =  \sum_i \frac{(O_{i} -C_{i})^2}{\sigma_{i}^2} 
\end{equation}
with $O_i$ the data points, $C_i$ the fit values and $\sigma_i$ the error of each data point. The quantity $Q(A_V)$ is a penalty function that prevents the fitted HST amplitudes to diverge too far from the expected values (i.e. the $g$-band amplitude of each Cepheid). It is defined as:
\begin{equation}
Q(A_V) = \frac{(A_{V, \, \rm fitted} - A_{V, \, \rm ground})^2}{\sigma_A^2}
\end{equation}
The dispersion in the difference in amplitudes is set to $\sigma_A = 0.030 \, \rm mag$ from the ground based sample. Finally, the errors on each apparent magnitude are estimated from  a $\chi^2$ distribution assuming $\chi^2 < \chi^2_{\rm min} + 1$ \citep{Press1992}. Figure \ref{fig:example_LCs} shows a few examples of light curves obtained from the template fitting procedure.

In the following we consider two different subsamples. The \textbf{gold sample} includes Cepheids for which the phase is known with good confidence from ground-based light curves: these Cepheids have at least a valid $g$-band light curve or a valid $i$-band light curve (or both ideally). In the case where only one light curve is available among $g$ and $i$, amplitudes and phases in the missing band can be easily recovered from the relations adopted above. %Eq. \ref{eq:amp_ratio_VI} and Eq. \ref{eq:phase_lag_VI} respectively.
Cepheids of the gold sample must also have a phase uncertainty of $\sigma_{\phi} < 0.05$ to allow for a precise rephasing of HST observations. As their phase and amplitude are assumed to be known from the ground, the grid-search is performed on a limited range of parameters: across [$A_V-0.4; A_V+0.4$] in amplitude and [$\phi_V -0.05; \, \phi_V + 0.05$] in phase. The \textbf{silver sample} includes Cepheids with no $g$ and no $i$ light curves or with a larger phase uncertainty $\sigma_{\phi} > 0.05$. For these stars, the phasing is considered unknown and we perform the grid-search over [0; 1] in phase. The expected amplitude of these silver sample Cepheids is also unknown: as the mean V-band peak-to-peak amplitude of our sample is 0.8 mag, we perform the grid search within [0.3, 1.3] in amplitude, which corresponds to $0.8 \pm 0.5$ mag. After a visual inspection of each light curve, we find 9 Cepheids from the gold sample which appear to have an incorrect phasing (i.e. reaching the boundaries of the grid-search in $\phi_V$): for these stars we allow the search to cover [0; 1] in phase (but generally the final phase stays within $\pm 0.1$ from the first guess) and we find much lower $\chi^2$ and better fit quality. These nine stars are moved to the silver sample.  

Out of the 250 initial PHATTER Cepheids, we only keep the 220 which have optimal ground-based light curves \citep[Table 3 from][]{Pellerin2011}. We rejected 29 stars with only one epoch per filter or with multiple but very close epochs ($\Delta \phi < 0.01$), which did not allow the fit to converge successfully. We also excluded 26 stars for which the fit was not satisfactory and 11 stars which yielded a fitted $V$-band amplitude different by more than 0.5 mag from the expected amplitude from ground-based light curves. This leaves a total of 154 Cepheids which constitute the "gold+silver" sample. The final intensity-averaged mean magnitudes obtained for our sample of Cepheids in $F160W$, $F475W$ and $F814W$ are listed in Table~\ref{table:all_data} (Appendix \ref{Apdx:full_sample}). \\

\begin{figure*}[t]
\centering
\includegraphics[width=16.8cm]{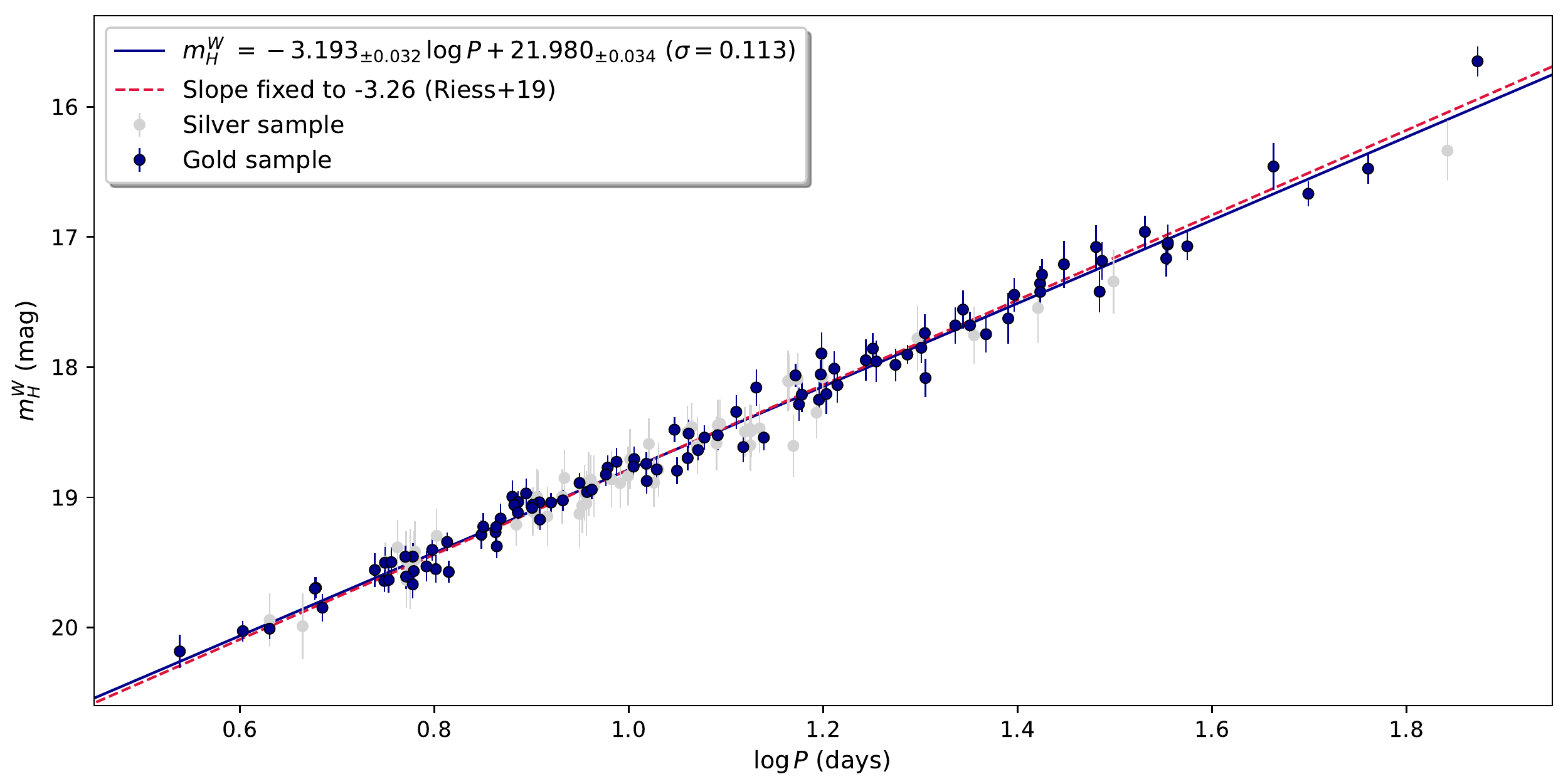} 
\caption{Period-luminosity relation in $m_H^W$ for M33 Cepheids. The dark solid line is the PL fit of the gold + silver sample assuming a free slope, and the red dashed line shows the same fit when the slope is fixed to $-3.26 \, \rm mag/dex$ \citep{Riess2019}. \\   }
\label{fig:PL}
\end{figure*}

%%%%%%%%%%%%%%%%%%%%%%%%%%%
%%%%%%%%%%%%%%%%%%%%%%%%%%%
\section{Period-Luminosity relation and distance to M33}
\label{sec:PL_distance}

\subsection{Photometric transformations to WFC3}

In order to derive the distance to M33, we will compare its PL relation with that established in the LMC, which has the most precise and Cepheid-independent distance measurement \citep{Pietrzynski2019}. The LMC PL relation \citep{Riess2019} is calibrated in the SH0ES photometric system based on HST/WFC3 filters ($F160W$, $F555W$ and $F814W$). On the other hand, the mean magnitudes obtained in the present work from PHATTER photometry were measured with the WFC3/IR camera for the $F160W$ filter and with ACS/WFC for the optical $F475W$ and $F814W$ filters. We transform the PHATTER color $(F475W-F814W)_{\rm ACS}$ into the SH0ES color $(F555W-F814W)_{\rm WFC3}$ using synthetic populations based on PARSEC isochrones generated by the CMD tool\footnote{\href{http://stev.oapd.inaf.it/cgi-bin/cmd}{http://stev.oapd.inaf.it/cgi-bin/cmd}} developed by \citet{Bressan2012} (version v3.7) for HST/WFC3 and HST/ACS bandpasses. We consider a population of Cepheid-like stars with ages of $1-500 \, \rm Myr$, $\log g < 2$, masses of $3-7M_{\odot}$ and temperatures of $4800-6500 \, \rm K$. We derive the following transformation with a scatter of 0.003 mag:
\begin{multline}
\label{eq:transform}
(F555W-F814W)_{\rm WFC3} = 0.065  \\ + 0.658 \, (F475W-F814W)_{\rm ACS}
\end{multline}
The mean Cepheid color of the sample is $(F475W - F814W)_{\rm ACS} = 1.41 \, \rm mag$ (sample standard deviation = 0.29 mag), for which  Eq.~\ref{eq:transform} gives $(F555W - F814W)_{\rm WFC3} = 0.99 \, \rm mag$ (sample standard deviation = 0.19 mag). In the following we adopt the near-infrared HST/WFC3 Wesenheit index defined in \citet{Riess2022} assuming the reddening law from \citet{Fitzpatrick1999} with reddening paramneter $R_V = 3.3$: 
\begin{equation}
m_H^W = F160W - 0.386\, (F555W-F814W)_{\rm WFC3}  
\end{equation}
We also derive optical Wesenheit magnitudes defined in \citet{Riess2019} as:
\begin{equation}
m_I^W = F814W - 1.3\, (F555W-F814W)_{\rm WFC3}  
\end{equation}
~

\begin{table*}[t]
\caption{Calibration of the PL relation in M33 ($m = \alpha \log P + \beta$) and resulting distance modulus. The second column gives the fitted PL slope $\alpha$. The third and fourth columns give the PL intercept $\beta$ when the slope is a free parameter and when the slope is fixed to the LMC value \citep{Riess2019} respectively. \\}
%\centering
\begin{tabular}{c c c c c c c c | c}
\hline
\hline
Band & $\alpha$  & $\beta_{\, \rm free}$ & $\beta_{\, \rm fixed}$ & $\sigma$ & $\chi^2_{\rm dof}$ & N$_{\rm stars}$ & Sample & $\mu_{\rm M33}$ (mag)    \\
\hline
$m_H^W$ & $-3.207 \pm 0.039$ & $21.993 \pm 0.041$ & $22.046 \pm 0.010$ & 0.110 & 0.96 & 99  & Gold   	    & 24.619 $\pm$ 0.030   \\
$m_H^W$ & $-3.193 \pm 0.032$ & $21.980 \pm 0.034$ & $22.048 \pm 0.008$ & 0.113 & 0.72 & 154 & Gold + Silver & \textbf{24.622 $\pm$ 0.030} \\
\hline
$m_I^W$ & $-3.167 \pm 0.046$ & $21.909 \pm 0.051$ & $22.065 \pm 0.014$ & 0.141 & 1.32 & 99  & Gold          & 24.617 $\pm$ 0.032 \\
$m_I^W$ & $-3.179 \pm 0.037$ & $21.933 \pm 0.041$ & $22.072 \pm 0.010$ & 0.130 & 1.01 & 154 & Gold + Silver & 24.624 $\pm$ 0.030 \\
\hline
~ \\
\end{tabular}
\label{table:PL_distances}
\end{table*}

\subsection{Count-rate Nonlinearity Correction}
\label{sect:CRNL}

The WFC3-IR instrument, which is used in the SH0ES distance ladder to measure nearby bright Cepheids as well as distant stars in supernovae host galaxies, is affected by count-rate nonlinearity (CRNL, or reciprocity failure). This effect dims faint sources relative to bright ones due to a decreased photon collection efficiency. Its most recent calibration gives a correction of $0.0077 \, \rm mag/dex$ \citep{Riess2019CRNL}. In order to derive the distance to M33, $m_H^W$ magnitudes in the LMC and in M33 must be corrected for the CRNL consistently. The PL intercept of $15.898 \, \rm mag$ calibrated in the LMC by \citet{Riess2019} does not include the CRNL term \citep[despite its note to the contrary, see footnote to table 5 of][]{Yuan2020}. Finally, M33 Cepheids are fainter than LMC Cepheids by about 2 dex \citep{Li2021}, so we add $0.015 \pm 0.005$ mag to the LMC intercept from \citet{Riess2019} to account for this difference (which is equivalent to subtracting 0.015 mag to our $m_H^W$ apparent magnitudes in M33). \\

\subsection{Geometric correction}

We take into account the position of each Cepheid relative to the center of M33 \citep[$\alpha=23.4625^\circ$, $\delta=30.6602^\circ$ from][]{VanDerMarel2019} by applying a geometric correction. Our HST sample is located very near the center of M33 and the galaxy has a moderate inclination, which limits the effects of projection and of reddening. We adopt an inclination angle of $i = 57\pm4 ^{\circ}$ and a position angle of $\rm PA = 22.5 ^{\circ}$ \citep[both from][]{Kourkchi2020b}, obtaining a mean correction of 0.0007 mag with a dispersion of 0.003 mag, with values ranging between -0.005 mag and +0.008 mag. A positive geometric correction corresponds to a Cepheid farther than the center of M33. \\

\subsection{Period-luminosity relation in M33}
\label{sect:PL}

In this section, we adopt the apparent Wesenheit $m_H^W$ mean magnitudes obtained from template fitting for our sample of M33 Cepheids. \corr{} We include an additional $0.07 \, \rm mag$ in quadrature to all magnitude errors to account for the finite width of the instability strip \citep{Riess2019}. The PL relation is then calibrated for the two subsamples defined in \S\ref{sec:template_fitting} (gold and silver samples). 

For the gold sample we obtain a slope of $-3.207 \pm 0.039 \, \rm mag/dex$ in $m_H^W$, which agrees well with that derived by \citet{Riess2019} in the LMC. The PL scatter is $0.110 \, \rm mag$ for a total of 99 stars. Including the silver sample yields a slightly shallower slope of $-3.193 \pm 0.032 \, \rm mag/dex$, still in good agreement with \citet{Riess2019}, and slightly raises the scatter to $0.113 \, \rm mag$. In the optical Wesenheit index $m_I^W$, we adopt the slope of $-3.31 \, \rm mag/dex$ as well as the LMC PL intercept of 15.935 from \citet{Riess2019} and we obtain a PL dispersion of 0.13 mag for the gold + silver sample, and 0.14 for the gold sample. We note that \citet{Li2021} obtained a PL scatter of $0.13 \, \rm mag$ in M31 for their gold sample with 42 Cepheids, which shows the great precision of our PL calibration. Our PL coefficients are listed in Table~\ref{table:PL_distances} and the PL relation is shown in Fig.~\ref{fig:PL}.    \\

\subsection{Distance to M33}
\label{sect:dist}

To obtain the distance modulus for M33 ($\mu_{\rm M33}$), we compare the intercept of our $m_H^W$ PL relation in M33 with that of the LMC obtained by \citet{Riess2019}, $m_H^W = 15.898-3.26 \log P$. We add the CRNL term of 0.015 mag to the LMC intercept to account for the difference in brightness between LMC and M33 Cepheids (see \S\ref{sect:CRNL}). We fix our PL slope to $-3.26$ for consistency with the LMC and we derive:
\begin{equation}
(\mu_{\rm M33} - \mu_{\rm LMC}) = (\beta_{\rm M33} - \beta_{\rm LMC}) +  \Delta m
\label{eq:dist}
\end{equation}
where $\mu_{\rm LMC} = 18.477 \pm 0.026 \, \rm mag$ is the most direct and precise geometric distance to the LMC available \citep{Pietrzynski2019}. The term $\Delta m$ is the correction for the difference in metallicity between M33 and LMC Cepheids:
\begin{equation}
\Delta m = - \gamma \, (\rm [O/H]_{M33} - [O/H]_{LMC})
\end{equation}
\citet{Romaniello2022} gives $\rm [O/H]_{LMC}=-0.32 \pm 0.01 \,  dex$ from a sample of 89 Cepheids. In M33 we use the metallicity gradient by \citet{Bresolin2011}
%\footnote{We selected the gradient from \citet{Bresolin2011} as it is near the median of the gradients listed in Table~\ref{table:M33_gradients}. The \citet{Toribio2016} gradient returns a very similar result.} 
which gives:
\begin{equation}
12 + \rm \log (O/H) = 8.50_{\pm 0.02} - 0.045_{\pm 0.006} \, R_{kpc}
\end{equation}
relative to 8.69 for solar \citep{Asplund2009}. For our HST sample, [O/H] metallicities range from -0.20 dex to -0.36 dex. We adopt the mean metallicity of $\rm [O/H]_{M33}=-0.27 \pm 0.03 \, dex$. Using the metallicity correction of $\gamma = -0.217 \pm 0.046 \, \rm mag/dex$ from \citet{Riess2022}, we obtain a correction of $\Delta m = 0.011 \pm 0.007 \, \rm mag$. Using the metallicity correction from \citet{Breuval2022} returns $\Delta m = 0.014 \, \rm mag$ which is very similar, but we adopt the former %$\gamma$ correction by \citet{Riess2022}
as it is more suited for measurements in the Wesenheit $m_H^W$ index. 
%We also note that using any other metallicity gradient from Table~\ref{table:M33_gradients} changes the final distance modulus by 0.002 mag at most. 
From Eq. \ref{eq:dist} we obtain a final distance modulus of $24.622 \pm 0.030$ mag based on the gold and silver samples combined in the NIR Wesenheit index $m_H^W$. Using only the pure gold sample results in a very similar value (see Table~\ref{table:PL_distances}). Finally, the optical Wesenheit index yields a very consistent distance of $24.617 \pm 0.032$ and $24.624 \pm 0.030$ mag from the gold sample only and gold + silver samples combined, respectively. We retain as final distance modulus the one from the gold + silver sample in the $m_H^W$ filter, $24.622 \pm 0.030 \, \rm mag$, as it is based on the most precise PL intercept. The full error budget is detailed in Table~\ref{tab:error_budget}.

In \S\ref{sec:template_fitting_procedure}, we adopted the phase lag from \citet{Inno2015} between $H$ and $V$ light curves ($\sim 0.08$, with a scatter of 0.03 in the relation), derived by using the mean magnitude along the rising branch as a reference for the phase. We made this choice for consistency, as we also use the NIR templates by \citet{Inno2015}. On the other hand, \citet{Soszynski2005} found a larger phase lag of $\sim 0.3$ (with a larger scatter of about $0.1$ mag) by using the phase at maximum brightness as a reference. 
%\scmod {We note that these two methods to phase the data are very different as shown by \citet{Inno2015}. Using the \citet{Soszynski2005} phase lag yields a distance modulus of $24.627 \pm 0.030$ mag for M33, which differs by only 0.006 mag from our final value.}
\citet{Inno2015} show that these two methods to phase the data are very different; using the \citet{Soszynski2005} phase lag with the \citet{Inno2015} template is formally inconsistent, but it leads to a difference of only 0.006 mag in the distance modulus.

We attempted to independently constrain the value of the metallicity effect of the PL relation, $\gamma$. However, the very narrow range of abundances spanned by these Cepheids yields uncertainties $\sigma(\gamma)$ of $0.24 - 0.42$~mag/dex which do not improve upon previous measurements \citep{Breuval2021, Breuval2022, Riess2022}. \\

\begin{table}[t!]
\caption{Error budget for the distance to M33.}
\centering
\begin{tabular}{l c c }
\hline
\hline
Error & Value & Source    \\
\hline
LMC DEBs                & 1.20 \% & \citet{Pietrzynski2019}  \\
LMC PLR mean            & 0.41 \% & \citet{Riess2019} \\
M33 PLR mean            & 0.38 \% & Measured here  \\ 
Metallicity correction  & 0.33 \% & \citet{Riess2022} \\
CRNL across 2 dex       & 0.23 \% & \citet{Riess2019CRNL}   \\
\hline
\textbf{Total}          & \textbf{1.38} \% \\
\hline
\end{tabular}
\label{tab:error_budget}
\end{table}

\begin{table*}[]
\caption{M33 distance modulus from the literature. The last column gives the LMC distance modulus adopted to obtain the distances given in the first column. \\}
\centering
\begin{tabular}{c l c c}
%~\\
\hline
\hline
$\mu_{\rm M33}$  & Reference & Method & $\mu_{\rm LMC}$ \\
\hline
$24.64 \pm 0.09$ & \citet{Freedman1991} & Cepheids 	& 18.50 \\
$24.56 \pm 0.10$ & \citet{Freedman2001} & Cepheids 	& 18.50 \\
$24.65 \pm 0.12$ & \citet{Macri2001PhD} & Cepheids 	& 18.50  \\
$24.50 \pm 0.06$ & \citet{McConnachie2004} & TRGB  & - \\
$24.67 \pm 0.08$ & \citet{Sarajedini2006} & RR Lyrae 	& 18.50 \\
$24.71 \pm 0.04$ & \citet{Rizzi2007} & TRGB & - \\
$24.53 \pm 0.11$ & \citet{Scowcroft2009} & Cepheids 	& 18.40 \\
$24.84 \pm 0.10$ & \citet{U2009} 		& TRGB	& 18.50 \\
$24.76 \pm 0.05$ & \citet{Pellerin2011} & Cepheids & 18.50      \\
$24.57 \pm 0.05$ & \citet{Conn2012} & TRGB & - \\
$24.62 \pm 0.07$ & \citet{Gieren2013} & Cepheids & 18.50 \\
%$24.67 \pm 0.07$ & \citet{DeGrijs2014} & Cepheids, RR Lyrae, TRGB & 18.50  \\
$24.62 \pm 0.06$ & \citet{Bhardwaj2016} & Cepheids   	& 18.47 \\
$24.80 \pm 0.06$ & \citet{Yuan2018} & Miras 	& 18.493 \\
$24.57 \pm 0.06$ & \citet{Zgirski2021} & JAGB & 18.477 \\
$24.67 \pm 0.05$ & \citet{Lee2022} & JAGB 	& - \\
$24.72 \pm 0.07$ & \citet{Lee2022} & TRGB	& -  \\
$24.71 \pm 0.04$ & \citet{Lee2022} & Cepheids 	& -  \\
$24.67 \pm 0.06$ & \citet{Ou2023} & Miras	& 18.49 \\
\textbf{24.622 $\pm$ 0.030} & \textbf{Present work} & \textbf{Cepheids} & \textbf{18.477} \\
\hline
~ \\
~ \\
\end{tabular}
\label{table:M33_distance}
\end{table*}

% \begin{table*}[t!]
% \caption{(Left): Metallicity gradients from the literature for M33, in the form: $12 + \log(\rm O/H) =$ $\beta + \alpha R_{\rm kpc}$. (Right): Metallicity effect ($\gamma$) derived assuming a given gradient, and metallicity range covered by the sample.}
% \centering
% \begin{tabular}{c c c c  | c c c c}
% \hline
% \hline
% $\beta$ & $\alpha$ & Reference & Method & $\gamma$ & $\Delta$[O/H] & [O/H] range  \\
% (dex) & (dex/kpc) &  &  & (mag/dex) & (dex) & (dex)  \\
% \hline
% $8.44_{\pm 0.06}$ & $-0.031_{\pm 0.013} $ & \citet{Magrini2009} & Planetary Nebulae & $-0.126_{\pm 0.417}$ & $0.11$ & [$-0.37 ; -0.26$]  \\
% $8.50_{\pm 0.02}$ & $-0.045_{\pm 0.006} $ & \citet{Bresolin2011} & \ion{H}{2} regions 		& $-0.075_{\pm 0.277}$ & $0.15$ & [$-0.36 ; -0.21$]  \\
% $8.76_{\pm 0.07}$ & $-0.048_{\pm 0.019} $ & \citet{Toribio2016} & \ion{H}{2} regions (RLs) & $-0.084_{\pm 0.274}$ & $0.15$ & [$-0.11; +0.05$] \\
% $8.52_{\pm 0.03}$ & $-0.053_{\pm 0.010} $ & \citet{Toribio2016} & \ion{H}{2} regions (CELs)& $-0.062_{\pm 0.239}$ & $0.18$ & [$-0.37; -0.19$] \\
% $8.46_{\pm 0.02}$ & $-0.025_{\pm 0.004}$ & \citet{Lin2017} & \ion{H}{2} regions 			& $-0.138_{\pm 0.496}$ & $0.09$ & [$-0.32; -0.23$] \\
% $8.59_{\pm 0.02}$ & $-0.037_{\pm 0.007}$ & \citet{Rogers2022} & \ion{H}{2} regions 		& $-0.090_{\pm 0.337}$ & $0.13$ & [$-0.24; -0.11$] \\
% \hline
% ~ \\
% \end{tabular}
% \label{table:M33_gradients}
% \end{table*}

\subsection{Comparison with the literature}
\subsubsection{Cepheids, RR Lyrae, Miras}

\begin{figure}[t]
\centering
\includegraphics[width=8.4cm]{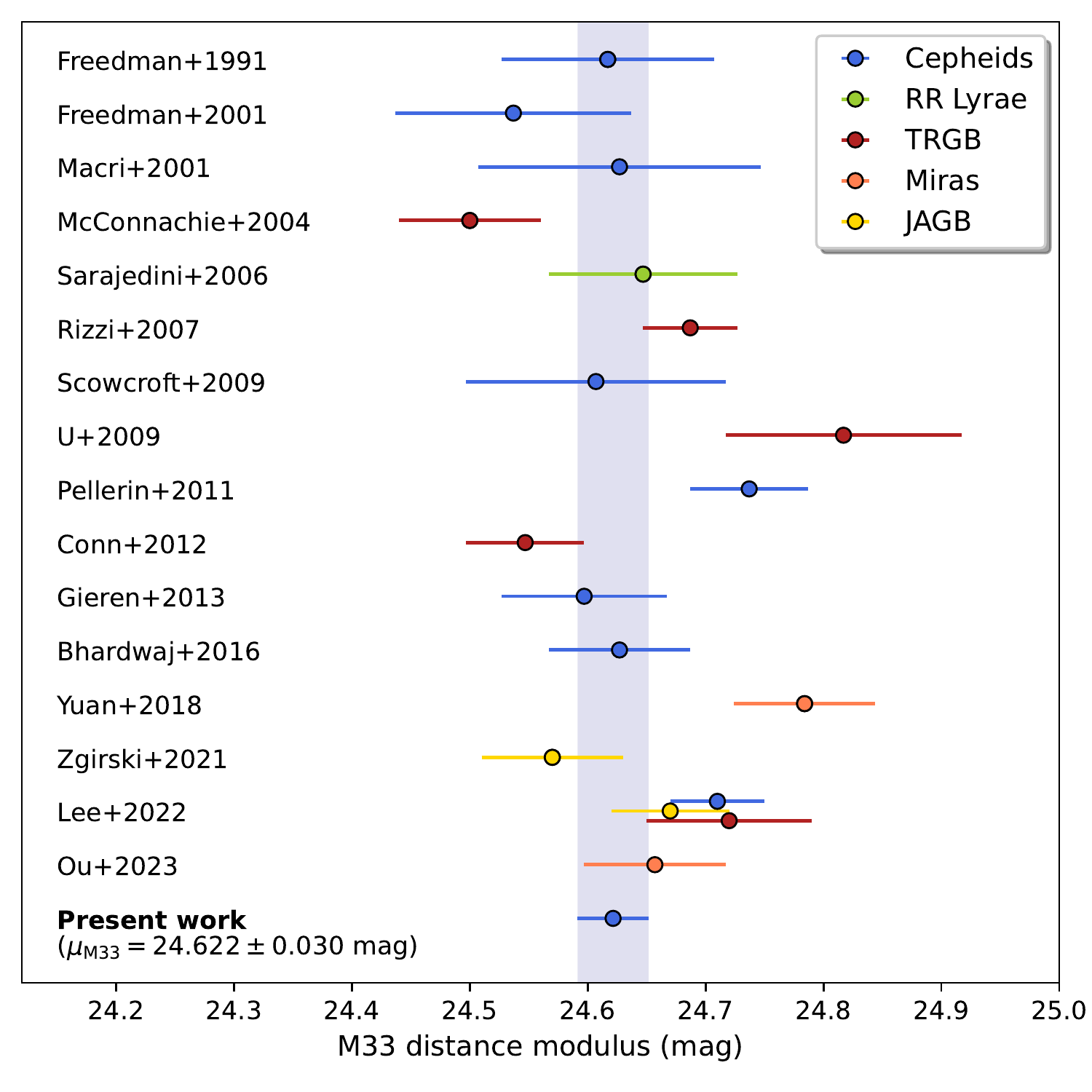} 
\caption{M33 distance modulus from the present work compared with values from the literature. All values shown here are rescaled to the recent LMC distance modulus of 18.477 from \citet{Pietrzynski2019} (see their original adopted LMC distance in Table~\ref{table:M33_distance}).}
\label{fig:distance_M33}
\end{figure}

Figure \ref{fig:distance_M33} shows our final distance modulus for M33 and compares it with other values from the literature based on various indicators (listed in Table~\ref{table:M33_distance}, all corrected to a common LMC distance modulus of 18.477 mag \citep{Pietrzynski2019}. Our distance agrees very well with other estimates based on Cepheids, especially with \citet{Freedman1991}, \citet{Macri2001PhD}, \citet{Scowcroft2009} and \citet{Bhardwaj2016}. In particular, Cepheids appear to provide the most consistent distance measurements among all other distance indicators. The error of the \citet{Pellerin2011} distance was revised to 0.05 mag to include the systematic uncertainties from the photometric comparison with \citet{Massey2006} in their section 3. The Cepheid distance to M33 by \citet{Lee2022} is larger than our value by 0.10 mag ($1.7 \sigma$). This difference matches the size and direction of the metallicity dependence of Cepheids, $\sim$ -0.2 mag/dex \citep{Breuval2022}, where metal poor Cepheids are fainter.   (It is the wrong direction for crowding which, if uncorrected, would make ground-based observations of Cepheids appear too bright).  The \citet{Lee2022} sample is in the outer regions of M33, around 5 kpc from the center, which corresponds to a metallicity of about [O/H] $\sim -0.4 \, \rm dex$, somewhat metal poor. They derive their distance to M33 relatively to the absolute PL calibration by \citet{Monson2012}, based on Milky Way Cepheids which are metal rich with [O/H] $\sim 0.1 \, \rm dex$. This 0.5 dex difference in metallicity produces an expected difference of $\sim$ 0.10 mag, bringing it into agreement with the study here.  An alternative to correcting for metallicity is to use a reference with a similar metallicity as M33, i.e., the metal poor Cepheids in the LMC, $\sim$ -0.3 dex, and the geometric DEB distance as a reference for the \citet{Lee2022} sample which we find yields 24.65 mag in good agreement with our result.  We also find good agreement with the RR Lyrae distance by \cite{Sarajedini2006}. Finally, the two Mira-based distances \citep{Yuan2018, Ou2023} differ by 0.13 mag, which can be attributed to the use of different data sets and methodologies, differences in periods and possible calibration systematics between ground and space-based data. \\

%: Their sample is therefore more metal-poor than ours: the gradient by \citet{Bresolin2011} gives a mean metallicity [O/H] of $-0.28$ dex and $-0.42$ dex at 2 kpc and 5 kpc respectively. Assuming the metallicity effect of $-0.217$ mag/dex by \citet{Riess2022} (metal-poor Cepheids are fainter), this metallicity difference shifts their distance modulus by 0.030 mag. Correcting the \citet{Lee2022} distance modulus from the metallicity effect would reduce their value to about 24.71 mag ($1.8 \sigma$ from our result). }

\subsubsection{Tip of the Red Giant Branch}

Distance moduli based on TRGB show a larger dispersion than Cepheid-based measurements with differences as large as 0.34 mag between different studies. For example, \citet{McConnachie2004} derive a distance modulus of $24.50 \pm 0.06$ mag based on an annulus region between $0.5-0.8^{\circ}$ in the outer disk of M33. They investigate the impact of crowding and conclude that this effect is negligible in regions farther than 0.5$^{\circ}$ from the center of M33. They also rule out the possibility of contamination from AGB stars. More recently, \citet{Lee2022} selected a TRGB field in the southern part of M33 at a distance of about $0.25-0.45^{\circ}$ from the galactic center, and therefore no less likely to be affected by blending than the \citet{McConnachie2004} sample, yet obtained a much higher distance modulus of $24.72 \pm 0.06$ mag. We speculate some of this difference could be attributed in part to a difference in sample color, as the redder color cut from \cite{Lee2022} extends further into the region where the TRGB is fainter and may benefit from a color correction \citep[see ][]{Jang2017}. We note that although the mean color of the sample from \cite{Lee2022} is located at the boundary where \cite{Jang2017} states a color correction is not necessary, \cite{Jang2017} do not calibrate their color relation based on the mean color of their entire sample. It is also possible the different TRGB-based measurements are due to population differences as recently seen in \citet{Hoyt2023}, \citet{Anderson2023} and \citet{Wu2022} which all identify significant, intrinsic variations in the TRGB brightness with location, sub-RGB-population or the apparent ratio of RGB to AGB stars (that may relate to age or metallicity). \citet{Durbin2020} also identify additional systematics associated with the TRGB by comparing different calibration approaches. The $\chi^2$ of the TRGB measurements around their weighted mean value is 3.67, compared to 1.42 for Cepheid-based distances (see Table~\ref{tab:chi2_M33_distances_literature}). Finally, both measurements based on the JAGB method \citep{Zgirski2021, Lee2022} agree to $1\sigma$ with our value.  \\

\begin{table}[t!]
\caption{Statistics of the M33 distances from the literature based on various distance indicators. Column (2) is the weighted mean distance modulus for each indicator. Columns (3) and (4) respectively give the $\chi^2$ of the measurements from the literature around the weighted mean distance modulus and around the distance modulus of the present paper. Column (5) gives the number of estimates from the literature considered for each distance indicator.    \\}
\centering
\begin{tabular}{l c c c c }
\hline
\hline
Indicator & Mean $\mu_{\rm M33}$ & $\chi^2_{mean,\nu}$ & $\chi^2_{B23,\nu}$ & N   \\
(1) & (2) & (3) & (4) & (5) \\
\hline
Cepheids & 24.644 & 1.08 & 1.39 & 9   \\
TRGB     & 24.639 & 3.67 & 2.96 & 5   \\
JAGB     & 24.625 & 1.65 & 0.84 & 2   \\
Miras    & 24.721 & 2.24 & 3.84 & 2  \\
\hline
~ \\
\end{tabular}
\label{tab:chi2_M33_distances_literature}
\end{table}

\subsubsection{Eclipsing binaries}

A distance to M33 based on detached eclipsing binaries (DEBs) was published by \citet{Bonanos2006}. However, unlike the well-established LMC and SMC distances by \citet{Pietrzynski2019} and \citet{Graczyk2020} respectively, which are based on late-type DEBs (a purely empirical method calibrated geometrically through red giant interferometry), the M33 distance by \citet{Bonanos2006} relies on early-type DEBs which depend on surface flux calculated from non-local thermodynamic equilibrium models and is strongly affected by model uncertainties. Therefore we will not compare our result with this measurement as we limit our comparisons to empirical measures.

In the future, the ability to measure many primary distance indicators in the same host offers the best chance to identify and rectify differences between distance indcators.  M33 offers one of the best such opportunities. \\

\section{Photometric bias from Cluster Cepheids}
\label{sec:clusters}

\subsection{Motivation}

Contamination from crowded backgrounds such as star clusters can bias photometric measurements of Cepheids in nearby galaxies. Photometric measurements of extragalactic Cepheids are usually corrected for crowding effects by injecting artificial stars in the vicinity of Cepheids and by remeasuring their contribution \citep{Riess2009}. However, this test may not properly reproduce the impact of stars physically associated with Cepheids, which might be unresolved and whose light properties might differ from those of the background field stars. \citet{Anderson2018} found that blending due to cluster Cepheids was responsible for a 0.23\% overestimate of $H_0$, although cluster Cepheids are a relatively rare phenomenon. They concluded that chance superposition of Cepheids with clusters was not a limit for a 1\% measurement of the Hubble constant. In this section, we estimate the blending contribution from cluster Cepheids in M33: we measure the occurrence rate of Cepheids in clusters in M33 and we derive the typical flux contribution of the clusters in order to determine by how much they affect Cepheid photometry and our M33 distance modulus.   \\

\subsection{Crossmatch of Cepheid and cluster catalogs}
\label{sec:clusters_regular}

First, we estimate the number of M33 Cepheids located in or near clusters. \citet{Anderson2018} reported a fraction of 2.4\% cluster Cepheids in the M31 galaxy, lower than in the Milky Way, LMC and SMC (with 8.5\%, 11\% and 6\% respectively). In order to obtain the fraction of cluster Cepheids in M33, we crossmatch an initial sample of 609 fundamental mode Cepheids from \citet{Pellerin2011} with the catalog of 2137 star clusters in M33 from \citet{Johnson2022}. We adopt a separation of $\theta_{\rm sep} < 1.2 \, r_{\rm ap}$ as membership criteria \citep{Senchyna2015}, with $r_{ap}$ the mean cluster radius provided in \citet{Johnson2022}. From this crossmatch we find a total of 10 cluster Cepheids, listed in Table~\ref{tab:cluster_ceps_regular}.  \\

\begin{table}[t!]
\caption{Cluster Cepheids found by crossmatching Cepheids from \citet{Pellerin2011} and star clusters from \citet{Johnson2022} with $\theta_{\rm sep} < 1.2 \, r_{\rm ap}$, where $r_{\rm ap}$ is the average cluster radius.}
\centering
\begin{tabular}{c c c c}
\hline
\hline
Cepheid & Cluster & $r_{\rm ap}$ (\arcsec\ ) & $\theta_{\rm sep}$ (\arcsec\ )   \\
\hline
01334331+3043559 & J22-241 & 1.17 & 0.85 \\ 
01340959+3036215 & J22-477 & 1.55 & 1.46 \\ 
01340060+3050079 & J22-521 & 1.83 & 1.77 \\ 
01332060+3034584 & J22-665 & 1.49 & 0.12 \\ 
01342512+3034381 & J22-722 & 1.41 & 0.27 \\ 
01335311+3048343 & J22-836 & 1.43 & 0.49 \\ 
01335809+3045568 & J22-900 & 1.31 & 1.51 \\ 
01332768+3034238 & J22-1464 & 1.32 & 0.26 \\ 
01334212+3032109 & J22-1492 & 1.60 & 1.74 \\ 
01342784+3041012 & J22-1762 & 1.35 & 1.04 \\ 
\hline
\end{tabular}
\label{tab:cluster_ceps_regular}
\end{table}

\begin{figure}[t!]
\centering
\includegraphics[width=8.5cm]{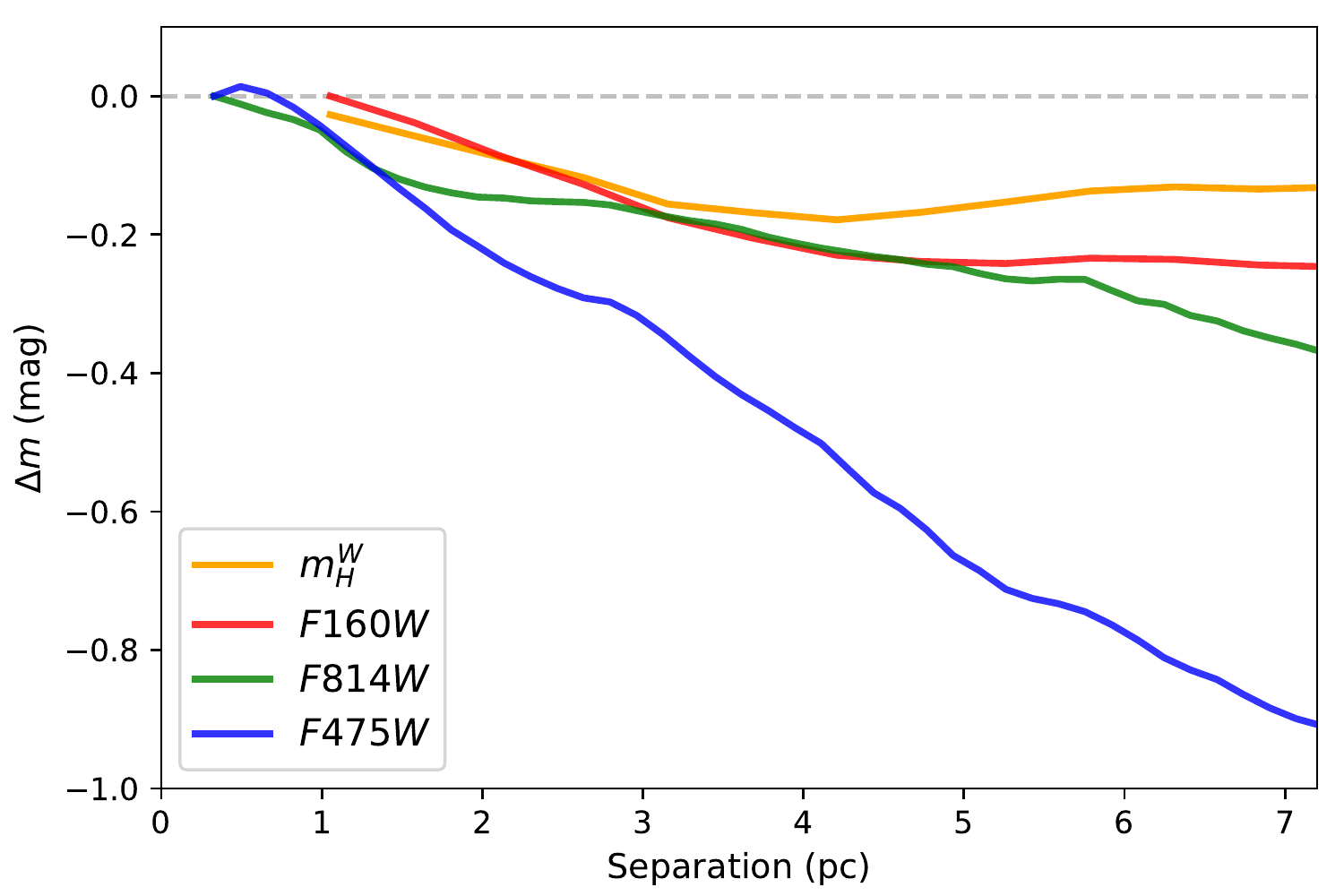} 
\caption{Average cluster contribution (curve of growth) for the 10 confirmed M33 cluster Cepheids (Table~\ref{tab:cluster_ceps_regular}) in the $m_H^W$, $F160W$, $F475W$, $F814W$ filters.}
\label{fig:curve_of_growth_regular}
\end{figure}

\subsection{Creation of stamp images}

We identified these 10 cluster Cepheids in the PHATTER mosaics and we produced stamp images (cutouts centered on each Cepheid) in each filter. The HST stamp images of the 10 crossmatched cluster Cepheids are shown in Fig.~\ref{fig:stamps_pairs} (Appendix \ref{Apdx:stamps}), where the clusters are easy to identify by eye. In particular, UV filters ($F275W$ and $F336W$) are well suited for detecting hot blue cluster stars, while background red giant stars with luminosity similar to that of the Cepheid may contribute more in the infrared. We note that some stamps are blank because the Cepheid is located outside the limit of the PHATTER fields in NIR and UV.  \\

\subsection{Visual inspection of stamp images}
\label{sec:clusters_extra}

In order to make sure that the cluster count is complete, we inspected each Cepheid stamp image for the presence of any additional undetected clusters. We report an additional 13 suspect cluster Cepheids, listed in Table~\ref{tab:cluster_ceps_extra} and shown in Fig.~\ref{fig:stamps_likely} (Appendix \ref{Apdx:stamps}). Three of them (01343182+3043050, 01343169+3043002, and 01340910+3036296) are listed in the \citet{Johnson2022} catalog but at a distance greater than $1.2 r_{\rm ap}$ from the Cepheid (on average at $\sim 2 r_{\rm ap}$), therefore they were not found by the crossmatch procedure. Two of them are also listed in the \citet{Sarajedini2007} catalog at about 1\arcsec\ and 4.5\arcsec\ from the Cepheid (this catalog does not provide the cluster radii). \\

\subsection{Flux contribution from the clusters}

We followed the approach used in \citet{Anderson2018} (see their \S3.2.1) to separate three contributions: the flux from the Cepheid, the flux of the cluster, and the background contribution. The average cumulative light contribution from clusters $\Delta m$ (or curve of growth) is obtained using a series of apertures of increasing radius, starting from the Cepheid in the center ($r=1$ pixel) to a radius of about 2\arcsec\ . We note that $\Delta m$ can be negative (if a light contribution from the cluster is detected) or positive (if the cluster flux is low or if its location is statistically sparser than the nearby environment of the Cepheid). At the distance to M33 found in \S\ref{sect:dist}, a separation of 1\arcsec\ corresponds to 4.1 pc along the major axis and 3.7 pc along the minor axis as projected on the plane of the disk. \citet{Anderson2018} find that the cluster light contribution in M31 flattens off at a separation of about 3.8 pc, which corresponds to approximately twice the average cluster radius.

Fig.~\ref{fig:curve_of_growth_regular} shows the average light contribution from the 10 confirmed M33 clusters (\S\ref{sec:clusters_regular}, Table~\ref{tab:cluster_ceps_regular}). Similarly to \citet{Anderson2018}, the contribution in the optical is significant, with $\Delta m$ around $-0.5$ mag in $F475W$ at a separation of 4 pc, and becomes lower towards the near infrared with $\Delta m = -0.20$ mag and $-0.17$ mag in $F814W$ and $m_H^W$ respectively. These values are about a factor of 2 smaller than those found by \citet{Anderson2018} in M31 at the same separation. 

\begin{table}[t!]
\caption{Cluster Cepheids found by inspecting visually the image cutouts from PHATTER mosaics centered on each Cepheid. The second column indicates if the Cepheid is found nearby a cluster listed in \citetalias{Johnson2022} or in \citetalias{Sarajedini2007}, and "$-$" indicates that no known cluster was found around this Cepheid in the literature. \\}
\centering
\begin{tabular}{c c c c}
\hline
\hline
Cepheid & Cluster & $r_{\rm ap}$ (\arcsec) & $\theta_{\rm sep}$ (\arcsec\ )   \\
\hline
01333896+3034140 & $-$ & $-$ & $-$  \\
01343182+3043050 & J22-40 & 1.60 & 3.73 \\
01333015+3038039 & $-$ & $-$ & $-$ \\ 
01333438+3035307 & $-$ & $-$ & $-$ \\
01333433+3034270 & $-$ & $-$ & $-$ \\ 
01343169+3043002 & J22-40 & 1.60 & 2.96 \\
01340910+3036296 & J22-27, S07-325 & 2.23 & 4.71 \\ 
% & S07-325 & $-$ & 4.55 \\
01341217+3036362 & $-$ & $-$ & $-$ \\
01340474+3049181 & S07-292 & $-$ & 1.12 \\
01333348+3033210 & $-$ & $-$ & $-$ \\
01334821+3038001 & $-$ & $-$ & $-$ \\
01342988+3047541 & $-$ & $-$ & $-$ \\
01340084+3049551 & $-$ & $-$ & $-$ \\
\hline
~ \\
~ \\
\end{tabular}
\label{tab:cluster_ceps_extra}
\end{table}

\begin{figure}[t!]
\centering
\includegraphics[width=8.5cm]{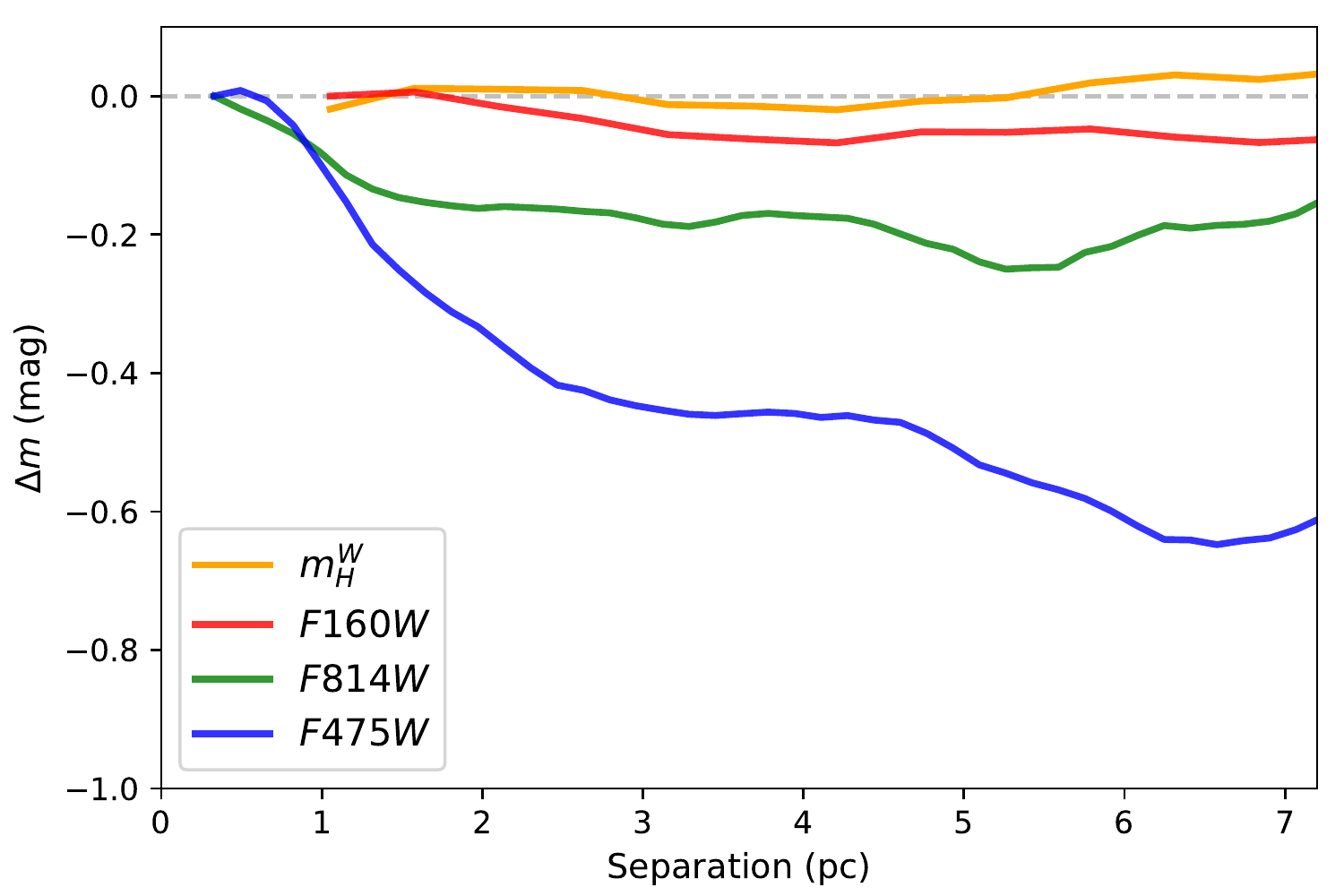} 
\caption{Average cluster contribution (curve of growth) for the 13 additional cluster Cepheids found by visual inspection (Table~\ref{tab:cluster_ceps_extra}) in the $m_H^W$, $F160W$, $F475W$, $F814W$ filters.}
\label{fig:curve_of_growth_extra}
\end{figure}

The contribution of the 13 additional cluster Cepheids obtained by visual inspection of the stamp images (\S\ref{sec:clusters_extra}, Table~\ref{tab:cluster_ceps_extra}) is represented in Fig.~\ref{fig:curve_of_growth_extra}. In all filters, it is lower than the contribution from the 10 confirmed clusters (Fig.~\ref{fig:curve_of_growth_regular}) with only $\Delta m = -0.02$ mag at 4 pc in $m_H^W$. This suggests that the 13 additional possible cluster Cepheids do not contribute to the contamination of the Cepheid flux.

Following \citet{Anderson2018}, we estimate the average photometric bias produced by the 10 confirmed cluster Cepheids from Table~\ref{tab:cluster_ceps_regular} by multiplying their occurence rate (1.6 \%) by their flux contribution in $m_H^W$, which gives a bias of 0.003 mag. This shows that cluster Cepheids do not impact the distance measurement of M33. If we conservatively assume that all 13 additional Cepheids from Table~\ref{tab:cluster_ceps_extra} are associated with clusters, we obtain a maximum occurence rate of 3.7\%, still lower than the fraction of cluster Cepheids in the Milky Way, LMC and SMC, and a correspondingly lower bias per cluster Cepheid (since the additional 13 do not produce a significant difference).    \\

%%%%%%%%%%%%%%%%%%%%%%%%%%%
%%%%%%%%%%%%%%%%%%%%%%%%%%%
\section{Summary}

We take advantage of the recently published high-quality PHATTER photometric survey of the M33 galaxy and we construct the Cepheid PL relation in the SH0ES near-infrared Wesenheit system ($m_H^W$). We use well-sampled ground-based light curves for the same Cepheid sample to recover the phases and amplitudes and we correct the random-epoch PHATTER measurements to mean magnitude. We also present new optical template light curves based on the same population of M33 Cepheids. These can be directly applied to fit sparsely-sampled light curves. 

We improve the uncertainty in the Cepheid distance to M33 to the 1.3\% level and we present the tightest PL relation to date in this galaxy, with a scatter of only 0.11 mag. In particular, the use of HST photometry allows to significantly reduce the effect of crowding over past studies based on ground-based observations. This new Cepheid distance provides groundwork for including M33 as an anchor galaxy in the empirical distance scale \citep{Riess2022}, with a similar role as the Milky Way, the LMC and NGC$\,$4258. In order to consider M33 as an independent anchor of the distance scale, a precise geometric calibration of its distance is required, using for example a large sample of late-type detached eclipsing binaries. Future facilities such as the ELTs and the {\it Roman} Space Telescope could enable significant improvements in that matter.

We discuss differences between past measurements of the distance to M33, especially based on the TRGB method, and we identify factors that explain these discrepancies such as the possible effect of blending and the choice of color cut when defining the red-giant branch in the color-magnitude diagram. Finally, we investigate the bias from cluster Cepheids and estimate that at most 3.7\% of M33 Cepheids are in these systems, resulting in a negligible contamination of 0.003 mag to our distance measurement. Our result, compared with other distance measurements from the literature, highlights the unprecedented reliability and precision of Cepheids as standard candles. \\

\section*{Acknowledgements}

%This work has made use of data from the European Space Agency (ESA) mission {\it Gaia} (\url{https://www.cosmos.esa.int/gaia}), processed by the {\it Gaia} Data Processing and Analysis Consortium (DPAC, \url{https://www.cosmos.esa.int/web/gaia/dpac/consortium}). Funding for the DPAC has been provided by national institutions, in particular the institutions participating in the {\it Gaia} Multilateral Agreement. 
We thank Abigail Lee for discussions about the TRGB and Cepheid distances to M33. L.B. is deeply grateful to Arshia Jacob for her constant support and kindness during the preparation of this paper. This research has made use of Astropy, a community-developed core Python package for Astronomy \citep{2018AJ....156..123A}, as well as of the SVO Filter Profile Service\footnote{\href{http://svo2.cab.inta- csic.es/theory/fps/}{http://svo2.cab.inta- csic.es/theory/fps/}}. Some of the data presented in this paper were obtained from the Mikulski Archive for Space Telescopes (MAST) at the STScI. \\

\bibliography{Breuval_2023}{}
\bibliographystyle{aasjournal}

%%%%%%%%%%%%%%%%%%%%%%%%%%%%%%%%%%%%%%%%%%%%%%%%%%%%%%%%%%%%%%%%%%%%%%%%%%%%

\newpage
\appendix

\section{Output parameters for HST M33 Cepheids}
\label{Apdx:full_sample}

\begin{longtable*}{l c c c c c c c c c c c c}
\caption{Full sample of HST M33 Cepheids used in this analysis and their mean magnitudes obtained from template fitting. The values given here do not include subtraction of 0.015 mag to correct CRNL for 2 dex between LMC and M33 Cepheids. They do not include addition of 0.069 mag to $m_H^W$ errors for intrinsic scatter. The sample names in the last column are: (G) = gold and (S) = silver. Cepheids marked with (*) are listed as cluster Cepheids in Sect. \ref{sec:clusters}.  } \\
\hline
\hline
ID  & RA & DEC & $\log P$ & $F160W$ & $F475W$ & $F814W$ & $m_H^W$ & sample   \\
\hline
\endfirsthead
\multicolumn{8}{c}{\textbf{Table~\ref{table:all_data}} (continued)} \\
\hline
ID  & RA & DEC & $\log P$ & $F160W$ & $F475W$ & $F814W$ & $m_H^W$ & sample   \\
\hline 
\endhead
\hline
\endfoot
\endlastfoot
01332610+3033200 & 23.3586 & 30.5555 & 1.3009 & $18.261_{\pm 0.091}$ & $20.678_{\pm 0.079}$ & $19.171_{\pm 0.024}$ & $17.853_{\pm 0.093}$ & G \\
01332735+3035515 & 23.3638 & 30.5976 & 1.0496 & $19.154_{\pm 0.071}$ & $21.093_{\pm 0.084}$ & $19.791_{\pm 0.046}$ & $18.798_{\pm 0.075}$ & G \\
01332768+3034238 $^{(*)}$ & 23.3652 & 30.5732 & 1.1715 & $18.402_{\pm 0.053}$ & $20.330_{\pm 0.035}$ & $19.103_{\pm 0.035}$ & $18.065_{\pm 0.054}$ & G \\
01332803+3032133 & 23.3667 & 30.5370 & 1.1975 & $18.575_{\pm 0.079}$ & $21.810_{\pm 0.113}$ & $19.868_{\pm 0.037}$ & $18.057_{\pm 0.085}$ & G \\
01332878+3034403 & 23.3698 & 30.5778 & 0.9494 & $19.179_{\pm 0.034}$ & $20.922_{\pm 0.038}$ & $19.891_{\pm 0.037}$ & $18.892_{\pm 0.037}$ & G \\
01332922+3031360 & 23.3716 & 30.5267 & 0.8148 & $19.920_{\pm 0.045}$ & $21.726_{\pm 0.062}$ & $20.463_{\pm 0.051}$ & $19.574_{\pm 0.049}$ & G \\
01332923+3037449 & 23.3717 & 30.6291 & 1.6632 & $16.942_{\pm 0.165}$ & $20.028_{\pm 0.050}$ & $18.235_{\pm 0.014}$ & $16.462_{\pm 0.166}$ & G \\
01332946+3035574 & 23.3726 & 30.5993 & 1.4868 & $17.655_{\pm 0.116}$ & $20.634_{\pm 0.125}$ & $18.886_{\pm 0.156}$ & $17.186_{\pm 0.127}$ & G \\
01332984+3034529 & 23.3742 & 30.5813 & 0.9080 & $19.398_{\pm 0.044}$ & $21.553_{\pm 0.087}$ & $20.243_{\pm 0.008}$ & $19.040_{\pm 0.049}$ & G \\
01333012+3036381 & 23.3754 & 30.6106 & 1.2546 & $18.325_{\pm 0.144}$ & $20.271_{\pm 0.025}$ & $18.926_{\pm 0.011}$ & $17.958_{\pm 0.144}$ & G \\
01333039+3035555 & 23.3765 & 30.5987 & 1.3438 & $17.940_{\pm 0.132}$ & $20.124_{\pm 0.015}$ & $18.725_{\pm 0.020}$ & $17.560_{\pm 0.132}$ & G \\
01333073+3034495 & 23.3779 & 30.5804 & 0.7492 & $19.809_{\pm 0.102}$ & $21.667_{\pm 0.013}$ & $20.567_{\pm 0.043}$ & $19.505_{\pm 0.103}$ & G \\
01333080+3031113 & 23.3782 & 30.5198 & 1.2438 & $18.372_{\pm 0.134}$ & $20.982_{\pm 0.174}$ & $19.409_{\pm 0.111}$ & $17.947_{\pm 0.144}$ & G \\
01333165+3039314 & 23.3817 & 30.6587 & 1.1194 & $18.743_{\pm 0.172}$ & $20.180_{\pm 0.026}$ & $19.316_{\pm 0.173}$ & $18.498_{\pm 0.178}$ & S \\
01333216+3039457 & 23.3839 & 30.6627 & 1.1751 & $18.712_{\pm 0.081}$ & $21.129_{\pm 0.223}$ & $19.564_{\pm 0.160}$ & $18.289_{\pm 0.107}$ & G \\
01333241+3031437 & 23.3849 & 30.5288 & 0.9012 & $19.506_{\pm 0.171}$ & $21.925_{\pm 0.108}$ & $20.460_{\pm 0.015}$ & $19.109_{\pm 0.173}$ & S \\
01333242+3034094 & 23.3849 & 30.5692 & 0.8484 & $19.545_{\pm 0.076}$ & $20.997_{\pm 0.127}$ & $20.087_{\pm 0.086}$ & $19.289_{\pm 0.085}$ & G \\
01333291+3035490 & 23.3870 & 30.5969 & 1.4807 & $17.557_{\pm 0.138}$ & $20.698_{\pm 0.181}$ & $18.914_{\pm 0.156}$ & $17.079_{\pm 0.151}$ & G \\
01333308+3042308 & 23.3877 & 30.7085 & 0.7790 & $19.828_{\pm 0.198}$ & $21.925_{\pm 0.115}$ & $20.678_{\pm 0.235}$ & $19.486_{\pm 0.209}$ & S \\
01333327+3037476 & 23.3885 & 30.6298 & 1.3551 & $18.266_{\pm 0.174}$ & $21.417_{\pm 0.394}$ & $19.512_{\pm 0.194}$ & $17.757_{\pm 0.207}$ & S \\
01333348+3033210 $^{(*)}$ & 23.3894 & 30.5558 & 0.9546 & $19.300_{\pm 0.200}$ & $21.277_{\pm 0.013}$ & $20.067_{\pm 0.057}$ & $18.968_{\pm 0.201}$ & S \\
01333424+3035009 & 23.3925 & 30.5835 & 1.1956 & $18.629_{\pm 0.082}$ & $21.051_{\pm 0.312}$ & $19.664_{\pm 0.171}$ & $18.252_{\pm 0.122}$ & G \\
01333438+3035307 $^{(*)}$ & 23.3931 & 30.5918 & 1.1995 & $18.459_{\pm 0.143}$ & $20.760_{\pm 0.263}$ & $19.367_{\pm 0.109}$ & $18.080_{\pm 0.160}$ & S \\
01333548+3044180 & 23.3977 & 30.7383 & 0.9782 & $19.173_{\pm 0.035}$ & $21.503_{\pm 0.167}$ & $20.039_{\pm 0.106}$ & $18.776_{\pm 0.061}$ & G \\
01333552+3033307 & 23.3978 & 30.5585 & 1.0616 & $18.869_{\pm 0.085}$ & $21.022_{\pm 0.011}$ & $19.710_{\pm 0.010}$ & $18.511_{\pm 0.085}$ & G \\
01333557+3036496 & 23.3980 & 30.6138 & 1.0288 & $19.166_{\pm 0.052}$ & $21.463_{\pm 0.211}$ & $20.072_{\pm 0.104}$ & $18.788_{\pm 0.079}$ & G \\
01333628+3037313 & 23.4010 & 30.6253 & 1.0713 & $19.110_{\pm 0.035}$ & $21.900_{\pm 0.037}$ & $20.144_{\pm 0.029}$ & $18.639_{\pm 0.037}$ & G \\
01333649+3030536 & 23.4020 & 30.5148 & 0.8802 & $19.243_{\pm 0.081}$ & $20.923_{\pm 0.227}$ & $20.045_{\pm 0.140}$ & $18.995_{\pm 0.106}$ & G \\
01333680+3034348 & 23.4031 & 30.5763 & 0.9571 & $19.254_{\pm 0.041}$ & $20.811_{\pm 0.166}$ & $19.752_{\pm 0.052}$ & $18.960_{\pm 0.060}$ & G \\
01333747+3031388 & 23.4059 & 30.5274 & 1.7605 & $16.988_{\pm 0.084}$ & $20.102_{\pm 0.136}$ & $18.180_{\pm 0.073}$ & $16.475_{\pm 0.093}$ & G \\
01333754+3033054 & 23.4062 & 30.5515 & 0.9531 & $19.297_{\pm 0.027}$ & $21.627_{\pm 0.070}$ & $20.234_{\pm 0.023}$ & $18.918_{\pm 0.033}$ & S \\
01333791+3033550 & 23.4078 & 30.5652 & 0.9521 & $19.418_{\pm 0.200}$ & $21.500_{\pm 0.075}$ & $20.205_{\pm 0.040}$ & $19.064_{\pm 0.201}$ & S \\
01333880+3037515 & 23.4115 & 30.6309 & 1.1389 & $18.981_{\pm 0.069}$ & $21.503_{\pm 0.025}$ & $19.875_{\pm 0.030}$ & $18.542_{\pm 0.070}$ & G \\
01334058+3045421 & 23.4189 & 30.7617 & 0.8682 & $19.584_{\pm 0.069}$ & $21.989_{\pm 0.211}$ & $20.444_{\pm 0.138}$ & $19.166_{\pm 0.094}$ & G \\
01334167+3043115 & 23.4234 & 30.7198 & 0.9064 & $19.402_{\pm 0.209}$ & $21.789_{\pm 0.027}$ & $20.397_{\pm 0.040}$ & $19.023_{\pm 0.209}$ & S \\
01334331+3043559 $^{(*)}$ & 23.4304 & 30.7322 & 1.3045 & $18.196_{\pm 0.123}$ & $20.940_{\pm 0.070}$ & $19.246_{\pm 0.092}$ & $17.741_{\pm 0.126}$ & G \\
01334390+3032452 & 23.4328 & 30.5458 & 1.8731 & $16.197_{\pm 0.091}$ & $19.803_{\pm 0.055}$ & $17.745_{\pm 0.031}$ & $15.649_{\pm 0.092}$ & G \\
01334456+3043132 & 23.4355 & 30.7203 & 1.3510 & $18.135_{\pm 0.076}$ & $20.891_{\pm 0.068}$ & $19.202_{\pm 0.031}$ & $17.681_{\pm 0.078}$ & G \\
01334582+3044207 & 23.4408 & 30.7390 & 1.1178 & $18.983_{\pm 0.094}$ & $21.239_{\pm 0.090}$ & $19.893_{\pm 0.013}$ & $18.616_{\pm 0.097}$ & G \\
01334596+3030303 & 23.4414 & 30.5084 & 0.5382 & $20.445_{\pm 0.087}$ & $22.468_{\pm 0.246}$ & $21.529_{\pm 0.081}$ & $20.181_{\pm 0.109}$ & G \\
01334654+3046449 & 23.4438 & 30.7791 & 1.2743 & $18.341_{\pm 0.104}$ & $20.247_{\pm 0.053}$ & $18.945_{\pm 0.026}$ & $17.985_{\pm 0.105}$ & G \\
01334681+3043335 & 23.4449 & 30.7260 & 0.7779 & $19.771_{\pm 0.064}$ & $21.715_{\pm 0.136}$ & $20.578_{\pm 0.112}$ & $19.457_{\pm 0.078}$ & G \\
01334691+3044112 & 23.4453 & 30.7365 & 0.7498 & $19.852_{\pm 0.132}$ & $22.002_{\pm 0.008}$ & $20.699_{\pm 0.019}$ & $19.496_{\pm 0.132}$ & S \\
01334720+3035365 & 23.4465 & 30.5934 & 1.4232 & $17.767_{\pm 0.110}$ & $20.319_{\pm 0.154}$ & $18.804_{\pm 0.008}$ & $17.357_{\pm 0.117}$ & G \\
01334821+3038001 $^{(*)}$ & 23.4507 & 30.6333 & 1.1780 & $18.557_{\pm 0.115}$ & $20.428_{\pm 0.020}$ & $19.166_{\pm 0.025}$ & $18.211_{\pm 0.115}$ & G \\
01334879+3049143 & 23.4531 & 30.8206 & 0.7486 & $19.980_{\pm 0.047}$ & $21.885_{\pm 0.071}$ & $20.668_{\pm 0.043}$ & $19.646_{\pm 0.052}$ & G \\
01334886+3034159 & 23.4534 & 30.5711 & 0.9011 & $19.351_{\pm 0.041}$ & $21.181_{\pm 0.191}$ & $20.118_{\pm 0.071}$ & $19.056_{\pm 0.066}$ & G \\
01334929+3032182 & 23.4552 & 30.5383 & 0.9912 & $19.302_{\pm 0.175}$ & $21.790_{\pm 0.062}$ & $20.264_{\pm 0.033}$ & $18.889_{\pm 0.176}$ & S \\
01334955+3047437 & 23.4563 & 30.7955 & 0.7788 & $19.902_{\pm 0.046}$ & $21.931_{\pm 0.078}$ & $20.721_{\pm 0.054}$ & $19.570_{\pm 0.052}$ & G \\
01334960+3044008 & 23.4565 & 30.7336 & 0.7978 & $19.742_{\pm 0.038}$ & $21.926_{\pm 0.113}$ & $20.697_{\pm 0.015}$ & $19.405_{\pm 0.048}$ & G \\
01334983+3037587 & 23.4575 & 30.6329 & 1.1105 & $18.690_{\pm 0.110}$ & $20.738_{\pm 0.020}$ & $19.470_{\pm 0.020}$ & $18.343_{\pm 0.110}$ & G \\
01335051+3047537 & 23.4603 & 30.7982 & 1.4233 & $17.912_{\pm 0.104}$ & $20.897_{\pm 0.116}$ & $19.077_{\pm 0.079}$ & $17.425_{\pm 0.110}$ & G \\
01335067+3034459 & 23.4609 & 30.5794 & 0.7803 & $19.842_{\pm 0.207}$ & $22.022_{\pm 0.340}$ & $20.689_{\pm 0.218}$ & $19.478_{\pm 0.231}$ & S \\
01335067+3047146 & 23.4610 & 30.7873 & 0.9875 & $19.088_{\pm 0.046}$ & $21.349_{\pm 0.240}$ & $20.036_{\pm 0.090}$ & $18.729_{\pm 0.080}$ & G \\
01335075+3035444 & 23.4613 & 30.5956 & 1.3676 & $18.191_{\pm 0.120}$ & $20.776_{\pm 0.020}$ & $19.121_{\pm 0.020}$ & $17.746_{\pm 0.120}$ & G \\
01335090+3033361 & 23.4619 & 30.5600 & 1.5745 & $17.638_{\pm 0.060}$ & $21.241_{\pm 0.201}$ & $19.103_{\pm 0.094}$ & $17.070_{\pm 0.082}$ & G \\
01335094+3031174 & 23.4621 & 30.5214 & 0.6782 & $19.998_{\pm 0.040}$ & $21.830_{\pm 0.055}$ & $20.726_{\pm 0.026}$ & $19.693_{\pm 0.043}$ & G \\
01335104+3043598 & 23.4626 & 30.7332 & 1.0915 & $18.935_{\pm 0.074}$ & $21.353_{\pm 0.137}$ & $19.833_{\pm 0.102}$ & $18.524_{\pm 0.086}$ & G \\
01335172+3050370 & 23.4654 & 30.8436 & 0.9564 & $19.405_{\pm 0.200}$ & $21.505_{\pm 0.483}$ & $20.209_{\pm 0.226}$ & $19.051_{\pm 0.242}$ & S \\
01335182+3033109 & 23.4657 & 30.5529 & 0.8132 & $19.753_{\pm 0.018}$ & $22.289_{\pm 0.009}$ & $20.764_{\pm 0.043}$ & $19.341_{\pm 0.021}$ & G \\
01335198+3048485 & 23.4664 & 30.8135 & 1.1251 & $18.920_{\pm 0.172}$ & $20.610_{\pm 0.215}$ & $19.459_{\pm 0.142}$ & $18.603_{\pm 0.184}$ & S \\
01335232+3046026 & 23.4678 & 30.7674 & 0.7752 & $19.957_{\pm 0.294}$ & $22.292_{\pm 0.024}$ & $20.793_{\pm 0.240}$ & $19.551_{\pm 0.300}$ & S \\
01335247+3038442 & 23.4685 & 30.6456 & 1.5544 & $17.443_{\pm 0.139}$ & $19.865_{\pm 0.080}$ & $18.448_{\pm 0.020}$ & $17.058_{\pm 0.141}$ & G \\
01335311+3048343 $^{(*)}$ & 23.4711 & 30.8095 & 0.9821 & $19.156_{\pm 0.174}$ & $20.813_{\pm 0.401}$ & $19.761_{\pm 0.138}$ & $18.864_{\pm 0.205}$ & S \\
01335345+3033085 & 23.4726 & 30.5523 & 1.0606 & $19.106_{\pm 0.060}$ & $21.593_{\pm 0.071}$ & $20.079_{\pm 0.078}$ & $18.696_{\pm 0.066}$ & G \\
01335428+3041107 & 23.4760 & 30.6863 & 1.4476 & $17.642_{\pm 0.165}$ & $20.302_{\pm 0.123}$ & $18.699_{\pm 0.064}$ & $17.210_{\pm 0.169}$ & G \\
01335478+3041061 & 23.4781 & 30.6850 & 1.5309 & $17.420_{\pm 0.105}$ & $20.191_{\pm 0.013}$ & $18.479_{\pm 0.006}$ & $16.960_{\pm 0.105}$ & G \\
01335479+3045181 & 23.4782 & 30.7550 & 0.7621 & $19.692_{\pm 0.198}$ & $21.582_{\pm 0.059}$ & $20.474_{\pm 0.033}$ & $19.385_{\pm 0.199}$ & S \\
01335482+3045317 & 23.4783 & 30.7588 & 1.5529 & $17.543_{\pm 0.099}$ & $19.167_{\pm 0.279}$ & $17.781_{\pm 0.018}$ & $17.166_{\pm 0.122}$ & G \\
01335502+3035372 & 23.4791 & 30.5936 & 1.0056 & $19.054_{\pm 0.068}$ & $21.186_{\pm 0.008}$ & $19.907_{\pm 0.022}$ & $18.704_{\pm 0.068}$ & G \\
01335514+3048309 & 23.4796 & 30.8085 & 0.9493 & $19.528_{\pm 0.200}$ & $21.798_{\pm 0.516}$ & $20.328_{\pm 0.282}$ & $19.130_{\pm 0.250}$ & S \\
01335523+3043429 & 23.4800 & 30.7285 & 1.4209 & $18.011_{\pm 0.232}$ & $20.786_{\pm 0.454}$ & $19.056_{\pm 0.117}$ & $17.547_{\pm 0.261}$ & S \\
01335612+3039029 & 23.4837 & 30.6508 & 1.0180 & $19.106_{\pm 0.056}$ & $21.335_{\pm 0.018}$ & $19.998_{\pm 0.018}$ & $18.741_{\pm 0.056}$ & G \\
01335615+3043424 & 23.4838 & 30.7284 & 1.0779 & $18.951_{\pm 0.058}$ & $21.309_{\pm 0.037}$ & $19.795_{\pm 0.077}$ & $18.541_{\pm 0.062}$ & G \\
01335645+3046435 & 23.4851 & 30.7788 & 0.8630 & $19.646_{\pm 0.075}$ & $21.755_{\pm 0.020}$ & $20.370_{\pm 0.024}$ & $19.269_{\pm 0.075}$ & G \\
01335646+3044420 & 23.4851 & 30.7449 & 0.7388 & $19.847_{\pm 0.103}$ & $21.739_{\pm 0.117}$ & $20.702_{\pm 0.105}$ & $19.559_{\pm 0.110}$ & G \\
01335737+3041133 & 23.4889 & 30.6870 & 1.0648 & $18.885_{\pm 0.172}$ & $21.487_{\pm 0.054}$ & $19.907_{\pm 0.046}$ & $18.459_{\pm 0.173}$ & S \\
01335760+3048341 & 23.4898 & 30.8094 & 0.6849 & $20.248_{\pm 0.054}$ & $22.664_{\pm 0.203}$ & $21.193_{\pm 0.107}$ & $19.849_{\pm 0.079}$ & G \\
01335761+3038053 & 23.4899 & 30.6348 & 1.0915 & $18.852_{\pm 0.172}$ & $21.290_{\pm 0.128}$ & $19.790_{\pm 0.230}$ & $18.446_{\pm 0.185}$ & S \\
01335809+3045568 $^{(*)}$ & 23.4919 & 30.7658 & 1.4986 & $17.818_{\pm 0.228}$ & $20.706_{\pm 0.204}$ & $18.935_{\pm 0.118}$ & $17.343_{\pm 0.236}$ & S \\
01335852+3043596 & 23.4937 & 30.7332 & 1.0470 & $18.905_{\pm 0.052}$ & $21.599_{\pm 0.182}$ & $20.026_{\pm 0.023}$ & $18.480_{\pm 0.070}$ & G \\
01335870+3033166 & 23.4945 & 30.5546 & 0.8863 & $19.371_{\pm 0.043}$ & $21.328_{\pm 0.058}$ & $20.094_{\pm 0.059}$ & $19.032_{\pm 0.048}$ & G \\
01335886+3037198 & 23.4951 & 30.6221 & 1.3903 & $18.027_{\pm 0.179}$ & $20.603_{\pm 0.114}$ & $19.116_{\pm 0.070}$ & $17.624_{\pm 0.182}$ & G \\
01335947+3032266 & 23.4977 & 30.5407 & 1.6990 & $17.186_{\pm 0.067}$ & $20.559_{\pm 0.035}$ & $18.604_{\pm 0.024}$ & $16.664_{\pm 0.068}$ & G \\
01335989+3037600 & 23.4994 & 30.6333 & 0.7779 & $19.916_{\pm 0.070}$ & $21.263_{\pm 0.121}$ & $20.381_{\pm 0.076}$ & $19.667_{\pm 0.079}$ & G \\
01340058+3036306 & 23.5023 & 30.6085 & 0.8024 & $19.643_{\pm 0.195}$ & $21.814_{\pm 0.119}$ & $20.535_{\pm 0.063}$ & $19.293_{\pm 0.198}$ & S \\
01340102+3043097 & 23.5041 & 30.7193 & 1.4843 & $17.804_{\pm 0.108}$ & $20.225_{\pm 0.351}$ & $18.811_{\pm 0.135}$ & $17.420_{\pm 0.144}$ & G \\
01340120+3048131 & 23.5048 & 30.8036 & 0.6645 & $20.350_{\pm 0.221}$ & $22.574_{\pm 0.358}$ & $21.258_{\pm 0.141}$ & $19.991_{\pm 0.242}$ & S \\
01340123+3031135 & 23.5050 & 30.5204 & 1.3965 & $17.746_{\pm 0.083}$ & $19.537_{\pm 0.253}$ & $18.430_{\pm 0.143}$ & $17.440_{\pm 0.111}$ & G \\
01340137+3040264 & 23.5056 & 30.6739 & 1.1737 & $18.586_{\pm 0.174}$ & $21.979_{\pm 0.165}$ & $20.129_{\pm 0.241}$ & $18.091_{\pm 0.189}$ & S \\
01340166+3031030 & 23.5067 & 30.5174 & 1.1982 & $18.359_{\pm 0.116}$ & $21.288_{\pm 0.331}$ & $19.544_{\pm 0.140}$ & $17.891_{\pm 0.148}$ & G \\
01340167+3049102 & 23.5068 & 30.8195 & 0.6306 & $20.286_{\pm 0.035}$ & $21.956_{\pm 0.064}$ & $20.970_{\pm 0.050}$ & $20.010_{\pm 0.041}$ & G \\
01340178+3039229 & 23.5073 & 30.6563 & 1.3358 & $18.084_{\pm 0.112}$ & $20.439_{\pm 0.133}$ & $18.936_{\pm 0.144}$ & $17.677_{\pm 0.123}$ & G \\
01340187+3041483 & 23.5077 & 30.6967 & 0.8015 & $19.859_{\pm 0.055}$ & $21.689_{\pm 0.209}$ & $20.572_{\pm 0.090}$ & $19.550_{\pm 0.080}$ & G \\
01340219+3037418 & 23.5090 & 30.6282 & 0.8642 & $19.655_{\pm 0.038}$ & $21.206_{\pm 0.156}$ & $20.198_{\pm 0.084}$ & $19.374_{\pm 0.059}$ & G \\
01340223+3042423 & 23.5092 & 30.7117 & 0.8943 & $19.331_{\pm 0.089}$ & $21.406_{\pm 0.012}$ & $20.082_{\pm 0.008}$ & $18.970_{\pm 0.089}$ & G \\
01340259+3036282 & 23.5107 & 30.6078 & 1.1346 & $18.887_{\pm 0.174}$ & $21.318_{\pm 0.085}$ & $19.763_{\pm 0.037}$ & $18.467_{\pm 0.176}$ & S \\
01340296+3047273 & 23.5122 & 30.7909 & 0.8503 & $19.550_{\pm 0.079}$ & $21.515_{\pm 0.064}$ & $20.337_{\pm 0.035}$ & $19.226_{\pm 0.081}$ & G \\
01340367+3045279 & 23.5152 & 30.7577 & 0.7919 & $19.820_{\pm 0.090}$ & $21.371_{\pm 0.102}$ & $20.329_{\pm 0.038}$ & $19.530_{\pm 0.094}$ & G \\
01340399+3036158 & 23.5165 & 30.6044 & 1.2114 & $18.389_{\pm 0.104}$ & $20.650_{\pm 0.206}$ & $19.251_{\pm 0.026}$ & $18.009_{\pm 0.117}$ & G \\
01340426+3041150 & 23.5176 & 30.6875 & 0.8822 & $19.441_{\pm 0.048}$ & $21.748_{\pm 0.177}$ & $20.336_{\pm 0.124}$ & $19.057_{\pm 0.073}$ & G \\
01340501+3035576 & 23.5208 & 30.5993 & 0.7702 & $19.777_{\pm 0.049}$ & $21.702_{\pm 0.053}$ & $20.524_{\pm 0.048}$ & $19.453_{\pm 0.052}$ & G \\
01340516+3038511 & 23.5214 & 30.6475 & 1.8421 & $16.855_{\pm 0.118}$ & $19.970_{\pm 0.634}$ & $18.018_{\pm 0.358}$ & $16.334_{\pm 0.219}$ & S \\
01340593+3038192 & 23.5246 & 30.6387 & 1.0305 & $19.197_{\pm 0.172}$ & $21.720_{\pm 0.317}$ & $20.192_{\pm 0.179}$ & $18.784_{\pm 0.195}$ & S \\
01340593+3039285 & 23.5246 & 30.6579 & 1.1639 & $18.491_{\pm 0.174}$ & $20.841_{\pm 0.468}$ & $19.413_{\pm 0.299}$ & $18.103_{\pm 0.224}$ & S \\
01340660+3038167 & 23.5273 & 30.6379 & 0.9338 & $19.195_{\pm 0.200}$ & $21.178_{\pm 0.190}$ & $19.912_{\pm 0.070}$ & $18.848_{\pm 0.207}$ & S \\
01340737+3030483 & 23.5305 & 30.5134 & 0.7710 & $19.957_{\pm 0.062}$ & $22.091_{\pm 0.109}$ & $20.796_{\pm 0.038}$ & $19.603_{\pm 0.069}$ & G \\
01340794+3038312 & 23.5330 & 30.6420 & 0.7710 & $19.805_{\pm 0.198}$ & $21.712_{\pm 0.196}$ & $20.526_{\pm 0.218}$ & $19.479_{\pm 0.212}$ & S \\
01340817+3039318 & 23.5339 & 30.6588 & 1.1246 & $18.857_{\pm 0.172}$ & $21.385_{\pm 0.080}$ & $19.963_{\pm 0.070}$ & $18.471_{\pm 0.175}$ & S \\
01340873+3045431 & 23.5363 & 30.7619 & 1.1311 & $18.535_{\pm 0.113}$ & $20.827_{\pm 0.135}$ & $19.432_{\pm 0.111}$ & $18.156_{\pm 0.121}$ & G \\
01340883+3039462 & 23.5367 & 30.6628 & 1.0050 & $19.132_{\pm 0.042}$ & $21.307_{\pm 0.007}$ & $19.944_{\pm 0.008}$ & $18.761_{\pm 0.042}$ & G \\
01340910+3036296 $^{(*)}$ & 23.5377 & 30.6082 & 1.1645 & $18.426_{\pm 0.174}$ & $20.352_{\pm 0.361}$ & $19.193_{\pm 0.202}$ & $18.107_{\pm 0.203}$ & S \\
01340936+3029565 & 23.5389 & 30.4990 & 0.9619 & $19.283_{\pm 0.026}$ & $21.328_{\pm 0.053}$ & $20.057_{\pm 0.033}$ & $18.935_{\pm 0.030}$ & G \\
01340959+3036215 $^{(*)}$ & 23.5398 & 30.6059 & 1.0206 & $18.932_{\pm 0.175}$ & $21.144_{\pm 0.206}$ & $19.881_{\pm 0.108}$ & $18.586_{\pm 0.185}$ & S \\
01341017+3044502 & 23.5423 & 30.7472 & 1.0185 & $19.253_{\pm 0.053}$ & $21.528_{\pm 0.143}$ & $20.131_{\pm 0.070}$ & $18.873_{\pm 0.067}$ & G \\
01341027+3034077 & 23.5426 & 30.5688 & 1.2866 & $18.334_{\pm 0.012}$ & $21.147_{\pm 0.079}$ & $19.530_{\pm 0.005}$ & $17.898_{\pm 0.023}$ & G \\
01341121+3047547 & 23.5466 & 30.7985 & 1.0938 & $18.753_{\pm 0.173}$ & $20.646_{\pm 0.018}$ & $19.484_{\pm 0.013}$ & $18.433_{\pm 0.173}$ & S \\
01341125+3041558 & 23.5467 & 30.6988 & 1.4251 & $17.682_{\pm 0.098}$ & $20.116_{\pm 0.051}$ & $18.659_{\pm 0.021}$ & $17.287_{\pm 0.099}$ & G \\
01341199+3035191 & 23.5498 & 30.5886 & 1.1258 & $18.879_{\pm 0.173}$ & $21.201_{\pm 0.114}$ & $19.759_{\pm 0.275}$ & $18.488_{\pm 0.189}$ & S \\
01341217+3036362 $^{(*)}$ & 23.5505 & 30.6100 & 1.2147 & $18.569_{\pm 0.112}$ & $21.245_{\pm 0.114}$ & $19.628_{\pm 0.061}$ & $18.133_{\pm 0.117}$ & G \\
01341217+3046411 & 23.5506 & 30.7781 & 1.5546 & $17.423_{\pm 0.092}$ & $19.816_{\pm 0.088}$ & $18.420_{\pm 0.052}$ & $17.043_{\pm 0.096}$ & G \\
01341226+3045507 & 23.5509 & 30.7641 & 1.0014 & $19.118_{\pm 0.175}$ & $21.574_{\pm 0.443}$ & $20.049_{\pm 0.282}$ & $18.706_{\pm 0.220}$ & S \\
01341241+3047206 & 23.5516 & 30.7890 & 0.6306 & $20.278_{\pm 0.193}$ & $22.276_{\pm 0.083}$ & $21.049_{\pm 0.077}$ & $19.941_{\pm 0.195}$ & S \\
01341259+3041262 & 23.5523 & 30.6906 & 0.9201 & $19.430_{\pm 0.026}$ & $21.757_{\pm 0.020}$ & $20.309_{\pm 0.035}$ & $19.037_{\pm 0.028}$ & G \\
01341330+3043069 & 23.5552 & 30.7185 & 0.9316 & $19.396_{\pm 0.200}$ & $21.851_{\pm 0.064}$ & $20.358_{\pm 0.019}$ & $18.992_{\pm 0.201}$ & S \\
01341343+3043340 & 23.5558 & 30.7261 & 0.9610 & $19.237_{\pm 0.178}$ & $21.420_{\pm 0.085}$ & $20.042_{\pm 0.012}$ & $18.862_{\pm 0.179}$ & S \\
01341356+3030274 & 23.5564 & 30.5076 & 0.8637 & $19.575_{\pm 0.035}$ & $21.689_{\pm 0.040}$ & $20.395_{\pm 0.015}$ & $19.221_{\pm 0.037}$ & G \\
01341383+3044184 & 23.5575 & 30.7384 & 1.0259 & $19.302_{\pm 0.172}$ & $21.796_{\pm 0.004}$ & $20.255_{\pm 0.015}$ & $18.886_{\pm 0.172}$ & S \\
01341387+3043240 & 23.5576 & 30.7233 & 0.9006 & $19.461_{\pm 0.023}$ & $21.780_{\pm 0.059}$ & $20.371_{\pm 0.040}$ & $19.078_{\pm 0.029}$ & G \\
01341389+3032122 & 23.5577 & 30.5366 & 0.9586 & $19.283_{\pm 0.200}$ & $21.671_{\pm 0.395}$ & $20.240_{\pm 0.297}$ & $18.894_{\pm 0.236}$ & S \\
01341396+3048375 & 23.5580 & 30.8104 & 0.7530 & $19.938_{\pm 0.066}$ & $21.839_{\pm 0.073}$ & $20.739_{\pm 0.080}$ & $19.634_{\pm 0.072}$ & G \\
01341425+3037138 & 23.5592 & 30.6205 & 0.9766 & $19.220_{\pm 0.030}$ & $21.670_{\pm 0.142}$ & $20.195_{\pm 0.086}$ & $18.820_{\pm 0.052}$ & G \\
01341471+3046097 & 23.5612 & 30.7694 & 1.0706 & $19.099_{\pm 0.172}$ & $22.208_{\pm 0.413}$ & $20.344_{\pm 0.227}$ & $18.600_{\pm 0.210}$ & S \\
01341554+3044116 & 23.5646 & 30.7365 & 0.9086 & $19.562_{\pm 0.038}$ & $21.784_{\pm 0.008}$ & $20.333_{\pm 0.005}$ & $19.168_{\pm 0.038}$ & G \\
01341590+3046436 & 23.5661 & 30.7788 & 0.9322 & $19.414_{\pm 0.034}$ & $21.830_{\pm 0.103}$ & $20.383_{\pm 0.012}$ & $19.021_{\pm 0.043}$ & G \\
01341889+3044412 & 23.5786 & 30.7448 & 1.3053 & $18.464_{\pm 0.128}$ & $20.761_{\pm 0.082}$ & $19.349_{\pm 0.012}$ & $18.080_{\pm 0.130}$ & G \\
01341986+3043020 & 23.5826 & 30.7172 & 0.9163 & $19.538_{\pm 0.200}$ & $21.899_{\pm 0.069}$ & $20.437_{\pm 0.300}$ & $19.142_{\pm 0.215}$ & S \\
01342038+3041010 & 23.5848 & 30.6836 & 1.2032 & $18.613_{\pm 0.137}$ & $20.968_{\pm 0.043}$ & $19.449_{\pm 0.050}$ & $18.202_{\pm 0.138}$ & G \\
01342055+3042440 & 23.5855 & 30.7122 & 0.7557 & $19.795_{\pm 0.085}$ & $21.595_{\pm 0.163}$ & $20.508_{\pm 0.051}$ & $19.494_{\pm 0.095}$ & G \\
01342102+3044152 & 23.5874 & 30.7375 & 0.9993 & $19.168_{\pm 0.175}$ & $20.997_{\pm 0.472}$ & $19.778_{\pm 0.172}$ & $18.833_{\pm 0.217}$ & S \\
01342122+3045369 & 23.5883 & 30.7602 & 1.2511 & $18.207_{\pm 0.086}$ & $20.391_{\pm 0.159}$ & $19.101_{\pm 0.081}$ & $17.854_{\pm 0.097}$ & G \\
01342142+3046180 & 23.5891 & 30.7716 & 0.6772 & $20.039_{\pm 0.058}$ & $22.129_{\pm 0.016}$ & $20.885_{\pm 0.007}$ & $19.698_{\pm 0.058}$ & G \\
01342241+3044080 & 23.5933 & 30.7355 & 0.8842 & $19.570_{\pm 0.145}$ & $21.678_{\pm 0.029}$ & $20.345_{\pm 0.016}$ & $19.206_{\pm 0.145}$ & S \\
01342254+3049055 & 23.5938 & 30.8182 & 1.2970 & $18.214_{\pm 0.185}$ & $20.823_{\pm 0.619}$ & $19.201_{\pm 0.038}$ & $17.777_{\pm 0.243}$ & S \\
01342419+3047389 & 23.6007 & 30.7941 & 0.8859 & $19.420_{\pm 0.111}$ & $21.220_{\pm 0.102}$ & $20.113_{\pm 0.052}$ & $19.114_{\pm 0.115}$ & G \\
01342472+3044311 & 23.6028 & 30.7420 & 0.7713 & $19.964_{\pm 0.198}$ & $21.907_{\pm 0.019}$ & $20.703_{\pm 0.002}$ & $19.633_{\pm 0.198}$ & S \\
01342913+3043388 & 23.6213 & 30.7274 & 1.1693 & $18.850_{\pm 0.174}$ & $19.794_{\pm 0.562}$ & $18.908_{\pm 0.197}$ & $18.600_{\pm 0.231}$ & S \\
01343020+3044567 & 23.6257 & 30.7491 & 0.6031 & $20.372_{\pm 0.029}$ & $22.500_{\pm 0.106}$ & $21.224_{\pm 0.034}$ & $20.023_{\pm 0.041}$ & G \\
01342988+3047541 $^{(*)}$ & 23.6244 & 30.7983 & 1.0902 & $19.015_{\pm 0.172}$ & $21.594_{\pm 0.318}$ & $19.990_{\pm 0.151}$ & $18.583_{\pm 0.194}$ & S \\
01335230+3045008 & 23.4678 & 30.7502 & 0.9642 & $19.338_{\pm 0.200}$ & $21.840_{\pm 0.293}$ & $20.273_{\pm 0.265}$ & $18.915_{\pm 0.224}$ & S \\
01334228+3037474 & 23.4260 & 30.6298 & 1.1931 & $18.734_{\pm 0.185}$ & $20.824_{\pm 0.025}$ & $19.411_{\pm 0.020}$ & $18.350_{\pm 0.185}$ & S \\
01341344+3033177 & 23.5558 & 30.5549 & 1.0603 & $18.898_{\pm 0.173}$ & $21.533_{\pm 0.028}$ & $20.041_{\pm 0.298}$ & $18.494_{\pm 0.189}$ & S \\
01333576+3033007 & 23.3988 & 30.5501 & 0.7801 & $19.765_{\pm 0.199}$ & $21.964_{\pm 0.352}$ & $20.697_{\pm 0.167}$ & $19.418_{\pm 0.222}$ & S \\
01333825+3042510 & 23.4092 & 30.7141 & 0.9057 & $19.378_{\pm 0.200}$ & $21.799_{\pm 0.069}$ & $20.407_{\pm 0.005}$ & $18.999_{\pm 0.201}$ & S \\
\hline
%\end{tabular}
%{\flushleft{ \textbf{Note:} (a) Does not include subtraction of 0.015 mag to correct CRNL for 2 dex between LMC and M33 Cepheids. \\ (b) Does not include addition of 0.069 mag to $m_H^W$ errors for intrinsic scatter. \par}}
\label{table:all_data}
\end{longtable*}

\newpage
\section{Postage stamps of cluster Cepheids}
\label{Apdx:stamps}

~ \\

%%% Confirmed Cepheids from cross-match:

\begin{figure*}[h]
\centering
\includegraphics[width=13.2cm]{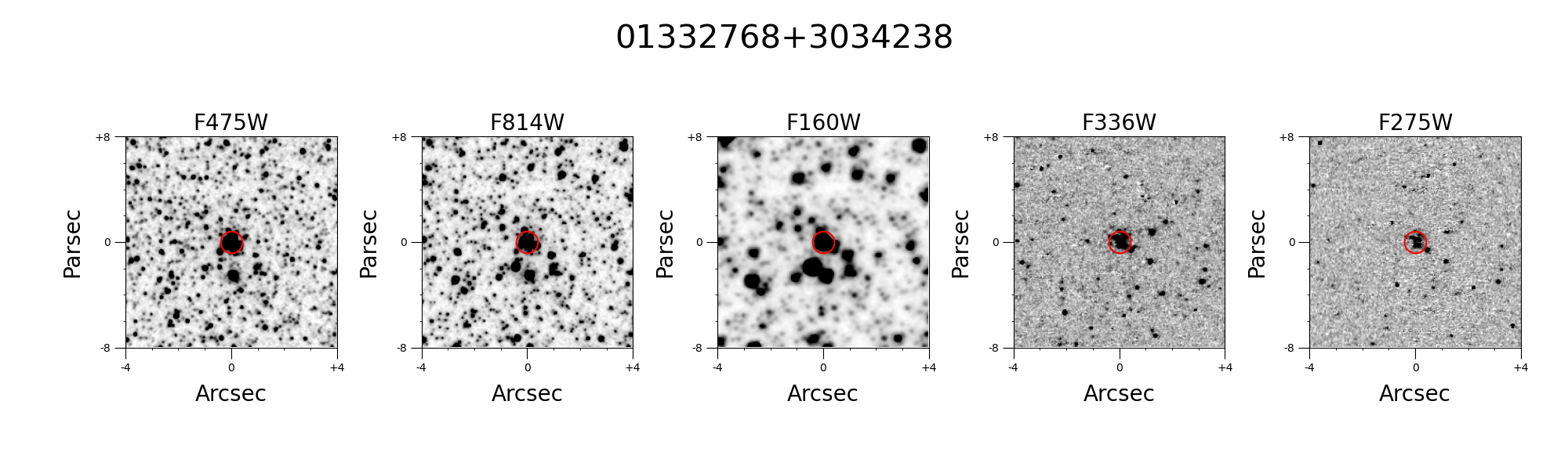}
\includegraphics[width=13.2cm]{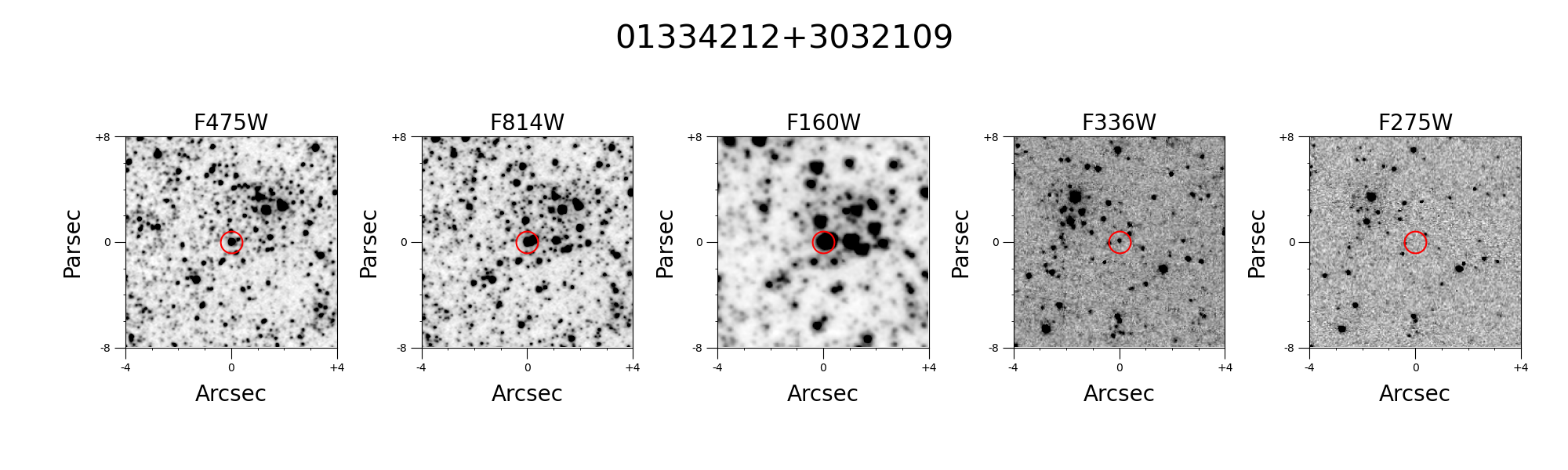} 
\includegraphics[width=13.2cm]{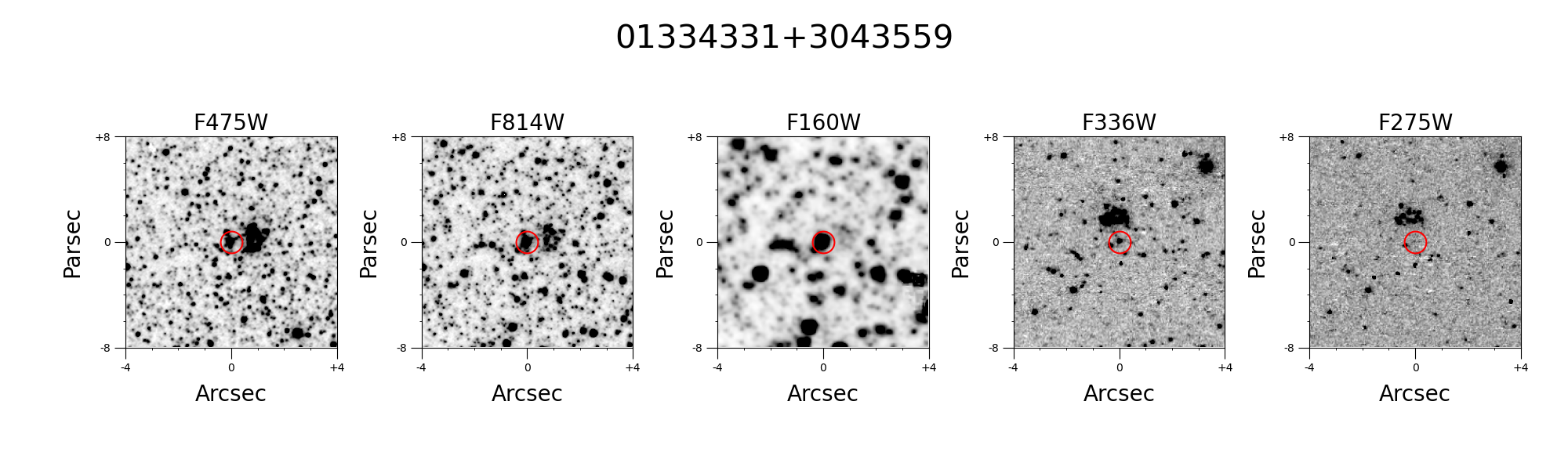}
\includegraphics[width=13.2cm]{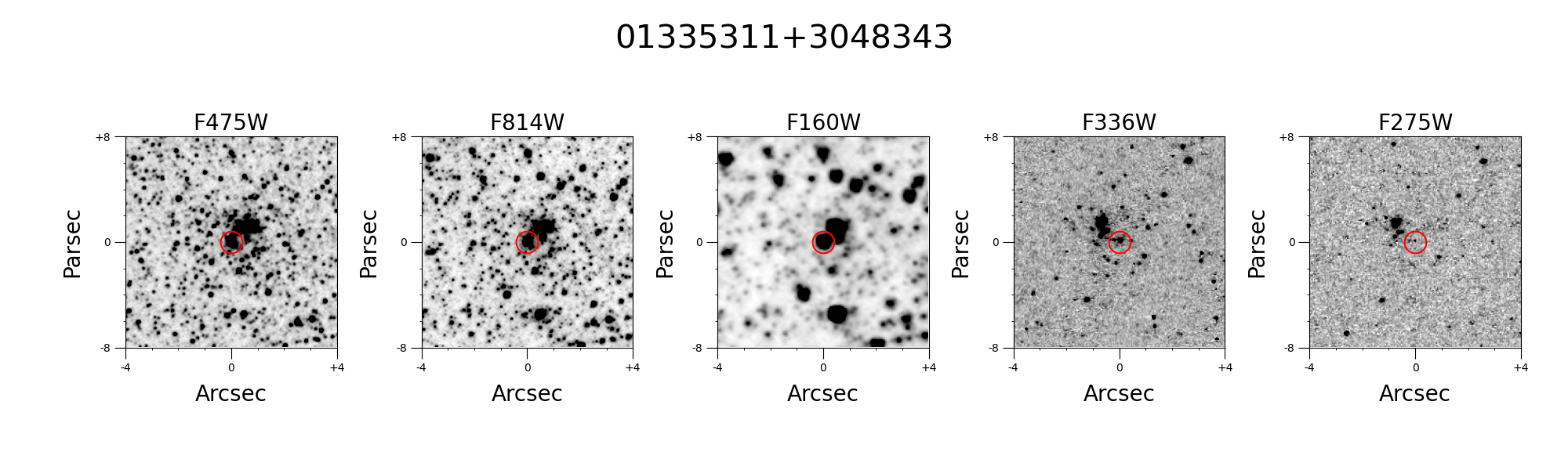} 
\includegraphics[width=13.2cm]{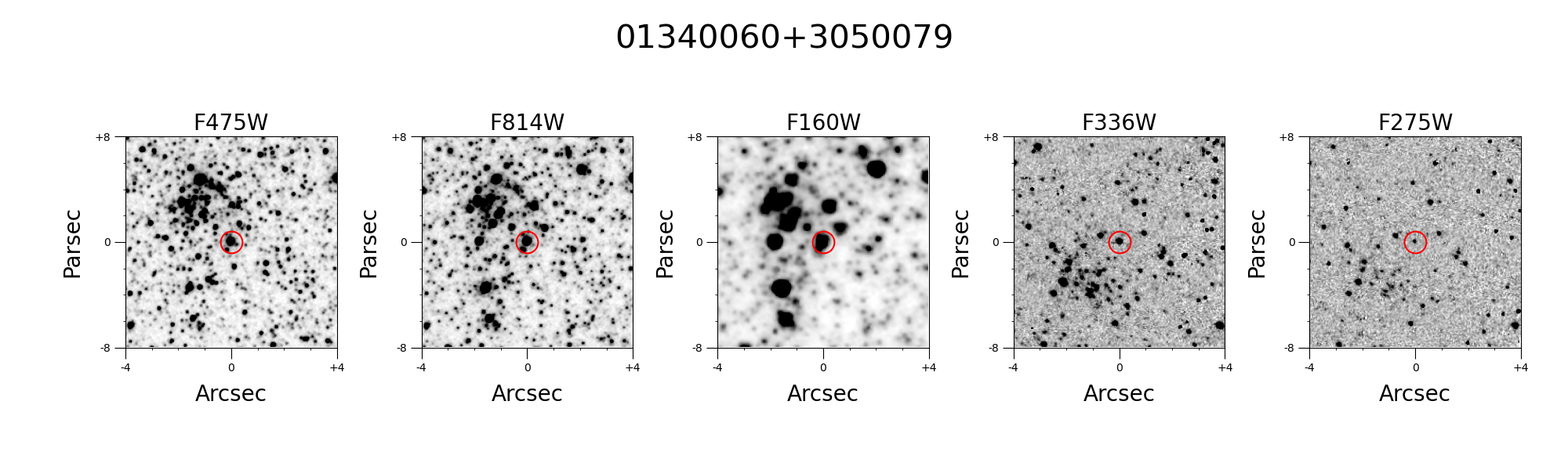}
\caption{Postage stamps in HST filters for the 10 Cluster Cepheids detected by the crossmatch procedure (\S\ref{sec:clusters}). }
 \label{fig:stamps_pairs}
 \end{figure*}

\begin{figure*}[t]
\centering
\includegraphics[width=15.0cm]{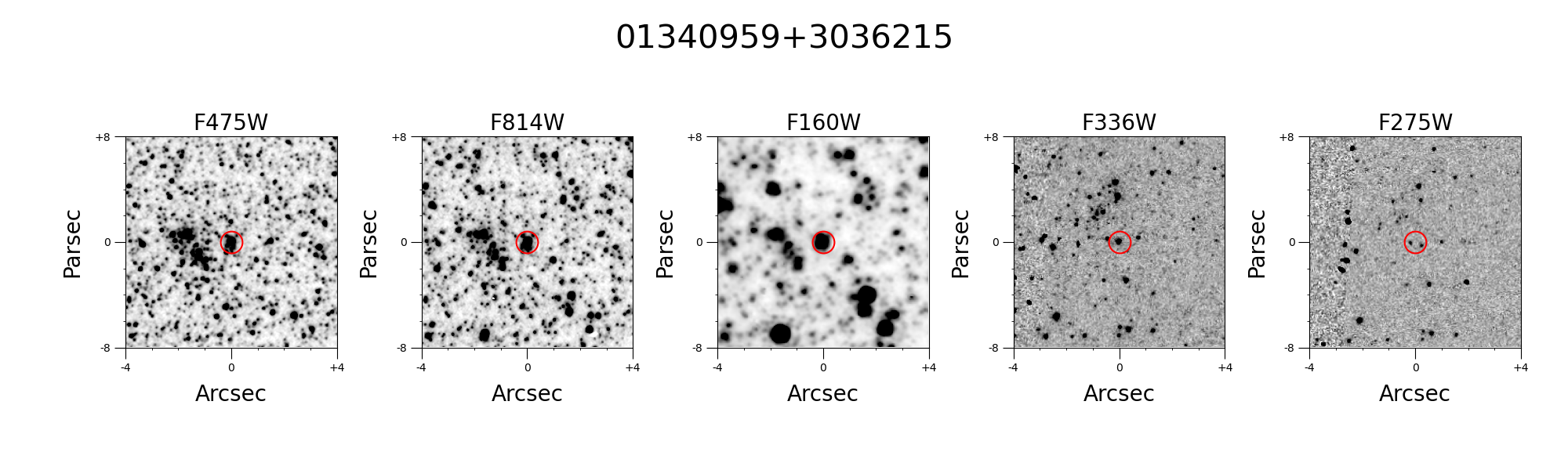} 
\includegraphics[width=15.0cm]{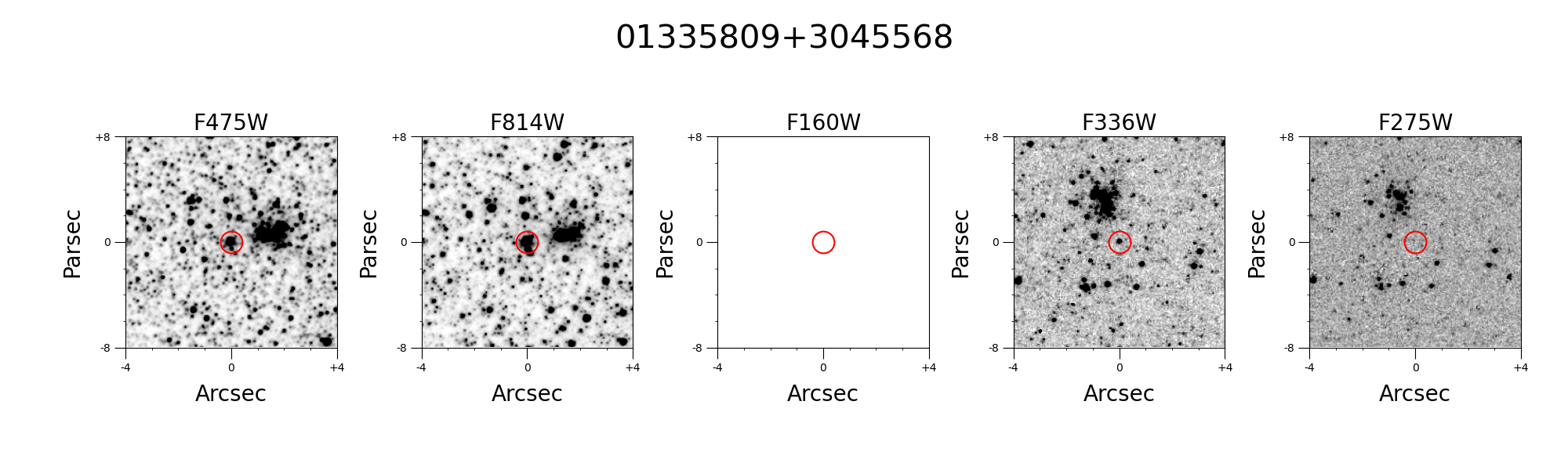}
\includegraphics[width=15.0cm]{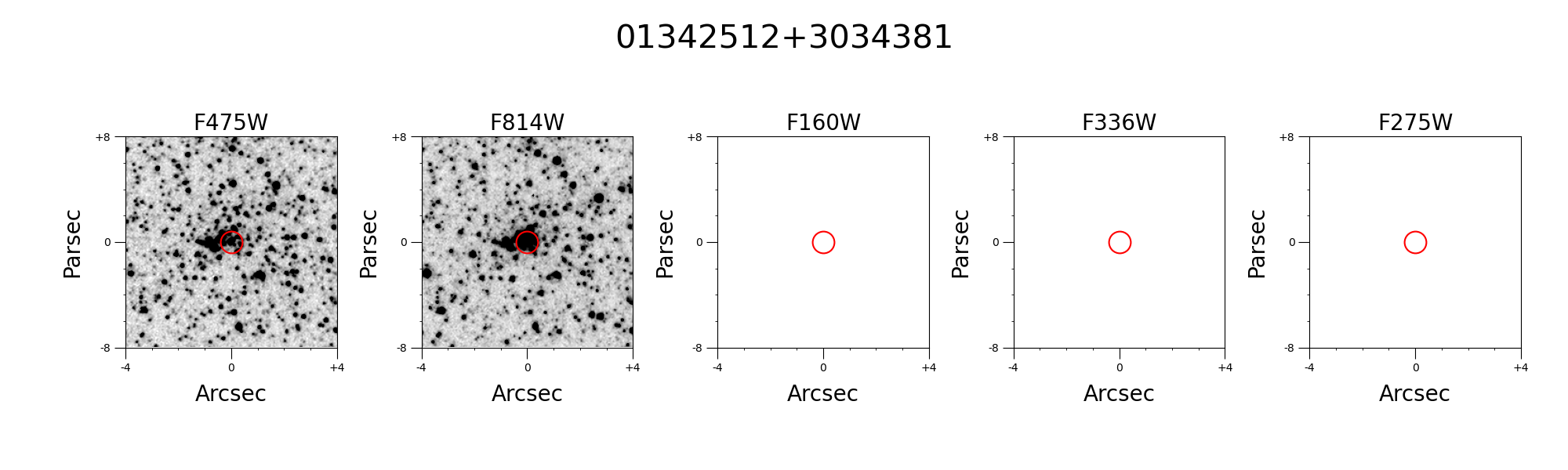} 
\includegraphics[width=15.0cm]{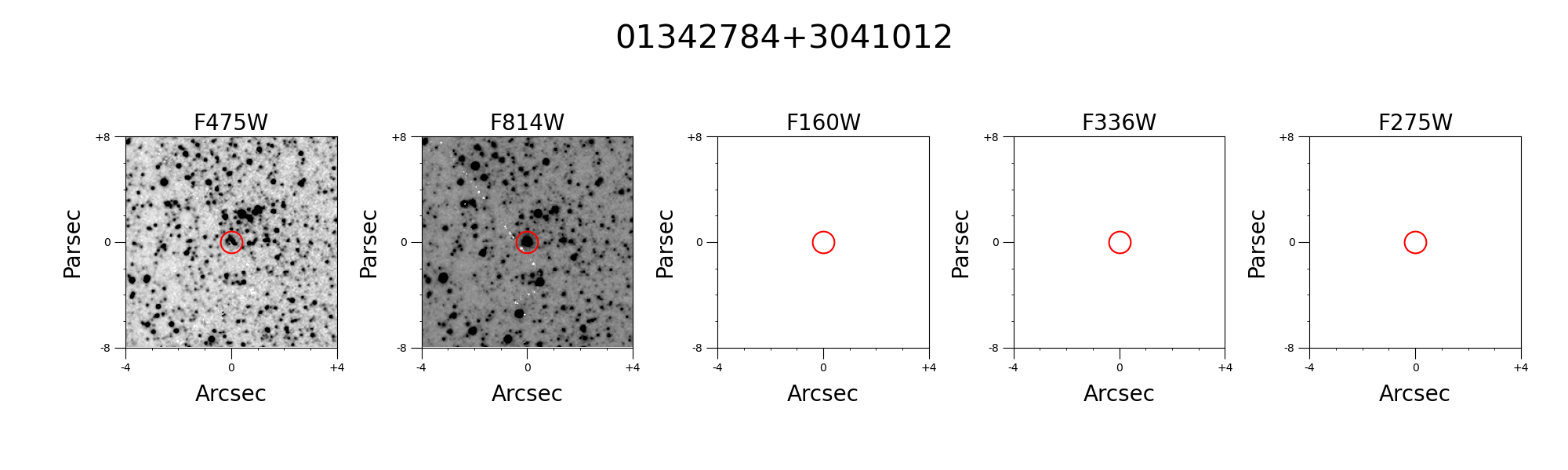}
\includegraphics[width=15.0cm]{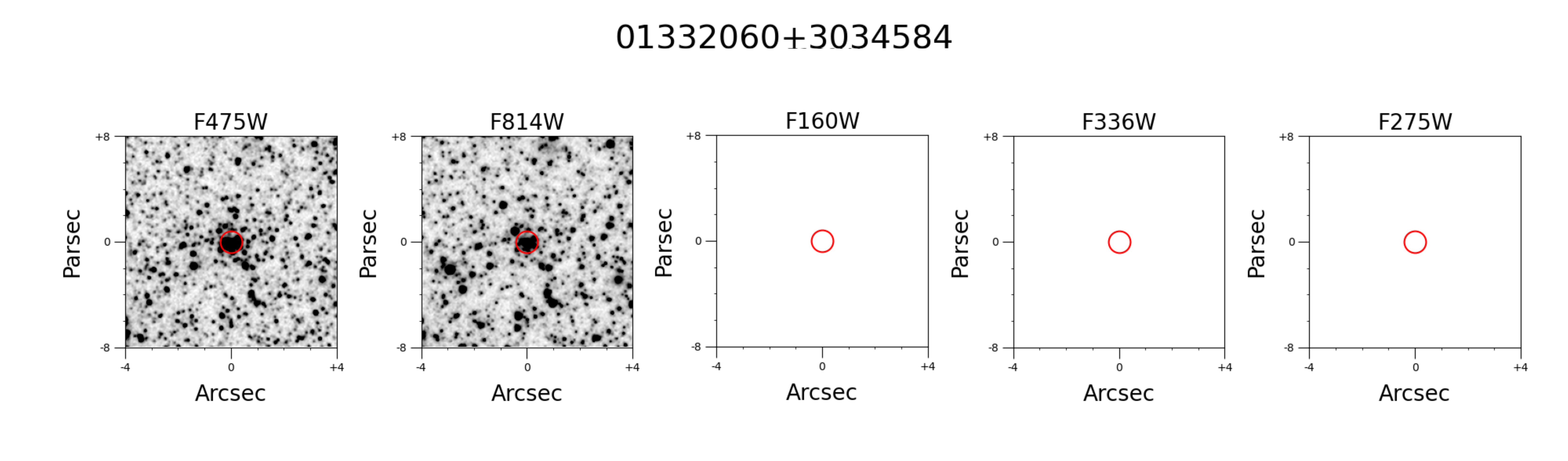} 
\caption{Continuation of Fig.~\ref{fig:stamps_pairs}. }
\label{fig:stamps_pairs2}
\end{figure*}

%%% Suspect Cepheids from stamps inspection:

\begin{figure*}[t!]
\centering
\includegraphics[width=15.0cm]{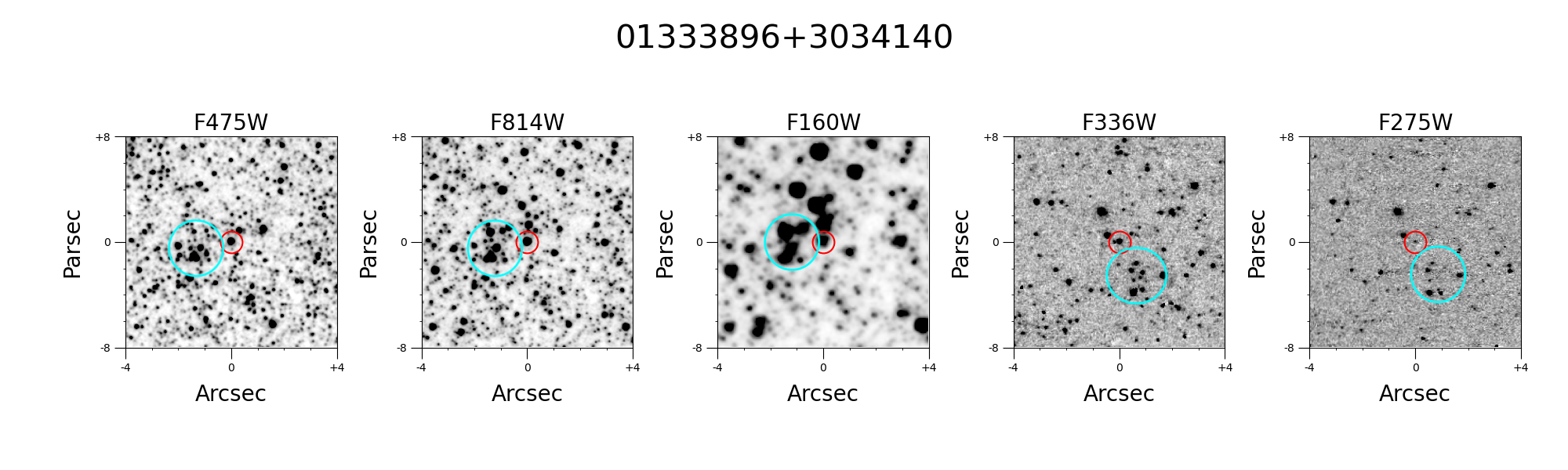}
\includegraphics[width=15.0cm]{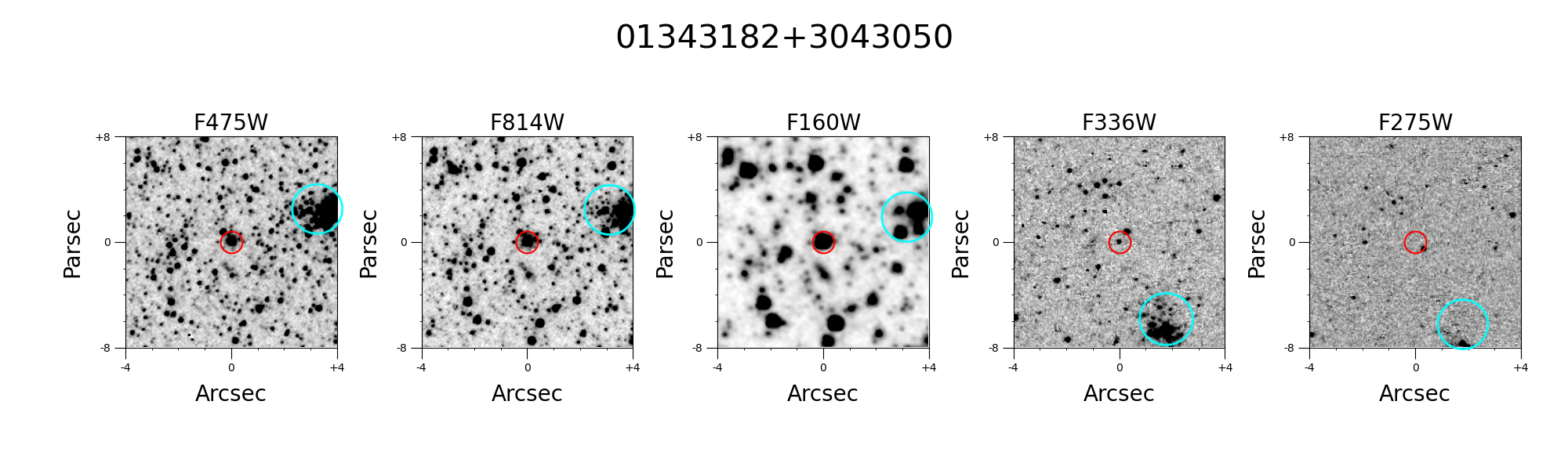} 
\includegraphics[width=15.0cm]{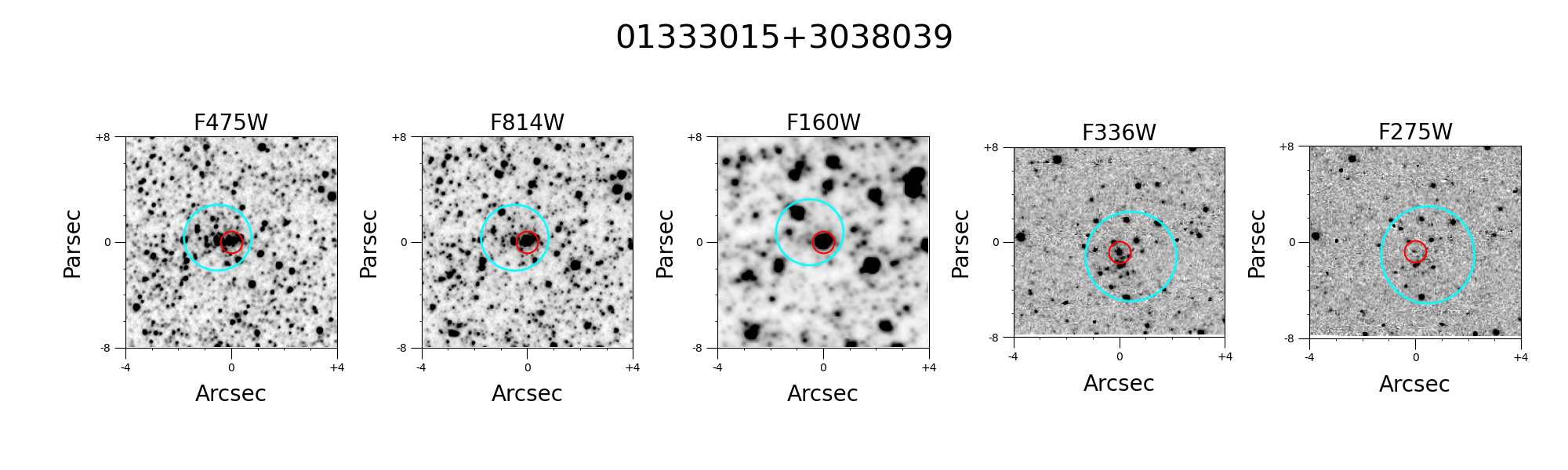}
\includegraphics[width=15.0cm]{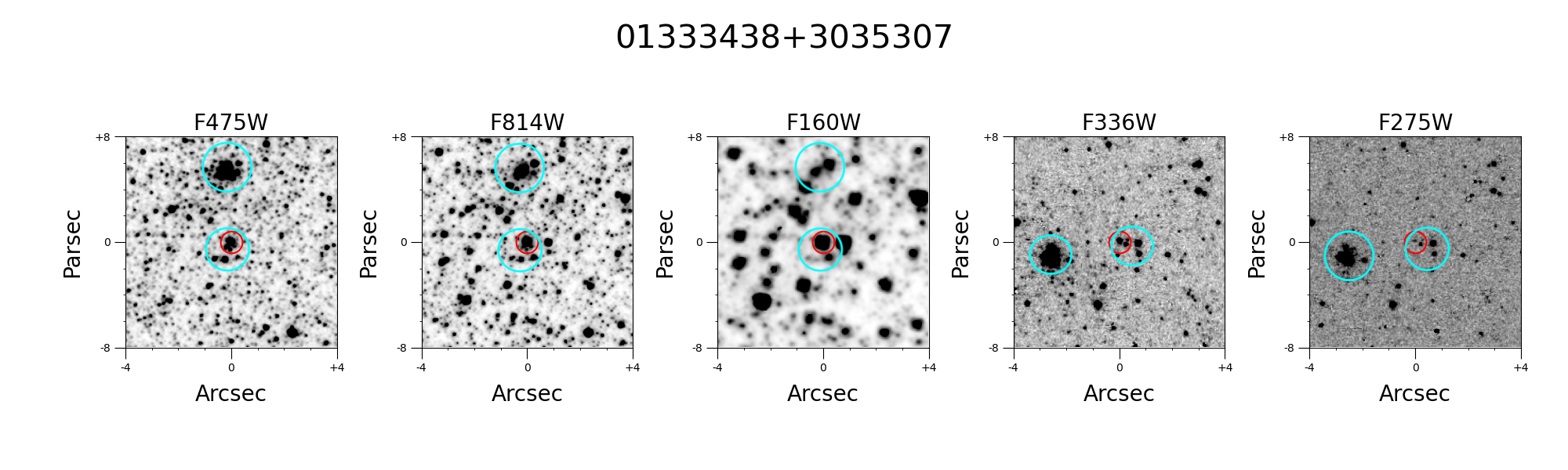} 
\includegraphics[width=15.0cm]{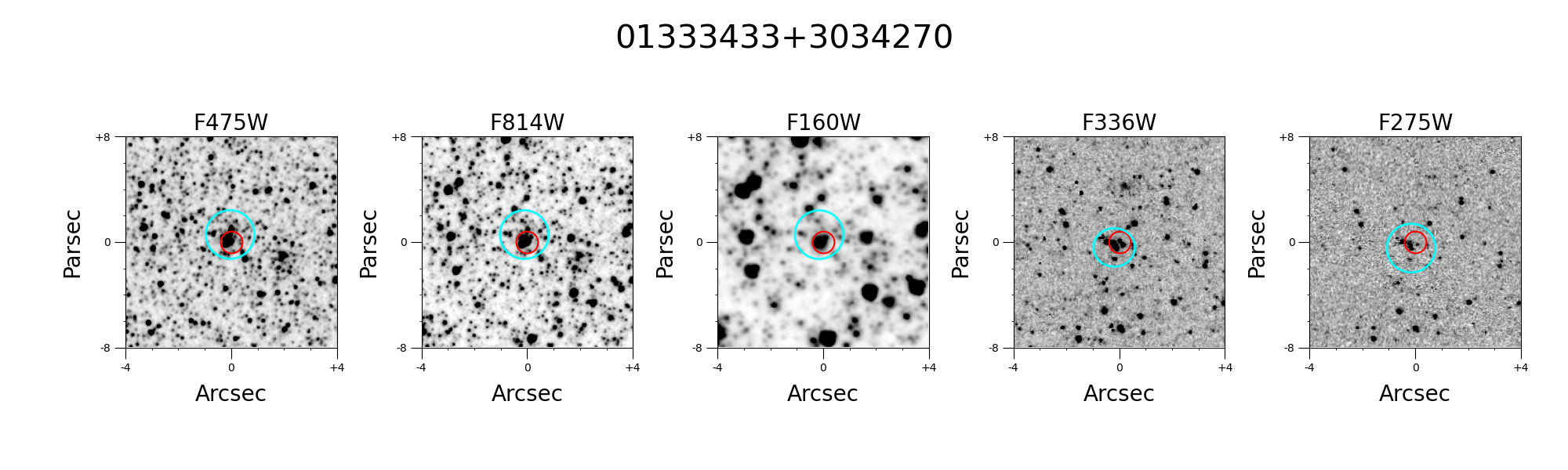}
\caption{Postage stamps in HST filters for the 13 additional cluster Cepheids found by visual inspection. The position of suspected clusters is shown in blue.}
\label{fig:stamps_likely}
\end{figure*}

\begin{figure*}[t!]
\centering
\includegraphics[width=15.0cm]{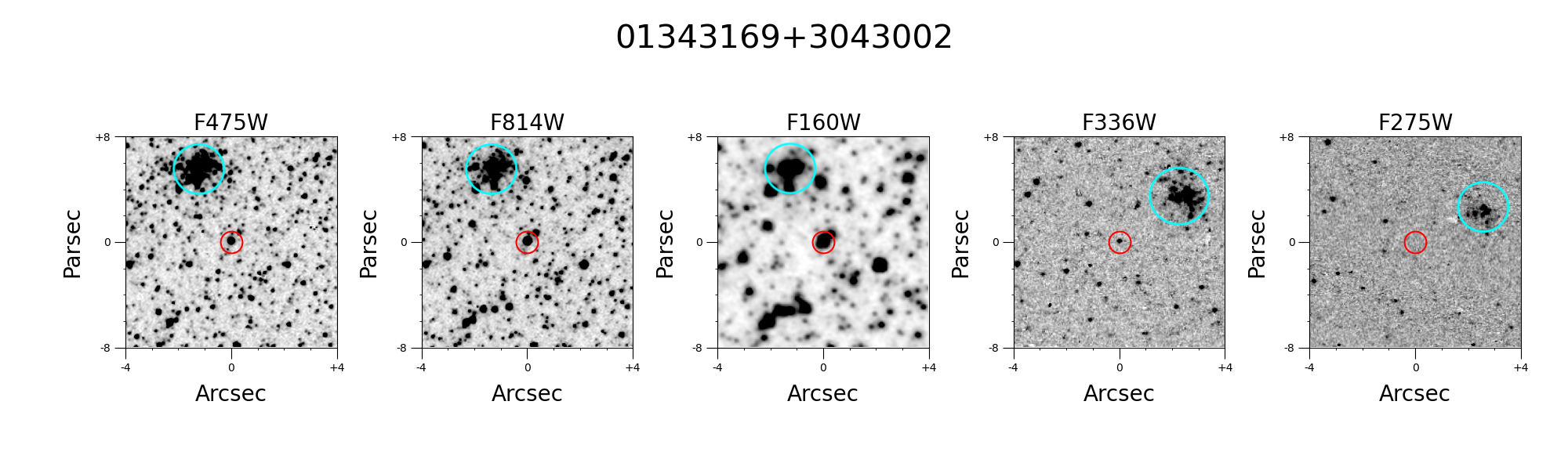} 
\includegraphics[width=15.0cm]{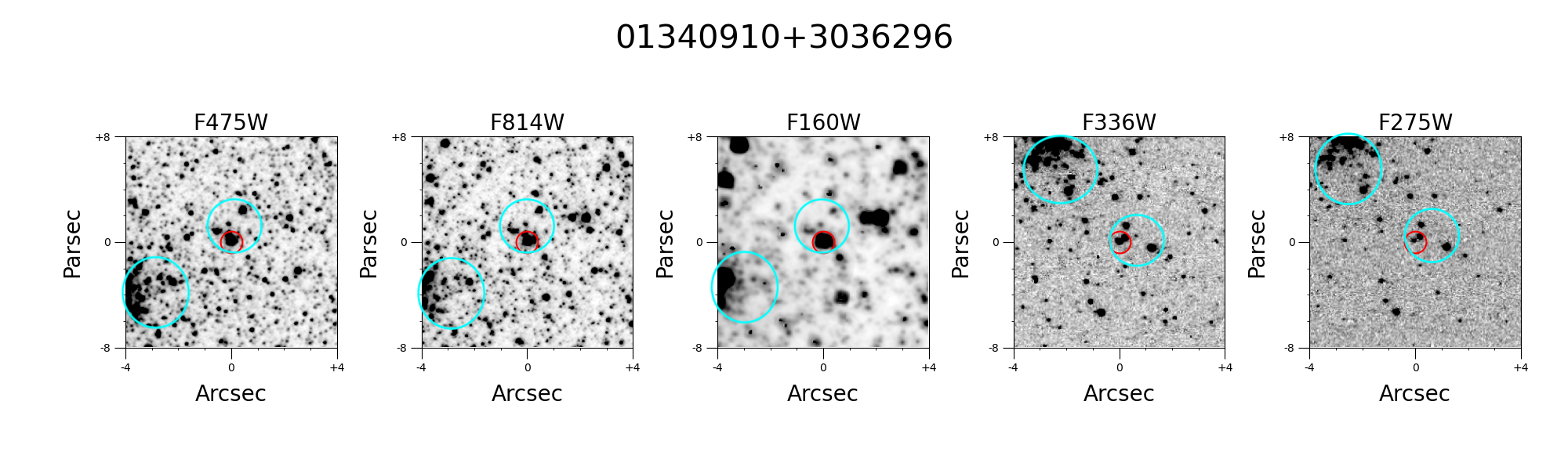}
\includegraphics[width=15.0cm]{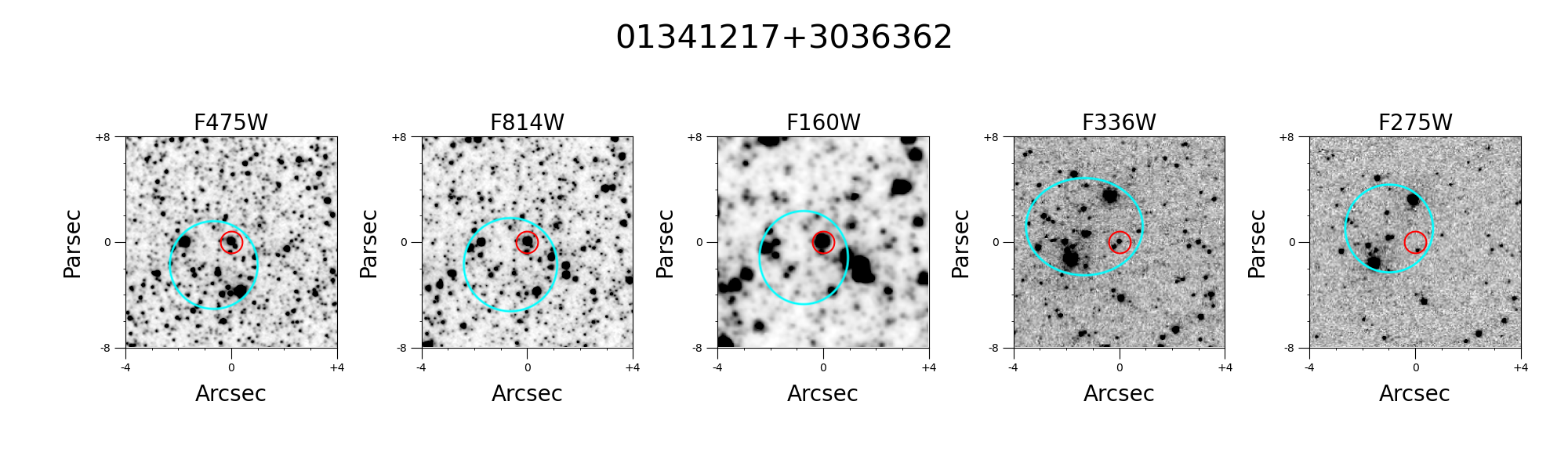} 
\includegraphics[width=15.0cm]{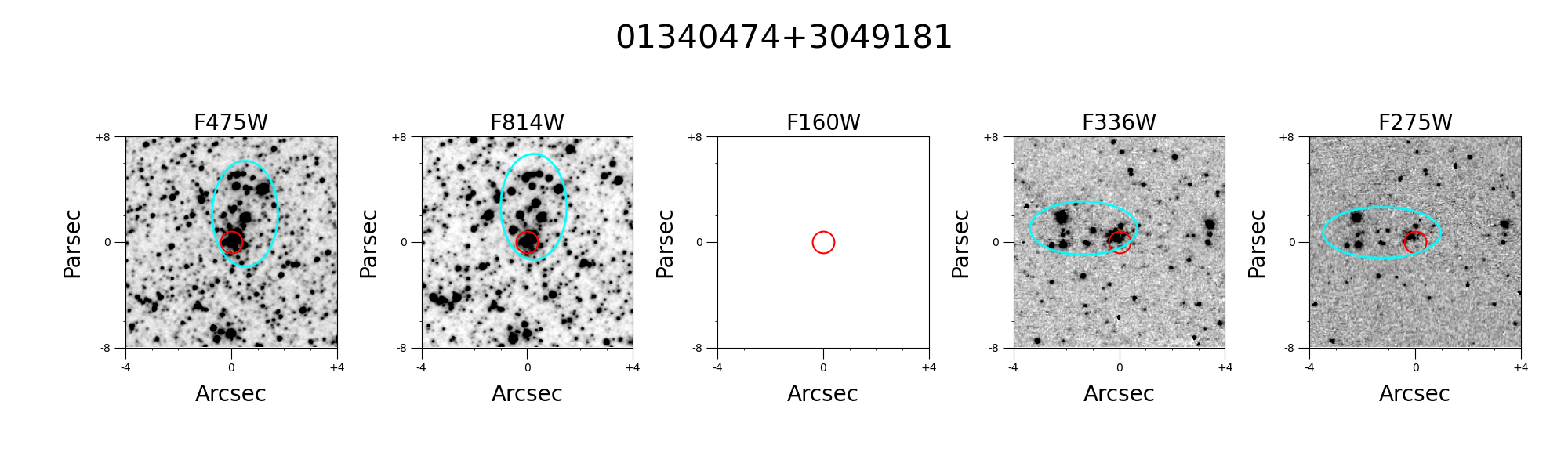}
\includegraphics[width=15.0cm]{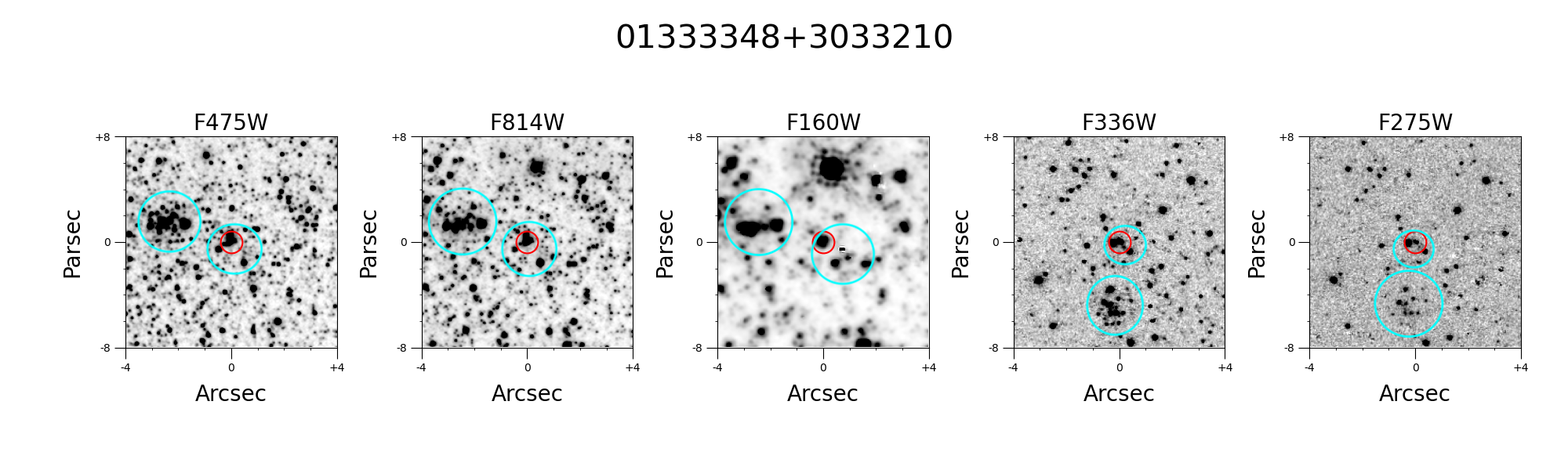} 
\caption{Continuation of Fig.~\ref{fig:stamps_likely}.}
\label{fig:stamps_likely2}
\end{figure*}

\begin{figure*}[t!]
\centering
\includegraphics[width=15.0cm]{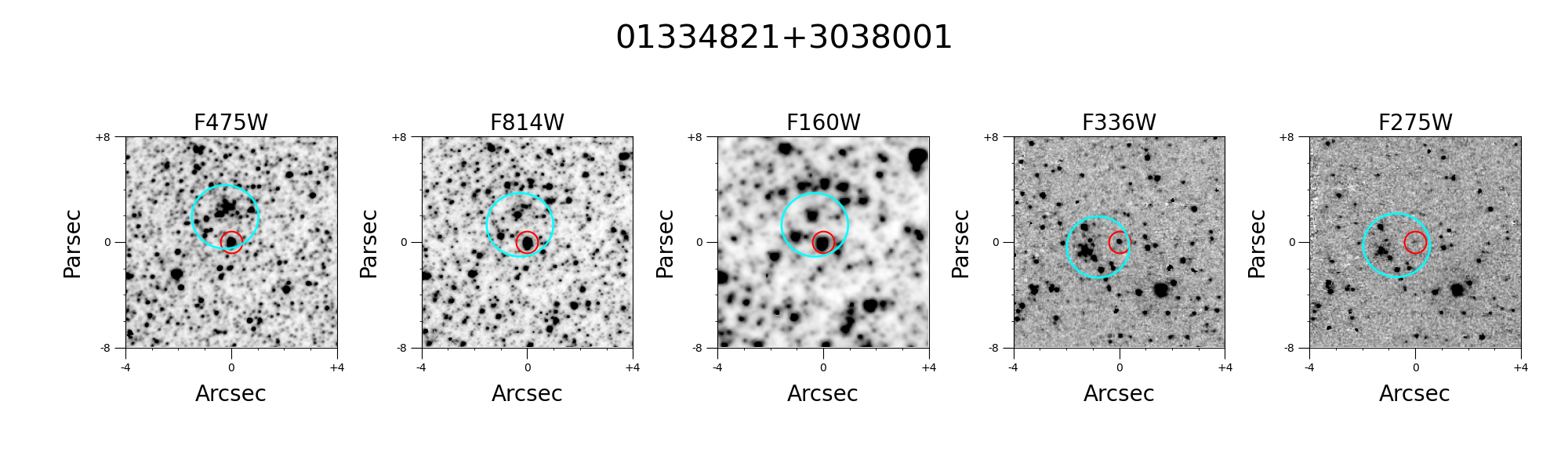}
\includegraphics[width=15.0cm]{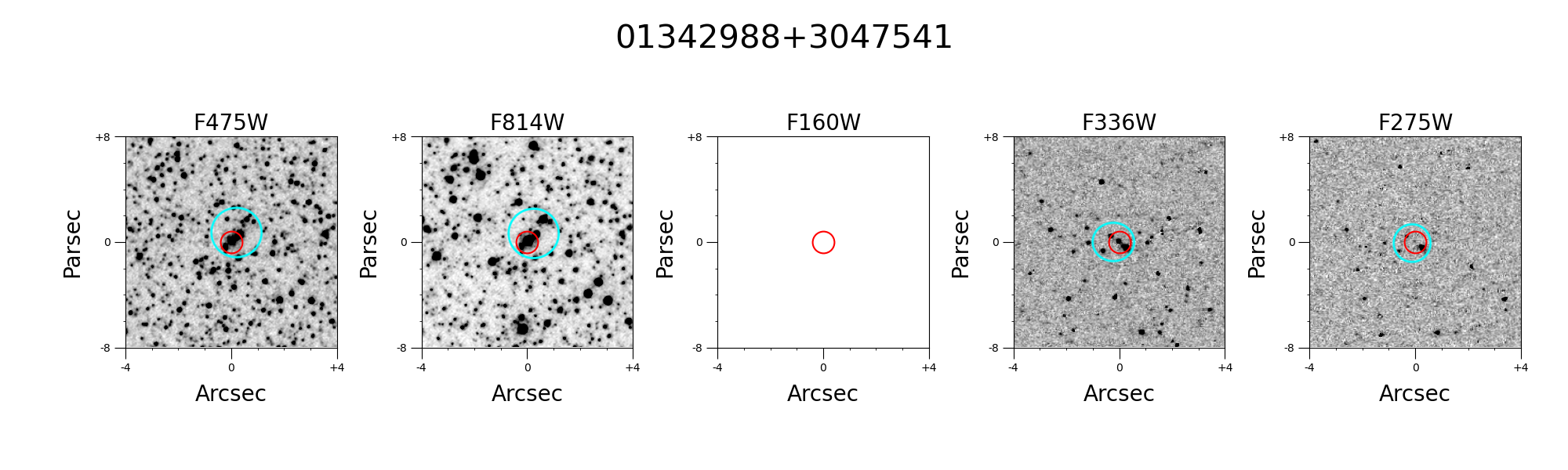} 
\includegraphics[width=15.0cm]{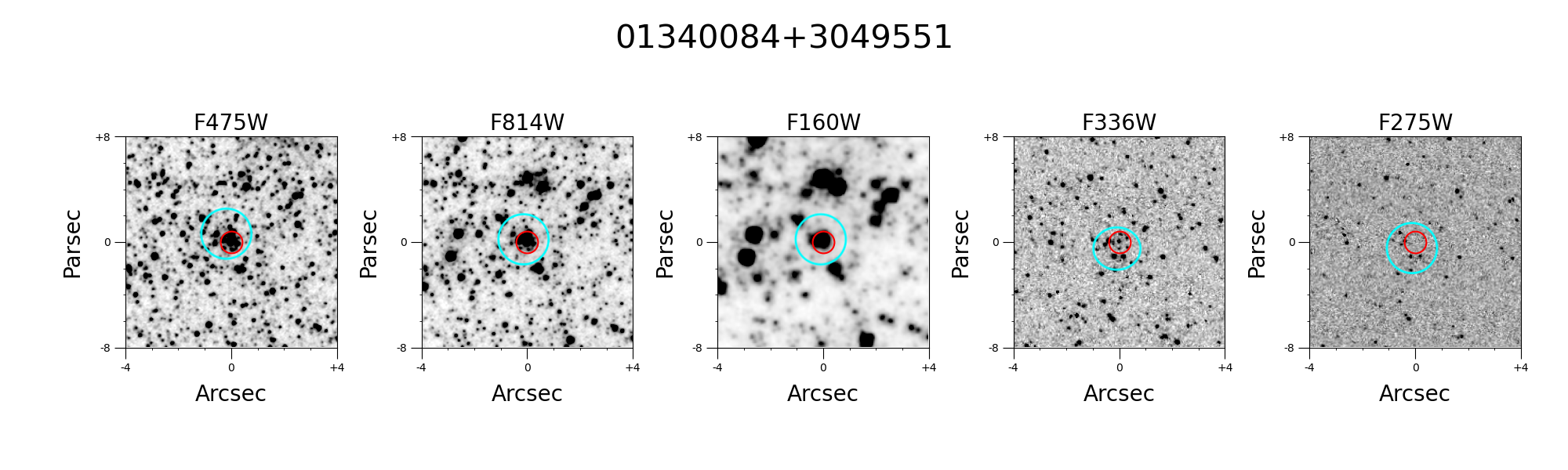}
\caption{Continuation of Fig.~\ref{fig:stamps_likely}.}
\label{fig:stamps_likely3}
\end{figure*}

\end{document}